\documentclass[twocolumn,numberedappendix,twocolappendix,appendixfloats]{openjournal_dhayaa}

\usepackage{amsmath}
\usepackage{fontawesome5}
\usepackage{color}
\usepackage{xcolor}
\usepackage{xspace}
\usepackage{enumitem}
\usepackage{multirow}
\usepackage{bigdelim}

\usepackage{natbib}
\setcitestyle{aysep={}}

\usepackage[colorlinks=true
  ,urlcolor=blue
  ,anchorcolor=blue
  ,citecolor=blue
  ,filecolor=blue
  ,linkcolor=blue
  ,menucolor=blue
  ,linktocpage=true
  ,pdfproducer=medialab
  ,pdfa=true
]{hyperref}

\usepackage{graphicx}	% Including figure files
\usepackage{amsmath}	% Advanced maths commands

\usepackage{fontawesome5}
\usepackage{color}
\usepackage{xcolor}

\newcommand{\eg}{{\sl e.g.}, }   
\newcommand{\ie}{{\sl i.e.}, }

\newcommand{\msun}{\ensuremath{\, {\rm M}_\odot}} 
\newcommand{\kpc}{\ensuremath{\, {\rm kpc}}}         
\newcommand{\mpc}{\ensuremath{\, {\rm Mpc}}}

\newcommand{\Mtwohc}{M_{\rm 200c}}
\newcommand{\Rtwohc}{R_{\rm 200c}}
\newcommand{\lambdacyl}{\lambda_{\rm cyl}}
\newcommand{\lambdatrue}{\lambda_{\rm tr}}
\newcommand{\Lstar}{L_\star}

\newcommand{\zspec}{z_{\rm spec}}

\newcommand{\decade}{DECADE\xspace}

\newcommand{\redmagic}{\textsc{redMaGiC}\xspace}
\newcommand{\redmapper}{\textsc{redMaPPer}\xspace}

\definecolor{orcidlogocol}{HTML}{A6CE39}
\definecolor{purple}{RGB}{128, 0, 128}
\definecolor{maroon}{RGB}{128, 0, 0}

\newcommand{\OrcidID}[1]{ \href[urlcolor = red]{https://orcid.org/#1}{\textcolor{lightgray}{\faOrcid}}}
\newcommand{\OrcidIDName}[2]{\href{https://orcid.org/#1}{#2}}

\defcitealias{y3-shapecatalog}{\textsc{GS21}}
\defcitealias{paper1}{\textsc{Paper I}}
\defcitealias{paper2}{\textsc{Paper II}}
\defcitealias{paper3}{\textsc{Paper III}}
\defcitealias{paper4}{\textsc{Paper IV}}

\newcommand*{\vcenteredhbox}[1]{\begingroup
\setbox0=\hbox{#1}\parbox{\wd0}{\box0}\endgroup}

\makeatletter
\g@addto@macro\frontmatter@authorformat{%
  \emergencystretch=3em\relax
}
\makeatother

\begin{document}

% Title of the paper, and the short title which is used in the headers.
% Keep the title short and informative.
\title{A DECADE of Giants: A catalog of 42,000 optically identified galaxy clusters from 13,000 $\boldsymbol{\deg^2}$ of the sky}

\shortauthors{Anbajagane, To, Gaido et al.}
\shorttitle{Galaxy clusters from 13,000 $\deg^2$ of DECam data}

\author{\OrcidIDName{0000-0003-3312-909X}{D. Anbajagane}$^{\star\dagger}$ (\vcenteredhbox{\includegraphics[height=1.2\fontcharht\font`\B]{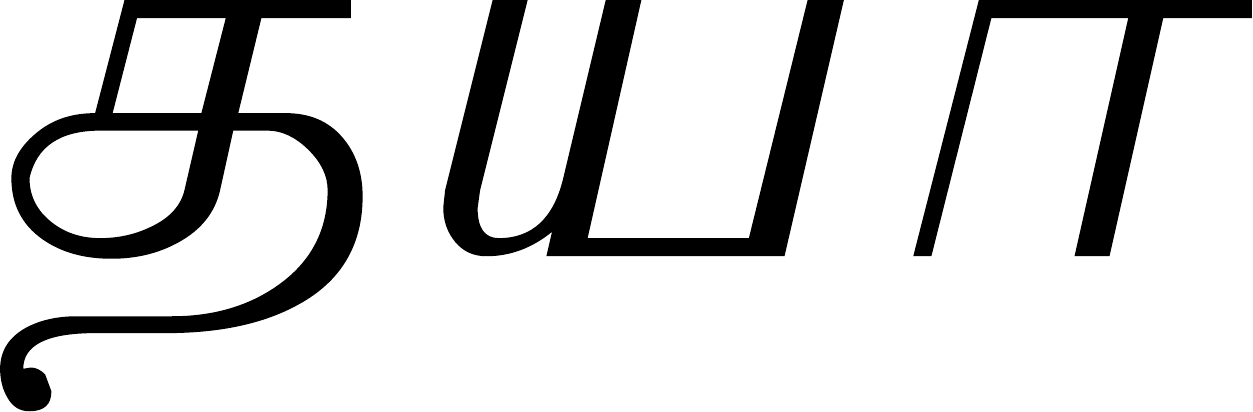}})}
\affiliation{Department of Astronomy and Astrophysics, University of Chicago, Chicago, IL 60637, USA}
\affiliation{Kavli Institute for Cosmological Physics, University of Chicago, Chicago, IL 60637, USA}
\thanks{$^\dagger$ NHFP Einstein fellow}

\author{\OrcidIDName{0000-0001-7836-2261}{C. To}$^\star$}
\affiliation{Department of Astronomy and Astrophysics, University of Chicago, Chicago, IL 60637, USA}
\affiliation{Kavli Institute for Cosmological Physics, University of Chicago, Chicago, IL 60637, USA}
\affiliation{NSF-Simons AI Institute for the Sky (SkAI), 172 E. Chestnut St., Chicago, IL 60611, USA}

\author{\OrcidIDName{0009-0001-3291-3010}{M. Gaido}$^\star$}
\affiliation{Kavli Institute for Cosmological Physics, University of Chicago, Chicago, IL 60637, USA}
\affiliation{Department of Physics, University of Chicago, Chicago, IL 60637, USA}

\author{\OrcidIDName{0000-0002-7887-0896}{C.~Chang}}
\affiliation{Department of Astronomy and Astrophysics, University of Chicago, Chicago, IL 60637, USA}
\affiliation{Kavli Institute for Cosmological Physics, University of Chicago, Chicago, IL 60637, USA}
\affiliation{NSF-Simons AI Institute for the Sky (SkAI), 172 E. Chestnut St., Chicago, IL 60611, USA}

\author{\OrcidIDName{0000-0001-8251-933X}{A.~Drlica-Wagner}}
\affiliation{Fermi National Accelerator Laboratory, P. O. Box 500, Batavia, IL 60510, USA}
\affiliation{Department of Astronomy and Astrophysics, University of Chicago, Chicago, IL 60637, USA}
\affiliation{Kavli Institute for Cosmological Physics, University of Chicago, Chicago, IL 60637, USA}
\affiliation{NSF-Simons AI Institute for the Sky (SkAI), 172 E. Chestnut St., Chicago, IL 60611, USA}

% Additional authors from both spreadsheets (alphabetical by surname)

\author{\OrcidIDName{0000-0002-6904-359X}{M.~Adamow}}
\affiliation{Center for Astrophysical Surveys, National Center for Supercomputing Applications, 1205 West Clark St., Urbana, IL 61801, USA}
\affiliation{Department of Astronomy, University of Illinois at Urbana-Champaign, 1002 W. Green Street, Urbana, IL 61801, USA}

\author{\OrcidIDName{0000-0001-8505-1269}{A.~Alarcon}}
\affiliation{Institute of Space Sciences (ICE, CSIC),  Campus UAB, Carrer de Can Magrans, s/n,  08193 Barcelona, Spain}

\author{\OrcidIDName{0000-0001-7774-2246}{M.~R.~Becker}}
\affiliation{Argonne National Laboratory, 9700 South Cass Avenue, Lemont, IL 60439, USA}

\author{\OrcidIDName{0000-0003-4383-2969}{C.~R.~Bom}}
\affiliation{Centro Brasileiro de Pesquisas F\'isicas, Rua Dr. Xavier Sigaud 150, 22290-180 Rio de Janeiro, RJ, Brazil}

\author{\OrcidIDName{0000-0002-3690-105X}{J.~A.~Carballo-Bello}}
\affiliation{Instituto de Alta Investigaci\'on, Universidad de Tarapac\'a, Casilla 7D, Arica, Chile}

\author{\OrcidIDName{0009-0005-1143-495X}{N.~Chicoine}}
\affiliation{Department of Physics and Astronomy, University of Pittsburgh, 3941 O’Hara Street, Pittsburgh, PA 15260}

\author{\OrcidIDName{0000-0003-1680-1884}{Y.~Choi}}
\affiliation{NSF NOIRLab, 950 N. Cherry Ave., Tucson, AZ 85719, USA}

\author{\OrcidIDName{0000-0001-8158-1449}{M.~Costanzi}}
\affiliation{Dipartimento di Fisica - Sezione di Astronomia, Università di Trieste, Via Tiepolo 11, 34131 Trieste, Italy}
\affiliation{Osservatorio Astronomico di Trieste, Via G. B. Tiepolo 11, 34143 Trieste, Italy}

\author{\OrcidIDName{0000-0001-6957-1627}{P.~S.~Ferguson}}
\affiliation{DIRAC Institute, Department of Astronomy, University of Washington, 3910 15th Ave NE, Seattle, WA, 98195, USA}

\author{\OrcidIDName{0000-0001-6134-8797}{M.~Gatti}}
\affiliation{Institut de Ci`encies de l’Espai, IEEC-CSIC, Campus UAB, Facultat de Ci`encies, Torre C5 par-2, Barcelona 08193, Spain}

\author{\OrcidIDName{0000-0003-4937-8453}{P.~A.~Giles}}
\affiliation{Department of Physics and Astronomy, University of Sussex, Brighton, BN1 9QH, UK}

\author{\OrcidIDName{0000-0002-4577-8217}{S.~Grandis}}
\affiliation{Universit\"at Innsbruck, Institut f\"ur Astro- und Teilchenphysik, Technikerstr. 25/8, 6020 Innsbruck, Austria}

\author{\OrcidIDName{0000-0003-3270-7644}{D.~Gruen}}
\affiliation{University Observatory, Faculty of Physics, Ludwig-Maximilians-Universität, Scheinerstr. 1, 81679 Munich, Germany}
\affiliation{Excellence Cluster ORIGINS, Boltzmannstr. 2, 85748 Garching, Germany}

\author{\OrcidIDName{0000-0002-4588-6517}{R.~A.~Gruendl}}
\affiliation{Center for Astrophysical Surveys, National Center for Supercomputing Applications, 1205 West Clark St., Urbana, IL 61801, USA}
\affiliation{Department of Astronomy, University of Illinois at Urbana-Champaign, 1002 W. Green Street, Urbana, IL 61801, USA}

\author{\OrcidIDName{0000-0002-4563-5101}{W.~G.~Hartley}}
\affiliation{Department of Astronomy, University of Geneva, ch. d’Ecogia 16, 1290 Versoix, Switzerland}

\author{\OrcidIDName{0000-0003-4394-7491}{G.~Herron}}
\affiliation{Department of Physics and Astronomy, Dartmouth College, Hanover, NH 03755, USA}

\author{\OrcidIDName{0000-0001-5160-4486}{David ~J.~ James}}
\affiliation{Astraveo LLC, PO Box 1668, Gloucester, MA 01931, USA}
\affiliation{Applied Materials, 35 Dory Road, Gloucester, MA 01930, USA}

\author{\OrcidIDName{0000-0001-6089-0365}{T.~E.~Jeltema}}
\affiliation{Department of Physics, University of California, Santa Cruz, Santa Cruz, CA 95064}

\author{\OrcidIDName{0000-0002-8998-3909}{N.~MacCrann}}
\affiliation{DAMTP, Centre for Mathematical Sciences, University of Cambridge, Wilberforce Road, Cambridge CB3 OWA, UK}
\affiliation{Kavli Institute for Cosmology Cambridge, Madingley Road, Cambridge CB3 0HA, UK}

\author{\OrcidIDName{0009-0005-0523-5540}{M.~Navabi}}
\affiliation{Department of Physics, University of Surrey, Guildford GU2 7XH, UK}

\author{\OrcidIDName{0000-0002-9144-7726}{C.~E.~Mart\'inez-V\'azquez}}
\affiliation{NSF NOIRLab, 670 N. A'ohoku Place, Hilo, Hawai'i, 96720, USA}

\author{\OrcidIDName{0000-0003-3519-4004}{S.~Mau}}
\affiliation{Department of Physics, Duke University Durham, NC 27708, USA}

\author{\OrcidIDName{0000-0002-4475-3456}{J.~McCullough}}
\affiliation{Department of Astrophysical Sciences, Princeton University, Peyton Hall, Princeton, NJ 08544, USA}

\author{\OrcidIDName{0000-0001-9649-4815}{B.~Mutlu-Pakdil}}
\affiliation{Department of Physics and Astronomy, Dartmouth College, Hanover, NH 03755, USA}

\author{\OrcidIDName{0000-0002-8282-469X}{N.~E.~D.~Noël}}
\affiliation{Department of Physics, University of Surrey, Guildford GU2 7XH, UK}

\author{\OrcidIDName{0000-0002-6021-8760}{A.~B.~Pace}}
\affiliation{Department of Astronomy, University of Virginia, 530 McCormick Road, Charlottesville, VA 22904, USA}

\author{\OrcidIDName{0000-0002-5933-5150}{J.~Prat}}
\affiliation{Departament de F\'{\i}sica, EEBE, Universitat Polit\`ecnica de Catalunya, c/ Eduard Maristany 16, E-08019 Barcelona, Spain}

\author{\OrcidIDName{0000-0001-5805-5766}{A.~H.~Riley}}
\affiliation{Lund Observatory, Division of Astrophysics, Department of Physics, Lund University, SE-221 00 Lund, Sweden}

\author{\OrcidIDName{0000-0002-9328-879X}{A.~K.~Romer}}
\affiliation{Department of Physics and Astronomy, University of Sussex, Brighton, BN1 9QH, UK}

\author{\OrcidIDName{0000-0003-4102-380X}{D.~J.~Sand}}
\affiliation{Steward Observatory, University of Arizona, 933 North Cherry Avenue, Tucson, AZ 85721-0065, USA}

\author{\OrcidIDName{0000-0003-3402-6164}{L.~Santana-Silva}}
\affiliation{Observatório do Valongo/UFRJ, Ladeira do Pedro Antônio, 43 - Centro, Rio de Janeiro - RJ, 20080-090, Brazil}

\author{\OrcidIDName{0000-0002-6002-4288}{L.~F.~Secco}}
\affiliation{Kavli Institute for Cosmological Physics, University of Chicago, Chicago, IL 60637, USA}

\author{\OrcidIDName{0000-0002-1831-1953}{I.~Sevilla-Noarbe}}
\affiliation{Centro de Investigaciones Energéticas Medioambientales y Tecnológicas CIEMAT, Av Complutense 40, 28040, Madrid, Spain}

\author{\OrcidIDName{0000-0002-6389-5409}{T.~Shin}}
\affiliation{Department of Physics, Carnegie Mellon University, 5000 Forbes Ave., Pittsburgh, PA 15213, USA}

\author{\OrcidIDName{0000-0003-1479-3059}{G.~S.~Stringfellow}}
\affiliation{University of Colorado Boulder, Boulder, CO 80309, USA}

\author{\OrcidIDName{0000-0003-0478-0473}{C.~Y.~Tan}}
\affiliation{Department of Physics, University of Chicago, Chicago, IL 60637, USA}
\affiliation{Kavli Institute for Cosmological Physics, University of Chicago, Chicago, IL 60637, USA}

\author{\OrcidIDName{0000-0002-5279-0230}{R.~Teixeira}}
\affiliation{Department of Physics, Duke University Durham, NC 27708, USA}

\author{\OrcidIDName{0000-0002-7904-1707}{H.-Y.~Wu}}
\affiliation{Department of Physics, Southern Methodist University, Dallas, TX 75205, USA}

\author{\OrcidIDName{0000-0001-6455-9135}{A.~Zenteno}}
\affiliation{Cerro Tololo Inter-American Observatory/NSF NOIRLab, Casilla 603, La Serena, Chile}

\author{\OrcidIDName{0000-0001-5969-4631}{Y.~Zhang}}
\affiliation{NSF NOIRLab, 950 N Cherry Avenue, Tucson, AZ, 85719}

\author{\OrcidIDName{0000-0002-7523-582X}{Z.~Zhang}}
\affiliation{Department of Physics, Stanford University, 382 Via Pueblo Mall, Stanford, CA 94305, USA}
\affiliation{SLAC National Accelerator Laboratory, Menlo Park, CA 94025, USA}

\author{\OrcidIDName{0000-0001-9789-9646}{J.~Zuntz}}
\affiliation{Institute for Astronomy, University of Edinburgh, Blackford Hill, Edinburgh EH9 3HJ, United Kingdom}

\email{$^{\star}$dhayaacosmo@gmail.com}

% Abstract of the paper
\begin{abstract}
Galaxy clusters are sensitive to a rich variety of physics spanning a wide dynamic range, from microphysical to cosmological scales. Optical datasets have been efficient both at identifying clusters and at studying their properties. We leverage 13,000 $\deg^2$ of the sky imaged by the Dark Energy Camera (DECam) and extract over 40,000 clusters using the \redmapper algorithm. Our catalog spans $0.2 < z < 0.9$ and $\Mtwohc \gtrsim 10^{14} \msun/h$, with four-band ($griz$) photometric redshifts that have a precision $\sigma_z / (1 + z) \leq 0.008$ below $z < 0.65$, and $\leq 0.011$ between $0.65 < z < 0.9$ and that exhibit an outlier rate of $f_{\rm out} < 0.2\%$. Our cluster finding is calibrated using spectroscopic data from the Dark Energy Spectroscopic Instrument, but we find that a larger spectroscopic sample does not improve the performance of our cluster redshift estimation. We cross-match our catalog to those produced from millimeter-wave and X-ray data, and extract scaling relations between various cluster observables. There is excellent agreement between our estimates for different subsets of the 13,000 $\deg^2$ dataset, and also with results from prior catalogs and simulations. We use our joint analysis of millimeter-wave and optical data to estimate the halo masses of individual clusters. We also provide lensing profiles, measured for each cluster using overlapping galaxy shape catalogs, to enable additional mass-calibration work from the community. We demonstrate from a suite of tests that spatial inhomogeneity in the quality of the imaging data does not have an appreciable systematic impact on the properties of this catalog. We also test a cylindrical richness observable, which is defined such that it is significantly simpler to model, and show that it has precision competitive with the traditional optical richness observable. All data products and measurements are made publicly available.
\end{abstract}

\section{Introduction}
Cosmic structure is a powerful probe of cosmological and astrophysical processes in our Universe.  Galaxy clusters --- gravitationally bound clusters of galaxies --- are uniquely powerful in constraining such processes \citep{Allen:2011:Clusters}. These clusters are ``extreme'' structures in our cosmological context, commonly viewed as the most massive gravitationally bound objects in our Universe, and their number densities are highly sensitive to the initial state and subsequent evolution of the cosmic density field \citep{Kravtsov:2012:Clusters}. Galaxy clusters are also astrophysically dynamic, with their matter distribution governed by processes ranging from cosmic-ray transport to star formation to large-scale collisionless shocks \citep[\eg][]{Kravtsov:2012:Clusters, Walker2019OutskirtsReview, Ruszkowski:2023:CRClusters}.

Optical datasets have long been at the forefront of cluster detection and characterization; the earliest written recognition of a galaxy cluster was provided in Charles Messier's catalog in 1781 \citep{Messier1781, Biviano:2000:history}. Optical cluster catalogs are assembled by identifying galaxy overdensities in optical surveys. One of the most notable early catalogs of massive galaxy clusters came from \citet{Abell:1958:clusters}, and more modern data sets such as the Sloan Digital Sky Survey \citep[SDSS,][]{York_2000} have enabled substantially larger and more homogeneous, algorithmically selected optical samples \citep{Koester:2007:maxBCG,Rykoff:2014:AlgorithmCatalog}. Modern cluster catalogs are a multi-wavelength enterprise, with precision samples being selected using optical, millimeter-wave (mm-wave), and X-ray observations \citep[\eg][]{Rykoff:2016:redmapper, Bleem:2026:Clusters, Bulbul:2024:eROSITA, Hilton:2026:ACTDR6}. These three classes of data trace different components of clusters and are highly complementary in the halo-mass and redshift ranges that they probe \citep{Allen:2011:Clusters}.

Clusters are identified in the mm-wave using the thermal Sunyaev Zeldovich effect \citep[tSZ,][]{Sunyaev1972SZEffect}, which is the inverse Compton scattering of Cosmic Microwave Background (CMB) photons off hot electron plasma in the cluster. This approach enables cluster detection up to $z \sim 2$ as the tSZ surface brightness does not dilute with redshift. However, the steep scaling of the integrated tSZ signal with mass, ${\rm tSZ} \propto M^{5/3}$, means only the most massive structures ($\Mtwohc \gtrsim 10^{14.5} \msun/h$) can be detected. X-ray datasets, on the other hand, are particularly well-suited for detecting lower-mass systems (down to group scales of $\Mtwohc \approx 10^{13.5} \msun/h$) at lower redshifts; high-redshift measurements are challenging as the X-ray surface brightness suffers from cosmic dimming, which scales as $(1 + z)^{-4}$. Optical datasets provide a unique bridge between the two --- they probe a wide range of halo masses $(\Mtwohc \gtrsim 10^{14} \msun/h$) at fairly high redshifts ($z \sim 0.7$).

In this work, we leverage imaging data from the Dark Energy Camera \citep[DECam,][]{Flaugher2005} processed by the Dark Energy Camera All Data Everywhere (DECADE) cosmic shear project \citep{paper1, paper5}. These data are publicly available\footnote{\url{https://datalab.noirlab.edu/data/delve}} under Data Release 3 (DR3) of the Dark Energy Local Volume Exploration (DELVE) survey \citep{Drlica-Wagner:2021}. DELVE DR3 also contains the Year 6 ``Gold'' cosmology data set \citep{Bechtol:2025:Y6GOLD} from the Dark Energy Survey \citep[DES,][]{DESDR2}. We can treat the DECADE and DES data on the same footing since both sets of imaging data were collected by the same instrument and were processed using the same pipelines, which we detail below. The galaxy cluster sample produced from the combined datasets is one of the largest galaxy cluster catalogs to date, containing over 42,000 (560,000) clusters with richness $\lambda > 20$ ($\lambda > 5$), where ``richness'' is analogous to cluster member galaxy counts and is defined further below. This represents nearly a three-fold increase compared to the cluster catalog from DES Year 3 (Y3), which has 16,000 clusters with $\lambda > 20$ \citep{DES:2025:6x2pt_Y6}. The wide-area coverage makes this catalog an apt avenue to prepare for analyses of data from the Vera C. Rubin Observatory \citep{Ivezi__2019}, the Nancy Grace Roman Space Telescope \citep{Spergel:2015:Roman}, and the \textit{Euclid} space mission \citep{Laureijs:2011:Euclid}. Furthermore, the DECADE data triple the overlap between existing photometric cluster catalogs and other spectroscopic, mm-wave, and X-ray datasets (\eg Figure 1 in \citealt{paper5}), significantly enhancing the opportunity for cross-correlation analyses.

The DECADE dataset also provides an opportunity to stress-test cluster-finding algorithms on unconventional imaging data. Unlike DES and SDSS, which use dedicated survey strategies and image processing choices that maximize homogeneity in observing conditions (sky background, seeing, depth, etc.), DECADE is an amalgamation of public archival data from DECam \citep{paper1}. While this archival data is predominantly comprised of survey programs such as DELVE \citep{Drlica-Wagner:2021, Drlica-Wagner:2022}, the Dark Energy Camera Legacy Survey \citep[DECaLS,][]{Dey:2019}, and the DECam eROSITA Survey \citep[DeROSITAS,][]{Zenteno:2025:Derositas}, it is also made of disparate programs from individual groups, with survey strategies that cater to a wide variety of science programs (\eg transient searches, cluster lensing, etc.). Optical cluster finding is generally designed for inhomogeneous data \citep[\eg][]{Koester:2007:maxBCG, Rykoff:2014:AlgorithmCatalog}, and this work provides a concrete test of these design features on a precision photometric dataset with large spatial inhomogeneities in imaging quality, depth, and other relevant survey properties. Thus, in this work, we pay particular attention to characterizing these cluster-finding algorithms, and to empirically quantifying any biases or trends in cluster detection and cluster properties as a function of these spatially varying quantities.

There are a variety of choices for optical cluster finding, of which we use the Red-sequence Matched-filter Probabilistic Percolation (\redmapper) algorithm \citep{Rykoff:2014:AlgorithmCatalog}. \redmapper performs a matched-filter detection step on a photometric galaxy catalog, by utilizing the fact that cluster member galaxies are preferentially quenched and therefore redder in their photometric colors \citep{Gladders:2000:CRS}. Several alternative methods have also been applied to wide-field photometric data, including the optimal matched-filter algorithm \textsc{AMICO} \citep{Bellagamba:2018:AMICO}, the wavelet- and photometric-redshift-based \textsc{WaZP} finder \citep{Aguena:2021:WaZP}, the red-sequence-based \textsc{CAMIRA} algorithm \citep{Oguri:2014:CAMIRA}, and the galaxy-overdensity method used to construct the Wen--Han catalogs \citep{Wen:2012:WHL}. We prefer the \redmapper algorithm given its heritage in cosmology efforts within DES \citep{Rykoff:2016:redmapper, 2021PhRvL.126n1301T, 2025PhRvD.112f3537T, DES:2025:6x2pt_Y6}, and given the extensive calibration and characterization done on the SDSS and DES \redmapper catalogs \citep[\eg][]{Rykoff:2008:XrayRichness, 2009ApJ...703..601R, Rykoff:2012:OpticalRichness, Rykoff:2014:AlgorithmCatalog, Rykoff:2016:redmapper, Giles:2022:XrayRichness, Kelly:2024:miscentering}, which provide useful reference points for comparing the results of this study. Our work focuses on a broad set of validation tests for our catalogs, comparing the performance of DES and DECADE to each other, as well as to results in the literature from prior catalogs and from simulations. We make extensive use of data from mm-wave and X-ray datasets in our validations.

This work is organized as follows: Section \ref{sec:Data} details the photometric and spectroscopic datasets used in the cluster finding, as well as the mm-wave and X-ray data used for validation. We then detail the \redmapper algorithm in Section \ref{sec:Clusterfinder}, including the additional cylindrical richness observable that we test in this work. Section \ref{sec:Validation} provides a list of validations of the catalog, including the redshift performance, distributions of cluster properties, multi-observable scaling relations, and miscentering. We then provide mass estimates and lensing profiles for our samples in Section \ref{sec:MassCalibration}. Finally, we conclude in Section \ref{sec:Summary}. Appendix \ref{appx:tSZModel} introduces the Bayesian population model used for our optical--tSZ scaling relation analysis. Appendix \ref{appx:Inhomog} details the empirical validation of \redmapper on inhomogeneous data, while Appendices \ref{appx:RichnessCompare}, \ref{appx:Y3Compare}, \ref{appx:SurveyArea}, and \ref{appx:SZdepth} discuss the different richness estimators, comparison to earlier DES catalogs, the effective survey area of optical data, and the noise variations of mm-wave data, respectively. Appendix \ref{appx:FullTriangle} presents the full posteriors of our scaling-relation analysis on the richness and tSZ data.

\section{Data}\label{sec:Data}

\begin{figure*}
    \centering
    \includegraphics[width=1.6\columnwidth]{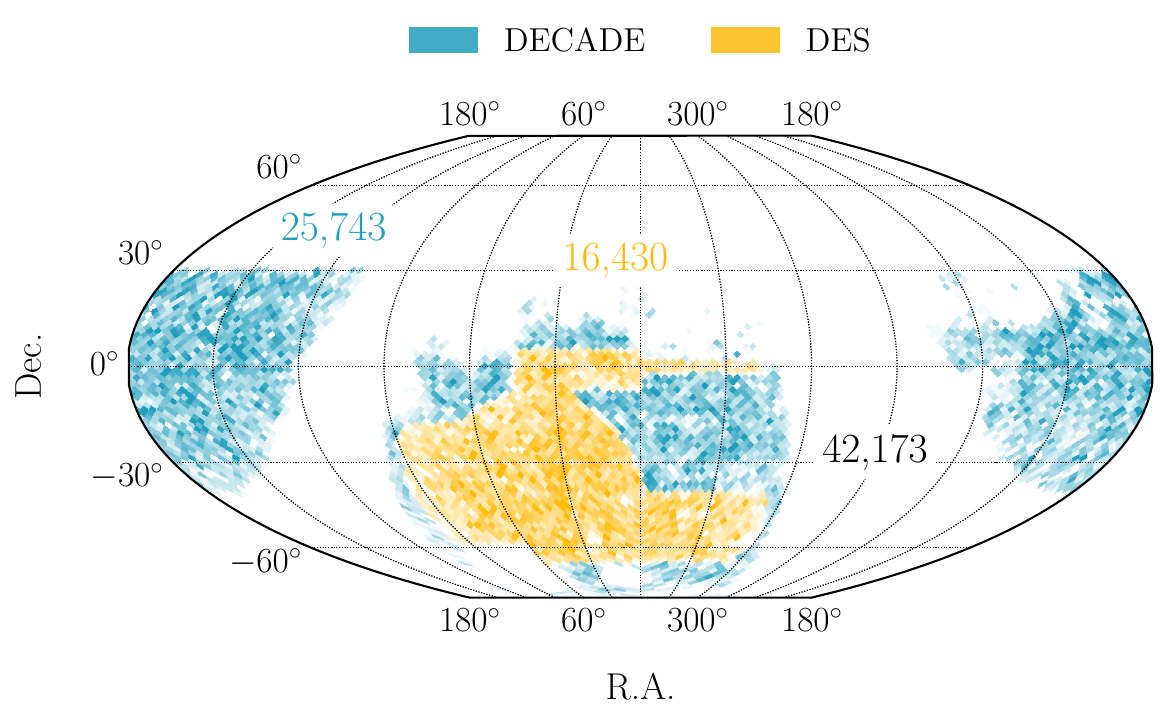}
    \caption{The number density of \redmapper clusters in the DECADE and DES patches of sky. Lighter colors represent fewer clusters. DECADE and DES roughly span 8,500 $\deg^2$ and 4,500 $\deg^2$ of the sky, respectively, after the foreground mask is applied (Section \ref{sec:data:photometric}).}
    \label{fig:Footprint}
\end{figure*}

We use a variety of different datasets in this work. First, the construction of an optical cluster catalog requires galaxy colors, as well as associated, high-accuracy redshifts for at least a subset of the galaxies. As a result, optical cluster finding often relies on both photometric and spectroscopic data. We detail the properties of our chosen catalogs in Section \ref{sec:data:photometric} and Section \ref{sec:data:spectroscopic}. In addition, to characterize the produced cluster catalogs, we make use of cluster measurements from the mm-wave and X-ray observations. These measurements are detailed further in Section \ref{sec:data:multiwavelength}.

\subsection{Photometric data} \label{sec:data:photometric}

As mentioned previously, the identification of galaxy clusters in optical surveys requires the photometric colors of galaxies. In this work, we use the DELVE DR3 photometric galaxy catalog, which is derived from coadded images produced with the DES Data Management Pipeline \citep[DESDM][]{Morganson:2018} at the National Center for Supercomputing Applications (NCSA). The image quality (depth, seeing, uniformity, etc.) is fairly homogeneous in the 4,500 $\deg^2$ spanned by DES, and is more inhomogeneous in the 8,500 $\deg^2$ region covered by DECADE, as the latter is an amalgamation of public archival data from the DECam community.\footnote{Roughly $10\%$ of the DECADE survey has superior imaging properties to DES due to the surplus of images taken in a given part of the sky (Figure 3 of \citealt{paper1}).} See \citet{paper1,paper3,paper5} for more details on the differences in image quality and survey properties between DECADE and DES. \citet{Morganson:2018} present an outline of the DESDM image processing pipelines, while \citet{Bechtol:2025:Y6GOLD} and \citet{Tan:2025:DR3} describe the latest variant as applied to the DES Y6 and DECADE analyses, respectively.

While existing analyses of DECADE data have used the \textsc{Metacalibration} \citep{Huff2017} galaxy catalogs, this work utilizes a catalog produced under a different analysis framework. In particular, we use the ``Single-Object Fitting'' (SOF) framework, as first mentioned in \citet{Sevilla-Noarbe2021} and later in \citet{Hartley:2020:DESY3Deep} and \citet{Bechtol:2025:Y6GOLD}. Here, objects are fit with a two-component light-profile model --- an exponential disk and a de Vaucouleurs bulge, with the ratio of sizes of the two components fixed\footnote{Varying this fraction leads to unconverged fits as the data are not sufficiently constraining to simultaneously determine this ratio while varying the other parameters of the light profile such as the galaxy fluxes, size, and shape \citep{Bechtol:2025:Y6GOLD}.} to unity, \ie equal sizes --- while varying the galaxy size, ellipticities, fluxes in the four bands, $griz$, and relative contribution of the bulge compared to the disk. Our primary measurements of interest are these object fluxes, and in particular, the photometric colors that can be constructed from them. In this work, we focus on the sequential colors, $g - r$, $r - i$, and $i - z$. Note that both frameworks (\textsc{Metacalibration} and SOF) use the same object detections --- the image detection process is listed in detail in \citet{paper1} --- and jointly analyze individual images while accounting for their associated PSFs. In the rest of our discussion to follow, a reference to ``galaxy catalog'' should be treated as the SOF galaxy catalog unless described otherwise. 

The photometric galaxy catalogs must be cleaned prior to its use in the cluster-finding algorithm. We perform the following selections on the catalogs:

\begin{itemize}
    \item $\texttt{EXT\_MASH} \geq 2$, which reduces contamination from potential stars \citep{Sevilla-Noarbe2021, Bechtol:2025:Y6GOLD}.

    \item $\texttt{FLAGS\_FOREGROUND} \,\&\, \neg 2 = 0$, to remove regions where galaxy measurements are contaminated by brighter objects, often associated with the Milky Way or nearby structures. See \citet{paper1} and \citet{Bechtol:2025:Y6GOLD} for details on these foreground masks in DECADE and DES, respectively. The operator $\neg 2$ removes one of the flags from the foreground bit mask (\ie we no longer consider that flag in our masking). See below for more details on this choice.

    \item $\texttt{FLAGS\_FOOTPRINT} == 1$, to only keep regions within the main survey footprint.

    \item $\texttt{FLAGS\_GOLD} == 0$, to remove objects that do not pass the quality cuts. See Section 4.3 and Table 4 in \citet{Bechtol:2025:Y6GOLD} for the cut definitions. The same cuts are applied to both DECADE and DES.
\end{itemize}

These selections mostly follow those described in the cosmology catalogs of each survey \citep[\eg][]{Yamamoto2025, paper1}. The only difference is in the foreground mask; in this work we remove a faint star flag, which is built using J-band luminosity from the 2MASS survey \citep{Skrutskie:2006:2Mass}, from the foreground mask. There are numerous stars used in this mask, which in turn generate significant small-scale masking structure in the data. This has a notable impact on cluster finding, as clusters have sizes of a similar order of magnitude as the chosen masking scale of a few arcminutes. We empirically found that the number density of the cluster catalog showed a distinct dependence on stellar density if we continued utilizing this mask. Given that cluster finding can effectively reduce stellar contamination, via its use of the red galaxy-based color selection \citep{Rozo:2016:RedmagicSV, Rykoff:2016:redmapper}, such stars are much less of a concern in our analysis than they are in other galaxy analyses. Therefore, we ignore this mask in order to significantly reduce sensitivity of our catalog to stellar density variations. Note that our foreground mask still removes the densest stellar regions \citepalias{paper1}.

Another useful input to the cluster finding process is an estimate of the imaging properties across the survey area. In particular, we make use of survey depth estimates in the different photometric bands, as well as maps of the seeing, sky background, airmass, and exposure time, computed for each of the four bands. These maps have all been produced using the \textsc{DECaSu} software,\footnote{\url{https://github.com/erykoff/decasu}} both for \decade and DES. We will also test our catalogs' dependence on two other maps: the \textit{Gaia} \citep{Gaia:2021:eDR3} stellar number density (computed using stars with $G < 21$\,mag) and the dust extinction map of \citet{Schlegel:1998:Dust}. The extinction map is used to deredden the fluxes that are provided to the cluster-finding algorithm. The stellar-density map is not used in the cluster finding, apart from its influence as a component of the foreground masking included in \texttt{FLAGS\_FOREGROUND} \citepalias{paper1}.

\subsection{Spectroscopic calibration} \label{sec:data:spectroscopic}

\begin{figure*}
    \centering
    \includegraphics[width=2\columnwidth]{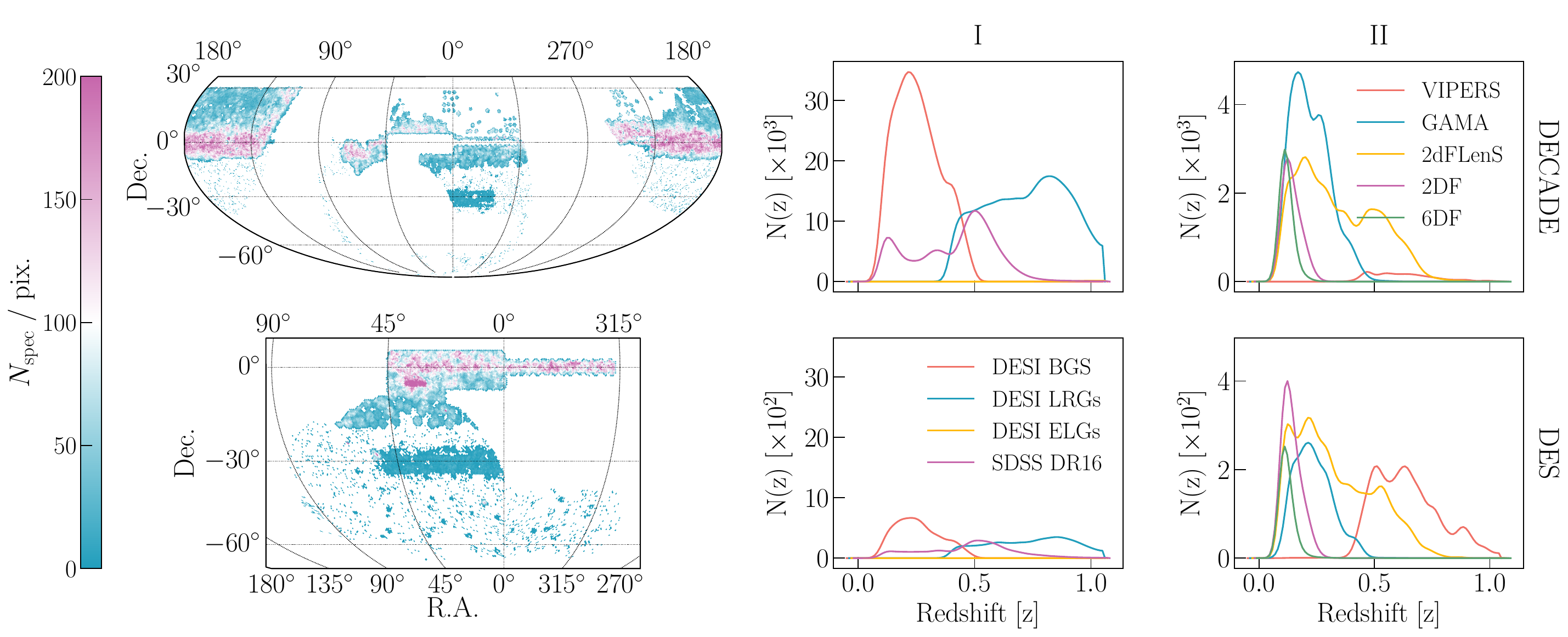}
    \caption{The spectroscopic calibration sample for DECADE (top) and DES (bottom). The middle and right columns show the redshift distributions from different spectroscopic samples (arbitrarily labeled as groups I and II for clarity). The ELG sample provides some matches, but it is subdominant relative to the other DESI and SDSS data; note that SDSS DR16 contains BOSS and eBOSS. The left column shows the area spanned by the calibration sample for each of DECADE and DES. The sample is predominantly in the Galactic north, where there is significant overlap with SDSS and DESI. The spectroscopic training catalog for DECADE is five times larger than that for DES, and this will grow with future DESI data releases.}
    \label{fig:specz_data}
\end{figure*}

The DECADE data have a significant overlap with Data Release 1 (DR1) from the Dark Energy Spectroscopic Instrument \citep[DESI,][]{DESI:2016:Part1} --- the joint coverage spans around 5,000 $\deg^2$ --- with significantly larger overlap expected with future DESI data releases. In comparison, DES alone contains 1,000 $\deg^2$ of overlap with DESI. However, as we will show below, this difference does not result in improved redshift precision/accuracy in \decade over that in DES as the quality of the redshift estimates is instead set by the properties of the red-sequence relation and by the photometric uncertainty of a given dataset. Therefore, the main benefit of the larger spectroscopic coverage in DECADE is a more precise characterization of the redshift uncertainties and bias.  We utilize three DESI samples for the calibrations: the Bright Galaxy Survey (BGS), Luminous-Red Galaxies (LRG), and Emission-Line Galaxies (ELG). See \cite{2023AJ....165...58Z} for definitions of these target selections. While the BGS and ELG samples contain a significant number of blue galaxies, the \redmapper algorithm automatically discards such objects using its (iteratively constructed) red-sequence model; see Section \ref{sec:ClusterFinding:Redmapper} for more details.

In addition to DESI, our sample also contains objects from the Baryon Oscillation Spectroscopic Survey \citep[BOSS,][]{Dawson:2013:BOSS} and extended BOSS \citep[eBOSS,][]{Dawson:2016:eBOSS}, collectively released under DR16 from the Sloan Digital Sky Survey \citep[SDSS,][]{York_2000}. These data contribute to the lower-redshift ($z \lesssim 0.5$) regime of the calibration sample, as they cover larger areas of the sky compared to DESI DR1. The BOSS and eBOSS data contribute a notable fraction of spectra in the combined sample --- even after accounting for duplicates with DESI DR1 (identified by position-matching within 1 arcsecond radius apertures) --- due to this difference in sky-area coverage. We use the Low Redshift (LOWZ), Constant Mass (CMASS), and LRG samples from BOSS and eBOSS.

Finally, we use an amalgamation of public redshift catalogs as listed in \citet{Gschwend:2018:LineA}, which we will denote as the ``LIneA'' catalog. This dataset has been used in the production of previous \redmapper catalogs as well as the redshift point estimates in the DES and DECADE public releases \citep{Sevilla-Noarbe2021, Bechtol:2025:Y6GOLD}. The primary contributors of the catalog (in terms of raw numbers) are the 2 degree field survey \citep[2DF,][]{Colless:2001:2dFGRS}, the 6 degree field survey \citep[6DF,][]{Jones:2009:6dFGS}, Galaxy And Mass Assembly survey \citep[GAMA,][]{Driver:2011:GAMA}, and the VIsible MultiObject Spectrograph Public Extragalactic Redshift Survey \citep[VIPERS,][]{Scodeggio:2018:VIPERS}. Once again, we explicitly account for duplicates using the one arcsecond position-matching method noted above. Less than $0.001\%$ of objects have ambiguous matching (\ie contain two matches within the one arcsecond aperture), and we simply discard such matches from the analysis. The final spectroscopic calibration sample in DECADE and DES has 2,070,692 and 459,256 galaxies, respectively; the former is nearly five times larger. Figure \ref{fig:specz_data} summarizes the spectroscopic coverage in this analysis, as well as the redshift distributions of the dominant contributors to our calibration sample. We have reliable calibration spectra out to $z < 1$ for both surveys.

Note that the spectroscopic samples are only used (i) to seed the red-sequence calibration step in \redmapper (Section \ref{sec:ClusterFinding:Redmapper}), and (ii) to quantify/calibrate the bias and uncertainty in the \redmapper redshift estimates. We never measure the auto/cross spatial correlations of these samples for calibrations --- as we do not perform any of the clustering-based redshift calibration commonly used in DES, DECADE, and other lensing surveys \citep[\eg][]{Gatti:2022:WzY3, Rau:2023:HSCY3, paper2, dAssignies:2026:DESY6} --- and therefore do not require any random-object catalogs for the spectroscopic samples.

\subsection{Multi-wavelength cluster catalogs}\label{sec:data:multiwavelength}

Throughout this work, we frequently validate the characteristics of our optically selected cluster catalog by matching it to catalogs selected from mm-wave or X-ray data. We will often refer to the mm-wave catalog as the tSZ-selected catalog; see \citet{Carlstrom2002SZReview} for a review of the SZ effect. Through this matching, we can use the scaling relations between different cluster observables to validate the performance of DECADE compared to DES, and to compare the behavior of both catalogs to those from prior catalogs and from simulated samples. Simultaneously, these derived relations will allow us to provide halo mass estimates for all optical clusters.

We rely on three different datasets for such analyses --- the measurements from the South Pole Telescope \citep[SPT,][]{Carlstrom2011} and Atacama Cosmology Telescope \citep[ACT,][]{ACT:2016} in the mm-wave, and from eROSITA \citep{Merloni:2012:Erosita} in the X-ray. Specifically, we use the SPT catalogs from \citet{Bleem:2015:SPT} and \citet{Bleem:2020:ECS},\footnote{The latest SPT catalog, presented in \citet{Bleem:2026:Clusters}, was published during the final stages of this work, and hence was not utilized in this analysis. We will instead employ it in future work on the DECADE and DES cluster catalogs.} the ACT catalog from \citet{Hilton:2026:ACTDR6}, and the eROSITA catalog from \citet{Bulbul:2024:eROSITA}. We use the signal-to-noise (SNR; also denoted as $\xi$) from the mm-wave data,\footnote{In both SPT and ACT, the cluster SNR is the maximum SNR from scanning over sky position and filtering scale. The choice of filter varies across the surveys \citep{Hilton:2026:ACTDR6, Bleem:2026:Clusters}. The SNR definitions are therefore not identical, but still similar.} and the integrated luminosity, $L_{500c}$, from the X-ray data. While both ACT and eROSITA provide cluster positions from optical follow-up of their clusters, we only use their mm-wave/X-ray based positions for the cross-matching as we want to characterize miscentering between the different center estimates (Section \ref{sec:Validation:Miscenter}). We also utilize the redshifts reported in these other catalogs, which are obtained through a variety of methods including spectroscopic follow-up and targeted red-sequence cluster finding.

We must define a methodology for matching optical clusters with their counterpart from a different wavelength. We follow \citet{Bleem:2020:ECS} in performing a position-based abundance matching. For two catalogs A and B, we rank-order them by their cluster observable. Then, the first object in catalog A is matched with the highest-ranked object in catalog B that also satisfies the following: (i) has a redshift estimate that is within $\Delta z = 0.15$ of the object in catalog A, and (ii) has a position within $1.5 \mpc/h$ (in physical, not comoving, separation) of the object in catalog A. Once a match is made, the object in catalog B is removed to avoid any duplicate matches. While the thresholds for the redshift/separation cuts can be varied, we found that allowing distances greater than $1.5 \mpc/h$ increased the fraction of random matches.

\section{Cluster finding methodology} \label{sec:Clusterfinder}

There are multiple ways of identifying clusters in a given dataset, of which we use the overdensities of red galaxies \citep{Gladders:2000:CRS, Gladders:2005:RCS} as implemented in the \redmapper\footnote{We use \url{https://github.com/DhayaaAnbajagane/redmapper} which is forked from the main repository, \url{https://github.com/erykoff/redmapper}.} framework. The key observable is the cluster richness, $\lambda$, which is a probabilistic count of red-sequence galaxy members of luminosity $L > 0.2 \Lstar$ within a given cluster. Here, $L$ is the luminosity of the galaxy in a given reference band ($z$-band, for our work), and $\Lstar$ is the characteristic ``knee'' of the luminosity function.  We also define a cylindrical richness observable \citep{2012MNRAS.426.2046A, 2017MNRAS.470.4767B, 2022MNRAS.515.4471W, Andylee} for each \redmapper cluster, to serve as an additional optical observable in the catalog. We detail the advantages of this latter observable in Section \ref{sec:ClusterFinding:cylinder}.

\subsection{The \redmapper algorithm}\label{sec:ClusterFinding:Redmapper}

\redmapper is an optical cluster finder that looks for spatial overdensities in the number of red galaxies through a matched-filter approach in position-magnitude-color space. See \citet{Rykoff:2014:AlgorithmCatalog} for a detailed introduction to the algorithm. \redmapper is particularly robust in self-calibrating its red-sequence and cluster-finding model to any photometric dataset. It does so by starting from a ``seed'' catalog of photometric objects with associated spectroscopic (or high-precision/accuracy) redshifts, and then iteratively building a red-sequence model while discarding low-likelihood cluster candidates from the cluster-finding process. We use the Python-based \redmapper v6.4.22+2, and follow the same settings employed on DES Y3. In particular, this includes an update to the percolation radius --- which is used to iteratively remove member galaxies from larger clusters so that these members are not assigned to smaller clusters --- from the relation $R_{\rm per} = 1.5 (\lambda / 100)^{0.2} \mpc/h$ to now $R_{\rm per} = 1.95 (\lambda / 100)^{0.45} \mpc/h$. This change was first implemented in \redmapper v6.4.22+1 as used in DES Y3. The updated relation is more consistent with the halo-exclusion radius computed in simulations, and matches the average halo mass of optically selected clusters as found in \citet{McClintock:2019:ClusterWL}. The existing \redmapper catalogs from DES were produced using the IDL-based \redmapper cluster finder, rather than the newer, Python-based one used here. The main difference between the two is in the assumed correlation in the scatter of different photometric colors in the red-sequence model. The IDL-based version fit the correlation coefficients (as a function of redshift) using the data, while the Python-based one fixes $\rho = 0.9$. The latter provides a more numerically stable result. The Python-based version also fits for errors in the color covariances, and therefore can accommodate any incorrect errors reported in the SOF fits to a galaxy.

The reader familiar with the \redmapper cluster finder can move on to the following sections. For the interested reader, we now describe the individual steps of \redmapper in slightly more detail; a complete description is provided in \citet{Rykoff:2014:AlgorithmCatalog} and \citet{Rykoff:2016:redmapper}. The \redmapper algorithm works in two stages, with each stage having iterative sub-stages within it. In the first stage, the redshift-dependent red-sequence model is built using the ``seed'' galaxies mentioned above. Notably, this uses a data-driven approach and does not rely on a stellar population synthesis model. The algorithm also estimates the intrinsic scatter in this red-sequence model as a function of redshift. These ``seed'' galaxies are considered as likely members of galaxy clusters, and therefore their sky positions and redshifts provide the three-dimensional positions of possible red-galaxy overdensities. \redmapper searches within a projected distance of $500 \kpc/h$ from each seed galaxy to find these overdensities. All photometric red galaxies within an overdensity are considered to be in the same cluster as the spectroscopic galaxy and can therefore be assigned the redshift of the latter. To minimize contamination, we only consider overdensities with $\lambda > 10$. The cluster finder is then run on this initial sample of red galaxies, \ie \redmapper considers each galaxy in this subsample as a cluster center and performs a matched-filter search. This step finds more cluster members, which are naturally all high-likelihood red-sequence galaxies. The cluster-finding procedure is performed three times in total. Figure 5 and Figure 6 in \citet{Rykoff:2014:AlgorithmCatalog} show that three iterations are sufficient for the red-sequence calibration to converge. In Section \ref{sec:Validation}, we will validate the performance of the redshift estimation using overlapping spectroscopic data. To avoid any circularity in the validation step, we calibrate the color--redshift relation using only half the overlapping area, and redshift validation is done using the other half.

Once this red-sequence model is calibrated, \redmapper proceeds with the final cluster-finding step. Cluster finding also occurs in three distinct sub-stages. The first stage, referred to as the First Pass run, considers each galaxy in the red-galaxy catalog as a cluster center and then performs cluster finding. Any center that results in $\lambda < 3$ is dropped from consideration for the next stages. The second stage, the Likelihood run, performs cluster finding on the remaining cluster centers/candidates and evaluates a log-likelihood for the cluster. This is primarily informed by the probability that the assigned member galaxies are indeed members of the cluster. Cluster candidates are then ranked based on their likelihoods. The last stage, the percolation run, performs cluster-finding in a serial manner. Cluster candidates are assigned members, and any satellite galaxies that are high-confidence members of a cluster are then masked in the catalog and are far less likely to be assigned as probabilistic members for clusters appearing later in the run. Due to the sorting step done in this stage, the high-likelihood clusters are given near full access to the catalog, while lower likelihood clusters start with masked versions.

There are three additional components of \redmapper that are useful to discuss. The first component is a masking/completeness correction. Survey datasets are incomplete both in the area they cover (\eg due to small-scale masking) and in the faintest magnitude at which they can reliably detect objects. So, cluster members that are otherwise present in the sky may not be identified from a given dataset. \redmapper accounts for this effect via a multiplicative scaling correction to the richness. This correction accounts for the fraction of a cluster that is masked, and for the fraction of the red-sequence luminosity function, with $L > 0.2 \Lstar$, that is below the detection threshold at the sky location of the cluster. Note that this means \redmapper can accomodate spatially varying depth in a survey. Indeed, in Appendix \ref{appx:Inhomog}, we show that the \redmapper richness and redshift estimates show no significant correlation with survey depth.

The second component is related to the background term. Not all red galaxies are expected to be cluster members, \ie they are part of the field and not satellites associated with larger halos. There are also slightly bluer galaxies that may be associated with the red-sequence sample as well. \redmapper estimates a background term using the input galaxy catalog (as defined for the final step of cluster finding) as a function of $z$-band magnitude, $\chi^2$, and redshift. The $\chi^2$ is the photometry-based distance between the galaxy and the red-sequence model, and serves as a simple way to build a color-dependent background estimate as a function of a single variable. The final probability that a given galaxy is a member of a specific cluster accounts for the probability of the galaxy belonging to the background population \citep[][see their Equation 1]{Rykoff:2014:AlgorithmCatalog}. Note that accounting for the background galaxy population is a standard ingredient in optical cluster finding and is not unique to \redmapper  \citep{Oguri:2014:CAMIRA,Bellagamba:2018:AMICO,Aguena:2021:WaZP}. Our description above serves as a brief introduction to the specific implementation within \redmapper.

Finally, \redmapper also generates a catalog of randoms. It does so by sampling random pairs of richness and redshift values using the real catalog, and randomly assigns the mock clusters into $50 \times 10^6$ locations in the survey. For a given mock cluster at an assigned location, the mock point is retained only if (i) the sampled redshift is below $z_{\rm max}$, the maximum redshift to which a red-sequence galaxy can be detected in a given sky location, and (ii) the fraction of masked pixels in the sky area spanned by the cluster is $f_{\rm mask} < 0.2$. Estimates of $z_{\rm max}$ and $f_{\rm mask}$ are provided via \redmapper, using the survey property maps and survey masks. See \citet{Rykoff:2014:AlgorithmCatalog,Rykoff:2016:DESCatalog} for more details. This step is repeated until $50 \times 10^6$ random points are generated. The final output is a catalog whose number density captures the spatially varying completeness of the \redmapper catalog, with a redshift and richness distribution that matches that of the real cluster catalog. In this work, the random catalog is used in our estimate of the lensing profiles around \redmapper clusters (Section \ref{sec:MassCalibration:WL}).

\subsection{Cylindrical richness as a cluster observable}\label{sec:ClusterFinding:cylinder}

The richness estimated by \redmapper is empirically known to be an excellent mass tracer \citep{2009ApJ...703..601R,Rykoff:2012:OpticalRichness}. However, it is challenging to model accurately. The \redmapper richness depends on several detailed properties of the galaxy population, including the relative locations of the red sequence and blue cloud in multi-dimensional color space, the intrinsic width of the red sequence, the luminosity function of the red-sequence galaxies, and the relative luminosities of central and satellite galaxies \citep[see, \eg][for a recent modeling attempt]{cardinal}. Consequently, several recent studies \citep{2020MNRAS.496.4468S,2022MNRAS.515.4471W,2024PhRvL.133v1002S,Andylee,2025arXiv251025706S} have approximated the \redmapper selection using galaxy counts within cylinders. This approximation is motivated by tests on realistic mock catalogs, which find that a cylinder-based richness reproduces much of the weak-lensing selection bias induced by the full \redmapper algorithm \citep{2022MNRAS.515.4471W}, and by observational measurements of the broad line-of-sight distribution of \redmapper member galaxies \citep{Andylee}. The remaining assumption is that the cylinder-based richness also reproduces the joint scatter between \redmapper richness and weak-lensing observables at fixed halo mass. Directly validating this assumption in the data is challenging because the true mass of an individual cluster is not observable, while its quantitative validation in simulations depends on high-fidelity modeling of galaxy colors, luminosities, and spatial distributions.

Motivated by this concern, we construct an alternative mass tracer for each \redmapper cluster. This tracer is easier to model, although a priori it can potentially be less optimal than the \redmapper richness. Following the aforementioned studies that approximate richness using galaxy counts within cylinders \citep{2020MNRAS.496.4468S,2024PhRvL.133v1002S,2025arXiv251025706S,Andylee}, we define a similar cylinder-based richness estimator, denoted as $\lambdacyl$.\footnote{Formally, our estimator is the counts of member galaxies associated with a cluster. It does not utilize membership probabilities or completeness corrections as defined for the fiducial \redmapper richness, $\lambda$. However, ``richness'' is the standard term for member galaxy-based mass tracers in optical datasets, and so for consistency in use of nomenclature, we refer to our estimator as a type of ``richness'' as well.} An important consideration is to measure the counts with a galaxy sample that can be reliably reproduced in simulations. Since cluster galaxy populations are dominated by red galaxies, luminous red galaxy (LRG) samples provide a natural choice: they can be selected robustly from broadband photometry \citep[\eg][]{2002AJ....123..485S,2001AJ....122.2267E,2023AJ....165...58Z} and have a long history of being modeled in simulations \citep[\eg][]{2009ApJ...707..554Z,2024MNRAS.530..947Y}. In this work, we use galaxies selected from the \redmagic algorithm \citep{Rozo:2016:RedmagicSV}. This is an LRG sample selected using the red-sequence color--redshift relation calibrated by \redmapper and is a widely employed sample in cosmological analyses of photometric surveys \citep{2018PhRvD..98d2006E,2022PhRvD.106d3520P}. Specifically, we define $\lambdacyl$ as the number of \redmagic galaxies within a cylinder centered on each cluster, with a line-of-sight separation of $\pm 100\mpc/h$ and a transverse radius that scales with $\lambdacyl$ according to the same richness--radius relation adopted by \redmapper. When cylinders overlap, we rank the clusters by $\lambdacyl$ and assign galaxies in the overlapping region to the cluster with the larger $\lambdacyl$, thereby approximately reproducing the percolation procedure used by \redmapper.

The principal advantage of this cylindrical richness estimator is that the \redmagic population can be relatively straightforward to populate in simulations, because its halo occupation distribution can be constrained empirically using galaxy-clustering measurements \citep[\eg][]{Cao:2026:HOD}. It is also significantly brighter, with $L > 0.5 \Lstar$ instead of the $0.2\Lstar$ limit used in \redmapper, which can greatly alleviate measurement/observational systematics in the sample. Consequently, $\lambdacyl$ can be modeled without requiring a high-fidelity reproduction of the full galaxy color and luminosity distributions needed to run \redmapper \citep[\eg][]{Cao:2026:HOD}. The trade-off is that $\lambdacyl$ may exhibit greater scatter in halo mass at fixed values for the cluster observable, $p(\Mtwohc\mid\lambdacyl)$, than the original \redmapper richness. However, we show empirically in Section~\ref{sec:MassCalibration:WL} that the corresponding degradation in mass-tracing performance is minimal. The \redmagic sample associated with the cluster catalogs/models produced here will be introduced separately (\textcolor{blue}{Anbajagane et al., in prep.}).

Note that throughout this procedure, our use of the original \redmapper catalog is limited to the centers of the identified clusters and their associated redshifts. We do not use any other quantities from \redmapper. While this means $\lambdacyl$ can in principle inherit some of \redmapper's complicated modeling dependencies and selection, we expect such effects to be inconsequential to the modeling accuracy. Cluster centering (and miscentering) in \redmapper is a well-studied process that can be calibrated empirically from the data, as we show below in Section \ref{sec:Validation:Miscenter}, as well as theoretically using simulations \citep[\eg][]{Zhang:2023:triaxiality,Payerne:2025:DC2,Yan:2020:miscentering}. The key challenges are modeling the cluster richness estimate and the percolation process (and their connection to weak lensing observables). These two aspects are precisely what is simplified by the use of the $\lambdacyl$ estimator above. A more complete solution would also determine cluster centers using counts in cylinders. However, this requires significant modifications and development of the existing pipelines. Our method here is instead a practical middle ground that alleviates the most pressing modeling challenges without having to build entirely new cluster-finding frameworks.

\section{Characterization of the cluster catalog} \label{sec:Validation}

We now present a variety of validations and characterizations of our cluster catalog, and pay special attention to comparisons between the DECADE and DES subsets. Section \ref{sec:Validation:Redshifts} lists the redshift performance, and Section \ref{sec:Validation:Dist} presents the distribution of multiple relevant properties from the catalogs. We discuss multi-property scaling relations, at varying levels of analysis sophistication, in Section \ref{sec:Validation:Scaling:Xray} and Section \ref{sec:Validation:Scaling:SZ}, and finally, extract cluster miscentering estimates in Section \ref{sec:Validation:Miscenter}.

\subsection{Accuracy of redshift estimates} \label{sec:Validation:Redshifts}

\begin{figure*}
    \centering
    \includegraphics[width=2\columnwidth]{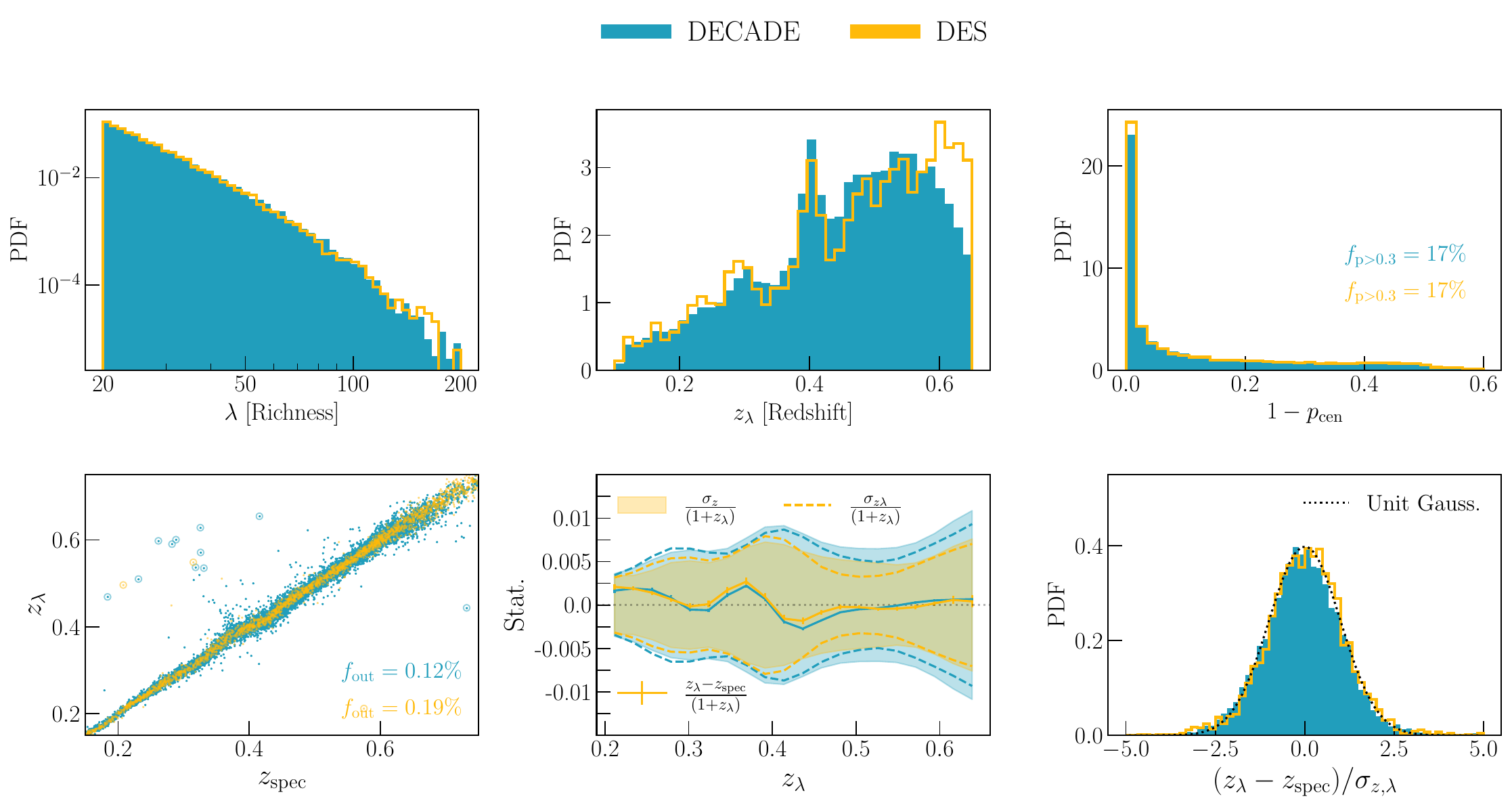}
    \caption{Characterization of the \redmapper catalogs from DECADE and DES. \textit{Top:} the distributions of richness (left), redshift (middle), and the probability that a different galaxy is the true central of the cluster (right). \textit{Bottom:} the comparison of photometric and spectroscopic redshifts for a subsample of clusters that have both estimates (left), the bias and uncertainty in the photometric redshifts (middle), and the histogram of the normalized residuals (right). In all cases, the DECADE and DES distributions are quite similar. The most visible difference is in the redshift distribution (top middle), which is expected as DECADE is a shallower dataset and therefore does not extend to as high a redshift as DES.}
    \label{fig:Catalog}
\end{figure*}

A key metric in a cluster catalog is the accuracy and precision of the cluster redshift estimates. In practice, optically identified clusters have fairly precise photometric redshifts as they average over the estimates of their member galaxies, \ie there is roughly a $1/\sqrt{N_{\rm gal}}$ suppression of the redshift uncertainties relative to that for a single galaxy. The lower panels in Figure \ref{fig:Catalog} detail our redshift performance. From left to right we show the outlier rate, the redshift uncertainty and bias, and the shape of the residuals' distribution. The calibration metrics --- such as redshift bias and scatter --- can be measured at high precision, especially in \decade, due to the availability of DESI DR1. As mentioned in Section \ref{sec:ClusterFinding:Redmapper}, spectroscopic data are used to both calibrate the color--redshift relation and to validate the final redshift estimates. To avoid circularity in the validation steps, only half the data was used for calibration and the other half is used for all redshift validation metrics discussed below.

The outlier rate --- where an outlier is defined as $|z_\lambda - z_{\rm spec}|/(1 + z_{\rm spec}) \geq 0.15$, following the standard choices in the literature \citep[\eg][]{Laigle2016, Alarcon:2021:PausCosmos, Weaver:2022:Cosmos} --- is less than 0.2\% in both datasets. Note that the estimate of the DES outlier rate is far noisier than that of DECADE as the former only have one-fifth as many clusters with $\zspec$ as the former. So the Poisson uncertainties on this fraction are far larger in DES. Nevertheless, both outlier rates are well below 1\%. We have visually checked the image cutouts of various redshift outliers, and find they are often caused by blending of different objects: for example, the spectroscopic sample has a small object at high-redshift that blends with the cluster at lower redshift, and so the redshift estimates are mis-assigned. Such blending is not expected to impact the calibration given its infrequent nature in the analysis, and only increases the redshift outlier rate. DECADE and DES exhibit similar redshift performance, as we will note further below. The scatter in the DECADE points appears larger than that of DES simply because the former is five times larger and therefore better samples the tails of the distribution.

Moving to the lower middle panel, we find the redshift uncertainties are $\sigma_z / (1 + z) < 0.008$ for both samples, and achieve $\sigma_z / (1 + z) \approx 0.005$ for some redshift ranges. The colored bands show the uncertainties estimated from matching to spectroscopic data ($\sigma_z$), while the dashed lines are the internal estimates provided by \redmapper ($\sigma_{z \lambda }$). There is good agreement across the two, showing the error estimates are well-calibrated. We see a small bias ($\Delta z / (1 + z) \approx 0.0025$) at $z = 0.35$ in both DECADE and DES where there is a known gap in the DECam filters for the $g$ and $r$ bands.\footnote{Including such a gap avoids the bright 557.7 nm green line in the atmosphere from atomic oxygen \citep{mclennan1925origin}, which would others greatly increase sky background and reduce object detection rates.} This has been previously noted in the DES Science Verification analyses of \citet[][see the bottom right panel of their Figure 3]{Rozo:2016:RedmagicSV} and in more recent DES \redmapper catalogs as well \citep{DES:2025:6x2pt_Y6}. Only $\approx 5\%$ of our cluster catalog falls in this redshift range, so the effect is overall a minor one. Most of our clusters are instead found at $z > 0.4$, where the measured bias is $|\Delta z/(1 + z)| \leq 0.001$.

This similarity in redshift performance across DES and DECADE is valuable to emphasize. The scatter in the photometric--spectroscopic redshift residuals is virtually the same, indicating that the size of the calibration sample is not limiting the precision of the redshift estimates. Indeed, this is to be expected as the dominant uncertainty in the \redmapper redshift estimate comes from photometric uncertainty in the galaxy colors and from the intrinsic scatter in the red-sequence color--redshift relation. The median redshift bias of DES and DECADE is virtually identical as well. Our redshift calibration results indicate that further increases to the size of the spectroscopic sample will have minimal impact on the redshift precision and accuracy for red-sequence cluster finding. This can be expected as \redmapper is largely a self-calibrated cluster-finding algorithm. While the spectroscopic sample is vital in providing \redmapper with an initial guess, the constraints from the photometric data rapidly dominate the final estimate in later iterations of \redmapper. \citet{Rykoff:2016:redmapper} further discuss how \redmapper only requires a small subset of objects with high-accuracy redshifts.

Finally, the lower right panel of Figure \ref{fig:Catalog} plots the distribution of residuals between the photometric and spectroscopic redshifts. The two distributions are consistent with a unit Gaussian, highlighting that the redshift errors provided by \redmapper are well calibrated and do not exhibit any significant, non-Gaussian tails.

\subsection{Distributions of redshift and richness}\label{sec:Validation:Dist}

While we have detailed the similarities in redshift performance across DECADE and DES, the distributions of various key properties in the catalogs are also similar, as shown by the top panels of Figure \ref{fig:Catalog}. The top-left panel shows the richness distributions in log-space, and both distributions follow the same shape/slope over a factor of ten in richness. The middle panel shows the redshift distribution, which varies a little. In particular, the DECADE distribution is suppressed at $z > 0.55$ relative to DES. This is natural given the DECADE data are shallower than DES \citep[][see their Figure 3]{paper1}, and so the effective area of the dataset shrinks towards higher redshift (Appendix \ref{appx:SurveyArea}). The distributions are otherwise in good agreement. Finally, the rightmost panel shows the probability of miscentering, \ie of a different galaxy in the cluster being the true central galaxy. We compute this as $p = 1 - p_{\rm cen}$, where the latter is the probability that the current central is the true one. These probabilities are computed internally by \redmapper; see Section 8 in \citet{Rykoff:2014:AlgorithmCatalog}. There is a clear peak at $p = 0$, with a tail to higher probabilities. About 17\% of the sample has probabilities of $p > 0.3$. The key point, once again, is that the distribution of probabilities is very similar between DES and DECADE.

\begin{figure}
    \centering
    \includegraphics[width=\columnwidth]{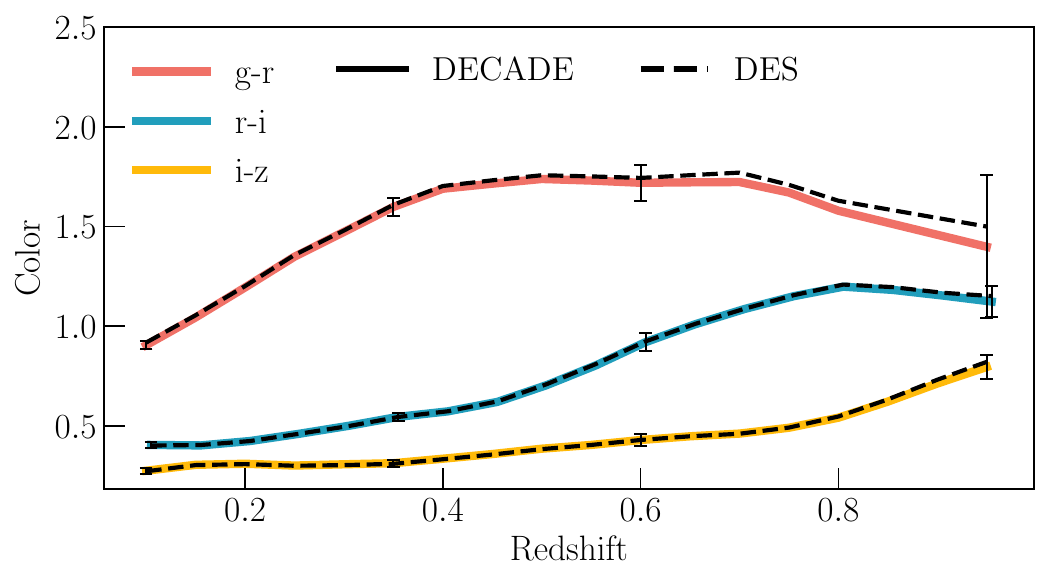}
    \caption{The red-sequence model for DECADE and DES. We show the color amplitude $A_a(z)$, as defined in Equation \eqref{eqn:redmapper:color}. Error bars show the intrinsic scatter in the red-sequence as calibrated in the model. The scatter for the DES model agrees with DECADE within 5\%, so we omit it from the plot for visibility. Overall, the two different DECam datasets (which were processed and analyzed independently) learn the same red-sequence relation, which we have also confirmed is consistent with other results from DECam-based photometry. The $r - i$ line (blue) is given a slight horizontal offset to improve visibility in the overlapping error bars.}
    \label{fig:RedSequence}
\end{figure}

Given the agreement above, one would expect the self-calibrated red-sequence model from each dataset to be similar. We demonstrate this by showing the color amplitude, $A_a(z)$, of the magnitude-dependent red-sequence relation in \redmapper,
\begin{equation}\label{eqn:redmapper:color}
    c_a(z) = A_a(z) + S_a(z) \times [m - m_{\rm piv}(z)]
\end{equation}
where $A_a$ is the intercept of the color-magnitude relation for color $a$, and $S_a$ is the slope of this relation with reference magnitude, $m$ (we use $z$-band). The pivot magnitude is determined internally within \redmapper, and roughly follows the median magnitude of the cluster member galaxies \citep{Rozo:2016:RedmagicSV}. Figure \ref{fig:RedSequence} shows that this amplitude, $A_a(z)$, is virtually identical across the two datasets. Any deviations are in regimes where the intrinsic scatter in the red-sequence relation (shown as black bars) is large. Given this relation is learned directly from the data, and must contend with the complexities latent in a given dataset (\eg photometric scatter and biases), the agreement of the two models is a strong indicator of consistent performance in cluster finding across the two surveys. We have also found the amplitude of the intrinsic scatter to be consistent across the two datasets, and so only show the scatter in the DECADE model for brevity. The similarity in scatter indicates that the dominant uncertainty in the red-sequence model is the intrinsic scatter and not the photometric uncertainty (which is larger in DECADE as it is shallower than DES on average). We have checked that our red-sequence model is qualitatively consistent with existing literature on the $griz$-band photometry: for example, see Figure 1 in \citet{Rykoff:2014:AlgorithmCatalog} for SDSS data and Figure 2 in \citet{DeRose:2022:Buzzard} for DES Year 3 data.

\subsection{X-ray scaling relations} \label{sec:Validation:Scaling:Xray}

\begin{figure}
    \centering
    \includegraphics[width=1\linewidth]{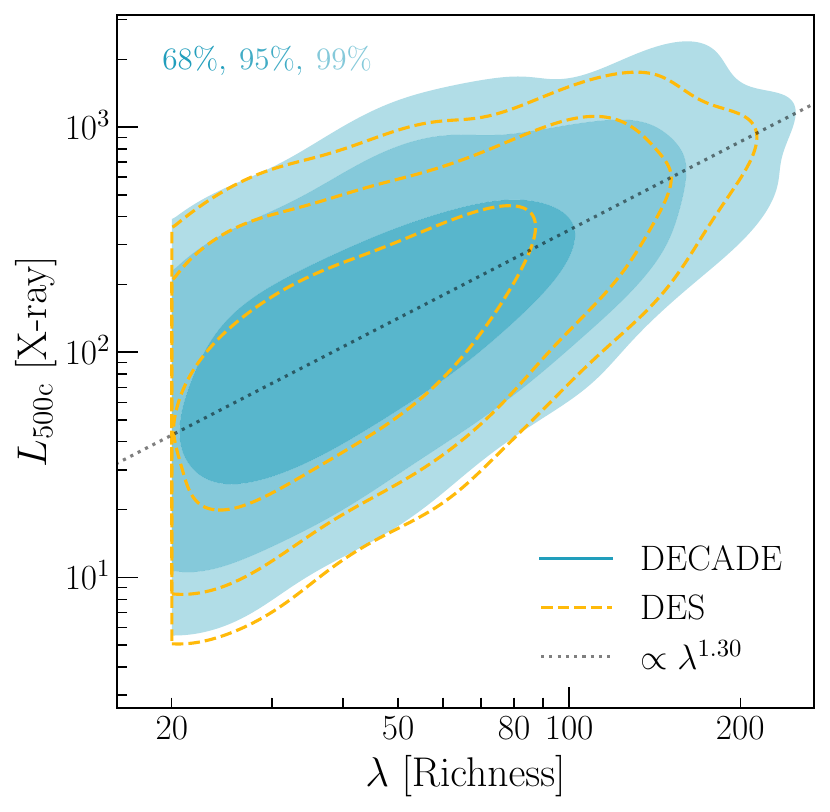}
    \caption{Contours of the X-ray luminosity and richness for the DECADE and DES samples, encompassing 68\%, 95\%, and 99\% of the probability density. A simple fit to both datasets (without accounting for any selection functions) yields a power-law slope of $1.3$, which is qualitatively consistent with existing results. Both DECADE and DES show behaviors consistent with one another.}
    \label{fig:Xrayscaling}
\end{figure}

We then characterize our optically selected cluster catalogs by cross-matching them with X-ray cluster catalogs and measuring scaling relations between the different cluster observables. We use the X-ray catalog from eROSITA \citep{Bulbul:2024:eROSITA} for this purpose. The exact matching procedure is discussed in Section \ref{sec:data:multiwavelength}. We use the integrated luminosity, $L_{\rm 500c}$, as our X-ray observable.

One of the most reliable ways to extract the richness--X-ray scaling relation from the data is through a Bayesian population forward model \citep[\eg][]{Costanzi:2021:DESxSPT, Grandis:2024:eROSITAMass}. However, we are unable to pursue this avenue as the X-ray selection function for eROSITA is not publicly available. While data-based alternatives are possible \citep[\eg][]{Grandis:2020:Selection}, this is beyond the scope of our work. Thus, we use the 2D distribution of richness and X-ray luminosity as our metric, and compare them across DECADE and DES to determine consistency in the two cluster samples. A Bayesian population analysis using X-ray data is left to future work.

Figure \ref{fig:Xrayscaling} shows the two distributions as level contours, with the DECADE and DES samples overlapping closely in richness--luminosity space. The scaling between X-ray luminosity and optical richness is qualitatively consistent with previous measurements using optically selected cluster samples \citep{Rykoff:2008:XrayRichness,Giles:2022:XrayRichness,
Upsdell:2023:XrayRichness,Kelly:2024:miscentering}.
The dotted black line shows the relation $L_{\rm 500c}\propto\lambda^{1.3}$, obtained from a simple linear fit to the combined DECADE and DES data. Under self-similar gravitational heating, the bolometric X-ray luminosity scales approximately as $L_{\rm X,bol}\propto M^{4/3}$ at fixed redshift \citep{Kaiser:1986:selfsimilar,Mantz:2016:scaling}. Assuming $\lambda\propto M$, this motivates $L_{\rm X,bol}\propto\lambda^{4/3}$ as a simple reference scaling. However, the eROSITA luminosity used here is measured in the soft $0.2$--$2.3\,{\rm keV}$ band
\citep{Bulbul:2024:eROSITA}, for which the precise self-similar scaling is somewhat shallower and depends on the cluster temperature and instrumental bandpass
\citep{Mantz:2016:scaling}. Given we have not accounted for any X-ray selection terms, we cannot accurately constrain the slope of the relation. Still, we note that we are qualitatively consistent with the aforementioned works, who quote slopes between $1.3$ and $1.8$ depending on the sample and analysis setup \citep{Upsdell:2023:XrayRichness, Kelly:2024:miscentering}.

We have also separately confirmed that these two-dimensional
distributions agree with those obtained using the targeted \textsc{eROMaPPer} optical measurements of the eROSITA clusters \citep{Kluge:2024:eROMapper}, \ie by using the \textsc{eROMaPPer} richness rather than those computed for DECADE and DES in this work.

\subsection{tSZ scaling relations}\label{sec:Validation:Scaling:SZ}

We then constrain the tSZ--richness scaling relation, using data from the SPT-SZ \citep{Bleem:2015:SPT}, SPT Extended Cluster Survey \citep[SPT-ECS,][]{Bleem:2020:ECS}, and ACT DR6 \citep{Hilton:2026:ACTDR6} datasets; see Section \ref{sec:data:multiwavelength} for more details. We analyze SPT and ACT separately. In all cases, we use the signal-to-noise, or $\xi$, as our tSZ observable. Unlike the above analysis of optical and X-ray data, we now forward model the optical and tSZ observables and use a full Bayesian population likelihood. We include the mass-observable scaling relations for both the tSZ and richness, accounting for correlations between the observables, and apply the selection functions from each dataset. This is done following the model presented in \citet{bocquet2024sptclustersdeshst} and references therein. Our full model is detailed in Appendix \ref{appx:tSZModel}.

We consider four analysis variants --- either DECADE or DES optical clusters matched with SPT or ACT tSZ clusters. The matching algorithm is detailed in Section \ref{sec:data:multiwavelength}. The parameters we constrain are the normalizations ($A_X$), slopes with log-mass ($B_X$), slopes with redshift ($C_X$), and log-scatter ($\sigma_X$), where $X$ is either richness or the SZ observable. We also constrain the cross-correlation, $\rho$, of the intrinsic scatter between the two observables. See Equations \eqref{eq:zeta_relation}, \eqref{eq:lambdatrue_relation}, and \eqref{eqn:cov_mat_relation} in Appendix \ref{appx:tSZModel} for more a detailed description of these parameters. In the case of SPT, we further introduce $\gamma_0^{\rm SPT}$ which is the noise level of the SPT-ECS survey relative to SPT-SZ. This is important as the tSZ observable, $\xi$, is a SNR and depends on the noise level of the data. The normalization of the tSZ--mass relation across two different patches of sky can therefore vary according to their relative noise levels. The $\gamma$ parameters account for this relative difference. We use the same technique for ACT, where we introduce $\gamma_0^{\rm ACT}$ and $\gamma_2^{\rm ACT}$. These are the noise levels in the highest-noise and lowest-noise subregions, relative to the median subregion (see Appendix \ref{appx:SZdepth}).

\begin{table*}
\centering
{
\renewcommand{\arraystretch}{1.5}
\begin{tabular}{l c c c c | l}
% \hline\hline
Parameter & DES $\times$ SPT & DES $\times$ ACT & DECADE $\times$ SPT & DECADE $\times$ ACT & Prior \\
\hline
& 275 matches & 1373 matches & 61 matches  & 1559 matches &  \\
& 3157 $\rm deg^2$ & 3978 $\rm deg^2$ & 1068 $\rm deg^2$  & 5229 $\rm deg^2$ &  \\
\hline\hline
$A_\lambda$ & $40.6 \pm 3.0$ & $41.1 \pm 1.1$ & $41.5^{+7.7}_{-7.4}$ & $42.3 \pm 1.7$ & $\mathcal{U}(20, 90)$ \\
$B_\lambda$ & $1.111 \pm 0.087$ & $1.142 \pm 0.036$ & $0.89 \pm 0.23$ & $1.187 \pm 0.045$ & $\mathcal{U}(0.4, 2)$ \\
$C_\lambda$ & $1.34 \pm 0.33$ & $0.86 \pm 0.16$ & $0.05 \pm 0.62$ & $1.25 \pm 0.17$ & $\mathcal{U}(-1, 3)$ \\
$\sigma_\lambda$ & $0.112^{+0.083}_{-0.075}$ & $0.075 \pm 0.050$ & $0.198^{+0.079}_{-0.094}$ & $0.157 \pm 0.101$ & $\mathcal{U}(0.01, 0.7)$ \\
$A_{\rm SZ}$ & $1.96 \pm 0.14$ & $3.45 \pm 0.13$ & $1.85 \pm 0.36$ & $3.57^{+0.26}_{-0.23}$ & $\mathcal{U}(0.5, 8)$ \\
$B_{\rm SZ}$ & $1.737 \pm 0.093$ & $1.811 \pm 0.045$ & $1.96 \pm 0.24$ & $1.908 \pm 0.049$ & $\mathcal{U}(1, 2.5)$ \\
$C_{\rm SZ}$ & $1.24 \pm 0.44$ & $0.36 \pm 0.25$ & $-0.45 \pm 0.95$ & $0.72 \pm 0.27$ & $\mathcal{U}(-2, 2)$ \\
$\sigma_{\rm SZ}$ & $0.373 \pm 0.081$ & $0.393^{+0.037}_{-0.034}$ & $0.33^{+0.18}_{-0.19}$ & $0.36^{+0.11}_ {-0.15}$ & $\mathcal{U}(0.05, 0.65)$ \\
$\rho$ & $0.06^{+0.33}_{-0.35}$ & $0.11^{+0.31}_{-0.34}$ & $0.03 \pm 0.34$ & $0.05 \pm 0.34$ & $\mathcal{U}(-0.5, 0.5)$ \\
\hline
$\gamma_{0}^{\rm SPT}$ & $0.812 \pm 0.045$ & -- & $0.576 \pm 0.093$ & -- & $\log\gamma \sim \mathcal{U}(-3, 3)$ \\
$\gamma_{0}^{\rm ACT}$ & -- & $0.802 \pm 0.021$ & -- & $0.757 \pm 0.022$ & $\log\gamma \sim \mathcal{U}(-3, 3)$ \\
$\gamma_{2}^{\rm ACT}$ & -- & $1.060 \pm 0.078$ & -- & $1.134 \pm 0.039$ & $\log\gamma \sim \mathcal{U}(-3, 3)$ \\
\hline
$\log_{10}M_{\rm 200c}^i$ & -- & -- & -- & -- & ${\rm HMF}(\Mtwohc, z)$ \\
\end{tabular}
}
\caption{The mean and standard deviation for the parameters in our Bayesian population model of richness and SZ cluster observables. This includes the joint scaling-relation parameters and per-region SZ noise-level corrections, $\gamma$, for each of the four possible combinations of optical and SZ survey datasets. See Equations \eqref{eq:zeta_relation}, \eqref{eq:lambdatrue_relation}, and \eqref{eqn:cov_mat_relation} in Appendix \ref{appx:tSZModel} for more a detailed description of these parameters. The values reported are mean and standard deviation of the posterior distribution. The posteriors are fairly Gaussian, with minimal overall correlations. However, the full triangle plots are still shown in Appendix \ref{appx:FullTriangle}. We also sample the halo masses for each object in the joint dataset as part of our analysis, where the prior distribution is the HMF at the redshift of the object. The table also shows how many clusters have measurements from both datasets, and the corresponding area of the overlap as determined using the surveys' masks.}
\label{tab:posterior_summary}
\end{table*}

\begin{figure}
    \centering
    \includegraphics[width=1\linewidth]{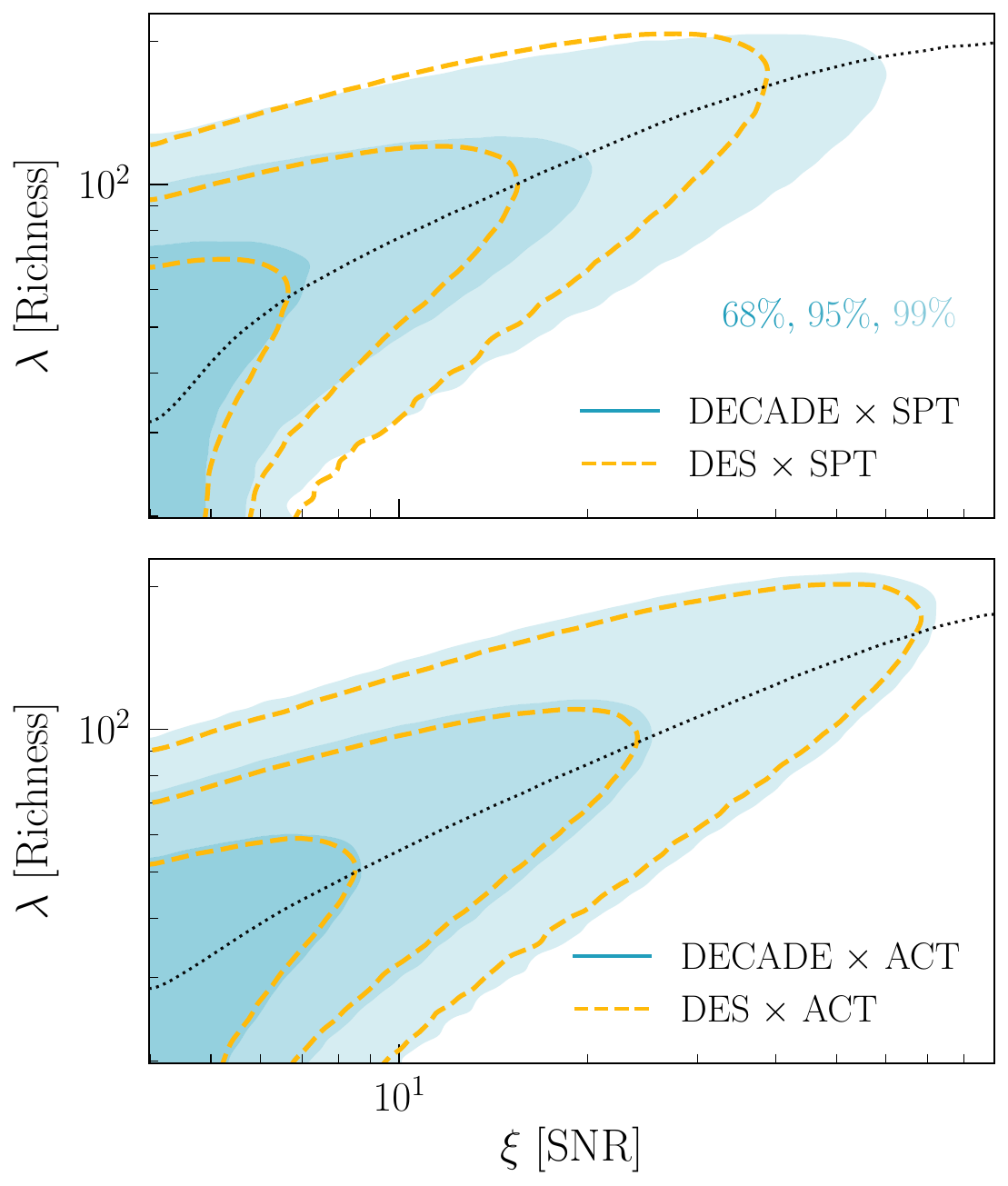}
    \caption{The distribution $P(\xi, \lambda)$ evaluated at $z = 0.6$. The contours for each data combination are obtained from the best-fit parameters in Table \ref{tab:posterior_summary}. DECADE and DES agree with each other in all cases. The $\xi$ measurements in ACT and SPT data have different normalization conventions and noise levels, so the x-axis of the distributions are offset relative to each other. The dotted line is obtained by finding the median of the distribution at fixed $\xi$.}
    \label{fig:tSZscaling}
\end{figure}

The primary goal of this section is to compare the scaling-relation behavior across the various dataset pairings, and to constrain the mass--richness relation so that mass estimates can be provided for individual clusters (see Section \ref{sec:MassCalibration:Scaling}). Our results for the four variants, alongside the chosen parameter priors, are provided in Table \ref{tab:posterior_summary}. The full posteriors are shown in Appendix \ref{appx:FullTriangle}. The priors follow the choices in \citet{bocquet2024sptclustersdeshst}. The posterior constraints on the scaling-relation parameters clearly demonstrate that the different data combinations are all within 1 to 2 $\sigma$ of each other and are therefore statistically consistent. This consistency is true regardless of the  pair of variants being compared. The only parameter that is statistically distinct is $A_{\rm SZ}$, where the values in the ACT analysis are higher than the SPT ones. This is expected as the noise levels and normalization convention are different across the two surveys.

The richness--mass scaling parameters are generally in agreement with results in the literature. The log-linear relation exhibits a slope with mass of $B_\lambda \approx 1$, following past works in observations and simulations \citep[\eg][]{Costanzi:2021:DESxSPT, Anbajagane2020StellarProp, DeRose:2022:Buzzard, DES:2025:6x2pt_Y6}. Similarly, the SZ--mass parameters are also consistent with existing results. In this case, we compare with SPT-based literature as reference results are not available for ACT. The SZ slope with mass broadly matches the self-similar scaling of $B_{\rm SZ} \approx 1.67$, while the slope with redshift matches $C_{\rm SZ} \approx 0.7$ \citep[\eg][]{Saro:2015:offsets, bocquet2024sptclustersdeshst}. The amplitude parameter, $A_{\rm SZ}$, for SPT is also consistent with that reported in \citet{bocquet2024sptclustersdeshst} for the SPT-SZ and SPT-ECS data. 

In Figure \ref{fig:tSZscaling}, we also show the joint probability, $P(\lambda, \xi)$, computed as
\begin{equation}
    P(\lambda, \xi) = \int d\Mtwohc P(\lambda, \xi | \Mtwohc) P(\Mtwohc),
\end{equation}
where $P(\lambda, \xi | \Mtwohc)$ is discussed in detail in Appendix \ref{appx:tSZModel}, and $P(\Mtwohc)$ is proportional to the halo mass function, also discussed in Appendix \ref{appx:tSZModel}. The above expression is evaluated at $z = 0.6$ for simplicity. Figure \ref{fig:tSZscaling} presents these likelihood contours, $P(\lambda, \xi)$, for each of the four combinations of data and find that the contours are qualitatively consistent with those in the literature \citep[\eg][see their Figure 10]{Bleem:2026:Clusters}.

Finally, the constraints on $\gamma_i$ generally deviate from $\gamma = 1$. The SPT-ECS field is found to be 25\% noisier (\ie $\gamma = 0.8$), consistent with Table 1 in \citet{Bleem:2020:ECS}. The DECADE $\times$ SPT analysis finds the field is noisier compared to DES $\times$ SPT. This is expected as most of the clusters in the former analysis are in the Northern Galactic Cap, where the SPT-ECS data are noisier relative to its Southern counterpart \citep{Bleem:2020:ECS}. Therefore, $\gamma$ is lower in the DECADE $\times$ SPT analysis than the DES $\times$ SPT one. The $\gamma$ parameters in the ACT analysis follow the expected qualitative trends from Appendix \ref{appx:SZdepth}, with the first (third) subregion being more (less) noisy. 

In summary, the results in Table \ref{tab:posterior_summary} show that the cross-matched variants of the catalogs exhibit behaviors consistent with each other, and consistent with data constraints and theoretical expectations from the literature.

\subsection{Miscentering}\label{sec:Validation:Miscenter}

For a given cluster, detections at different wavelengths---or even detections in independent surveys at the same wavelength---need not be assigned identical positions on the sky
\citep{George:2012:centering,Saro:2015:offsets,Rykoff:2016:redmapper,Zhang:2019:miscentering,Kelly:2024:miscentering}.
The separation between the cluster center in a catalog and the underlying true halo center, or the separation between cluster centers inferred from different observables, is commonly referred to as miscentering.
It can arise from observational or algorithmic effects, including misidentification of the central galaxy, masking or blending in optical data, and the finite angular resolution and noise of mm-wave observations \citep{Saro:2015:offsets,Rykoff:2016:redmapper,Zhang:2019:miscentering,Kelly:2024:miscentering}.
It can also reflect genuine astrophysical offsets between the galaxy, intracluster-gas, and dark-matter centroids in dynamically disturbed or merging systems
\citep{George:2012:centering,Seppi:2023:offsets}.

Characterizing the offset distribution therefore provides a useful validation test of the cluster finder's centering performance and an additional metric with which to compare the DECADE and DES catalogs produced in this work. Miscentering is also an important ingredient in analyses of projected halo profiles: averaging about displaced centers effectively convolves the centered profile with the offset distribution, smoothing the inner signal and potentially biasing inferred masses, concentrations, or other observables if it is neglected
\citep{Johnston:2007:miscentering,Simet:2017:lensing,McClintock:2019:ClusterWL, Anbajagane:2024:Shocks}.

Using our cross-matched catalogs, we quantify the miscentering between the \redmapper clusters and detections made in SPT, ACT, and eROSITA. Our model follows that of \citet[][see their Equation 24]{Bleem:2020:ECS}, and takes the form,
\begin{equation}\label{eqn:miscenter}
    P(x) = \rho \frac{1}{\sigma_\lambda} e^{-x/\sigma_\lambda} + (1 - \rho)\frac{x}{\tau^2} e^{-x/\tau} ,
\end{equation}
where $\rho$ is the fraction of well-centered objects, $\sigma_\lambda$ sets the miscentering scale for such objects, and $\tau$ is the scale for poorly centered objects (which form a thin tail to large offsets). Note that $\sigma_\lambda$ and $\rho$ do not have the same interpretations as they did in Table \ref{tab:posterior_summary}. We use the shorthand, $x = R_{\rm mis}/R_{\lambda}$, where the numerator is the miscentering offset and the denominator is the cluster radius defined within \redmapper, $R_{\lambda} = 1.0 (\lambda / 100)^{0.2} \mpc/h$ \citep{Rykoff:2014:AlgorithmCatalog}. The first term in Equation \eqref{eqn:miscenter} is an exponential while the second term is a Gamma distribution, $\Gamma(2, \tau)$. 

Equation \eqref{eqn:miscenter} represents the intrinsic miscentering of clusters, but does not account for the instrument beam and for uncertainty in the center identification. This is relevant for the mm-wave and X-ray data, which have significantly larger beams than the optical data. We account for this by convolving Equation \eqref{eqn:miscenter} with a circular 2D Gaussian of angular scale,
\begin{equation}
    \sigma_{\rm inst} = \frac{1}{\rm SNR}\sqrt{\theta_{\rm beam}^2 + \theta_c^2},
\end{equation}
which we convert to physical scale using the redshift of the cluster, and then normalize it by the radius, $R_\lambda$. The miscentering scale, $\sigma_{\rm inst}$, reduces with SNR as clusters detected at higher significance will have their centers more reliably identified. For SPT and ACT, we use $\xi$ as the SNR estimate, while for eROSITA, we compute SNR using the provided uncertainties on the luminosity, $L_{\rm 500c}$. We use the beam scales $\theta_{\rm beam} = 1.3\arcmin, 1.4\arcmin, 0.3\arcmin$ for SPT, ACT, and eROSITA, respectively. The cluster scale, $\theta_c$, is taken to be the angular extent of the cluster radius, $\Rtwohc$, at a given redshift.

We briefly comment on the sensitivity of the miscentering estimates to the matching procedure. As mentioned above, our procedure only keeps a matched pair if their positions are separated by $1.5 \mpc/h$ or less. This sets an upper limit on the largest miscentering separations that we allow. We have tested larger separation limits, and find the distribution of offsets exhibits an increase in probability for $R_{\rm mis} > 1.5 \mpc/h$ that is unphysical as it originates predominantly from increased random matches due to a larger search area. Thus, we fix the distribution to $R_{\rm mis} < 1.5 \mpc/h$, following the choices of \citet{Bleem:2020:ECS}.

\begin{figure}
    \centering
    \includegraphics[width=\columnwidth]{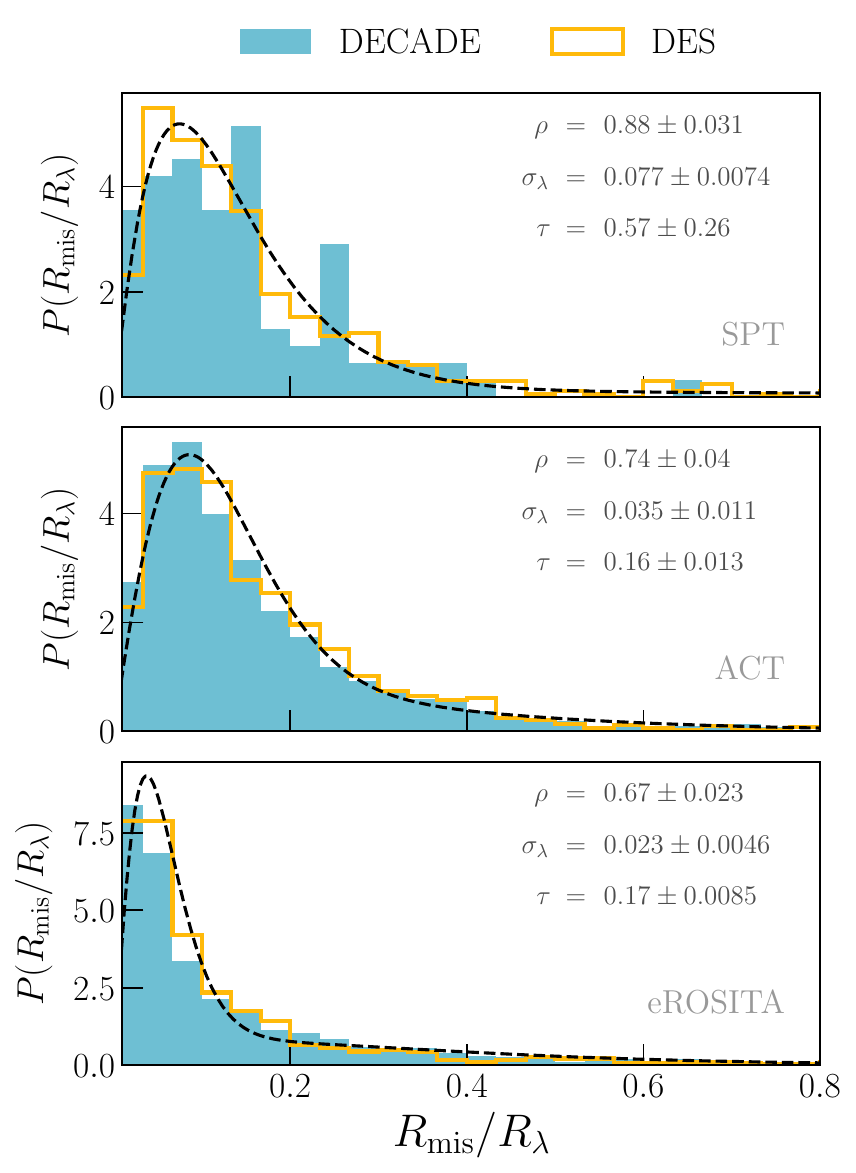}
    \caption{The miscentering estimates from DECADE and DES, alongside a joint fit to both datasets. There is broad agreement in the distributions between DECADE and DES, and both can be well fit by Equation \eqref{eqn:miscenter}. The medians of the model parameters' posteriors are listed in the top right of each panel, alongside their associated $1\sigma$ uncertainties. The eROSITA distribution appears qualitatively different at small separations in part because the beam for the X-ray data is three to four times smaller than that of the mm-wave data.}
    \label{fig:miscentering}
\end{figure}

Figure \ref{fig:miscentering} presents our measurements and the associated fits, for three reference catalogs in the cross-matches: SPT, ACT, and eROSITA. In the case of SPT, we recover parameters broadly consistent with \citet{Bleem:2020:ECS}. The distribution for ACT looks fairly similar to that of SPT, but indicates a mildly stronger tail as $\rho$ exhibits lower values. The eROSITA distribution is qualitatively different from the SPT and ACT distributions, particularly towards small scales, because the beam of the former dataset is significantly smaller than that of the latter datasets. In summary, the miscentering estimates show that the DES and DECADE datasets have comparable centering accuracy, and both also agree with prior results \citep[\eg][]{Zhang:2019:miscentering, Bleem:2020:ECS, Kelly:2024:miscentering}.

\section{Cluster mass estimates} \label{sec:MassCalibration}

Cluster masses are a vital quantity in studies of cluster cosmology and astrophysics \citep{Allen:2011:Clusters}. In Section \ref{sec:MassCalibration:Scaling} we use the scaling-relation constraints of Section \ref{sec:Validation:Scaling:SZ} to provide mass estimates for all clusters in our catalogs. The scaling relations where obtained with data from $0.2 < z < 0.65$, and clusters outside this interval are assigned masses by extrapolating the relation. We then measure weak-lensing profiles around each cluster. These measurements, binned in richness and redshift, are described in detail in Section \ref{sec:MassCalibration:WL}. Both the individual profiles and the binned measurements are made publicly available. While we do not fit lensing profiles to the measurements to directly constrain cluster masses, we discuss our estimators and validation process instead. More precise mass-calibration efforts with these lensing measurements are left to future work from the community. 

\subsection{Masses from the joint optical--tSZ analysis}\label{sec:MassCalibration:Scaling}

\begin{figure}
    \centering
    \includegraphics[width=\columnwidth]{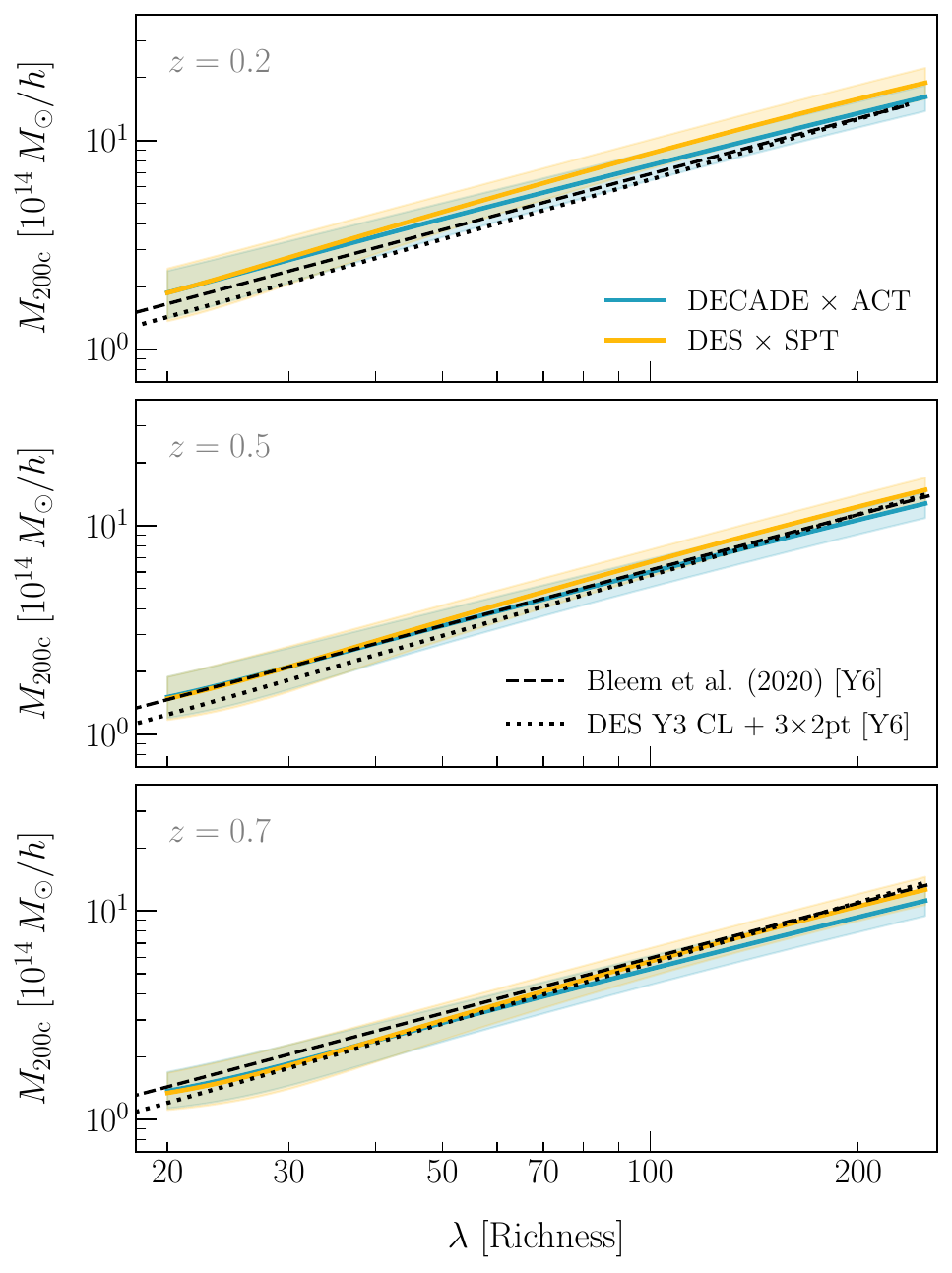}
    \caption{The mass--richness relation estimated from our joint optical--SZ analysis (Section \ref{sec:Validation:Scaling:SZ}) in comparison to other estimates in the literature; the SPT analysis of \citet{Bleem:2020:ECS} and the DES Y3 analysis of \citet{DES:2025:6x2pt_Y6}. We convert the mass definitions of the latter two works to match the $\Mtwohc$ relation used here, and we also account for differences in the \redmapper richness estimators relative to that of our work (Appendix \ref{appx:RichnessCompare:IDL}). There is broad agreement across the different constraints.}
    \label{fig:richness_relation}
\end{figure}

The mass estimates in our catalogs are obtained from the scaling relations extracted in Section \ref{sec:Validation:Scaling:SZ} via the optical--SZ joint analysis. The constrained richness--mass relation, $P(\lambda \,|\, \Mtwohc, z)$, can then be inverted to obtain $P(\Mtwohc \,|\, \lambda, z)$ using the Bayesian expression
\begin{equation}\label{eqn:MassPosterior}
    P(\Mtwohc \,|\, \lambda, z) = \frac{P(\lambda \,|\, \Mtwohc, z)P(\Mtwohc \,|\, z)}{P(\lambda \,|\, z)}
\end{equation}
where $P(\Mtwohc | z)$ is proportional to the halo mass function, and $P(\lambda | z)$ is effectively a normalization constant and cancels when normalizing the distribution to $\int d\Mtwohc P(\Mtwohc | \lambda, z) = 1$. Using Equation \eqref{eqn:MassPosterior}, we can condition on the observed richness of a given cluster and assign it a mass posterior. We compress this posterior into the mean and the $1\sigma$ uncertainty, and report them as part of our catalog for each detected cluster.

Figure \ref{fig:richness_relation} presents the resulting relation, $\langle \Mtwohc \,|\, \lambda, z \rangle$ for a few redshifts. For brevity, we only show the relation calibrated as part of the DES $\times$ SPT and DECADE $\times$ ACT analyses. We compare against the relations presented in \citet{Bleem:2020:ECS} and \citet{DES:2025:6x2pt_Y6}. The former used an optical--SZ joint analysis as well, while the latter used measurements of cluster lensing, cluster clustering, and galaxy clustering for the calibration. Both works utilized an older \redmapper catalog whose richness estimator is not the same as the one used in this work.\footnote{Particularly, these catalogs were derived from the IDL variant of \redmapper (v6.4.22+2). Our work uses the latest Python-based version, with version tags listed in Section \ref{sec:ClusterFinding:Redmapper}, which has some modifications to the matched filter used in the richness estimator.} As a result, we do not expect precise quantitative agreement. To remedy this, we build a data-driven relation connecting the two richness estimators (Appendix \ref{appx:RichnessCompare:IDL}) and apply it to the richness variable in the mass--richness relations taken from the literature. Figure \ref{fig:richness_relation} shows our relations are indeed in good agreement with the results from the literature. It also shows that the relations derived from the different combinations of DES, SPT, DECADE, and ACT are consistent, as is expected from the results noted in Table \ref{tab:posterior_summary}.

In practice, the cluster mass-assignment step follows a three-tier system. We first assign masses to all clusters using the mass--richness relation derived from the DES $\times$ SPT analysis. For the subset of clusters that have both richness and SZ measurements, we have sampled their mass posteriors during the analysis of Section \ref{sec:Validation:Scaling:SZ}, and so directly use this posterior instead. Finally, if a DECADE or DES cluster was part of both the SPT and ACT analyses, then we use the mass posterior associated with the SPT analysis given its more homogeneous noise properties. The released catalogs contain flags for identifying the origin of the mass estimate for a given cluster.

\subsection{Weak lensing profiles}\label{sec:MassCalibration:WL}

\begin{figure*}
    \centering
    \includegraphics[width=2\columnwidth]{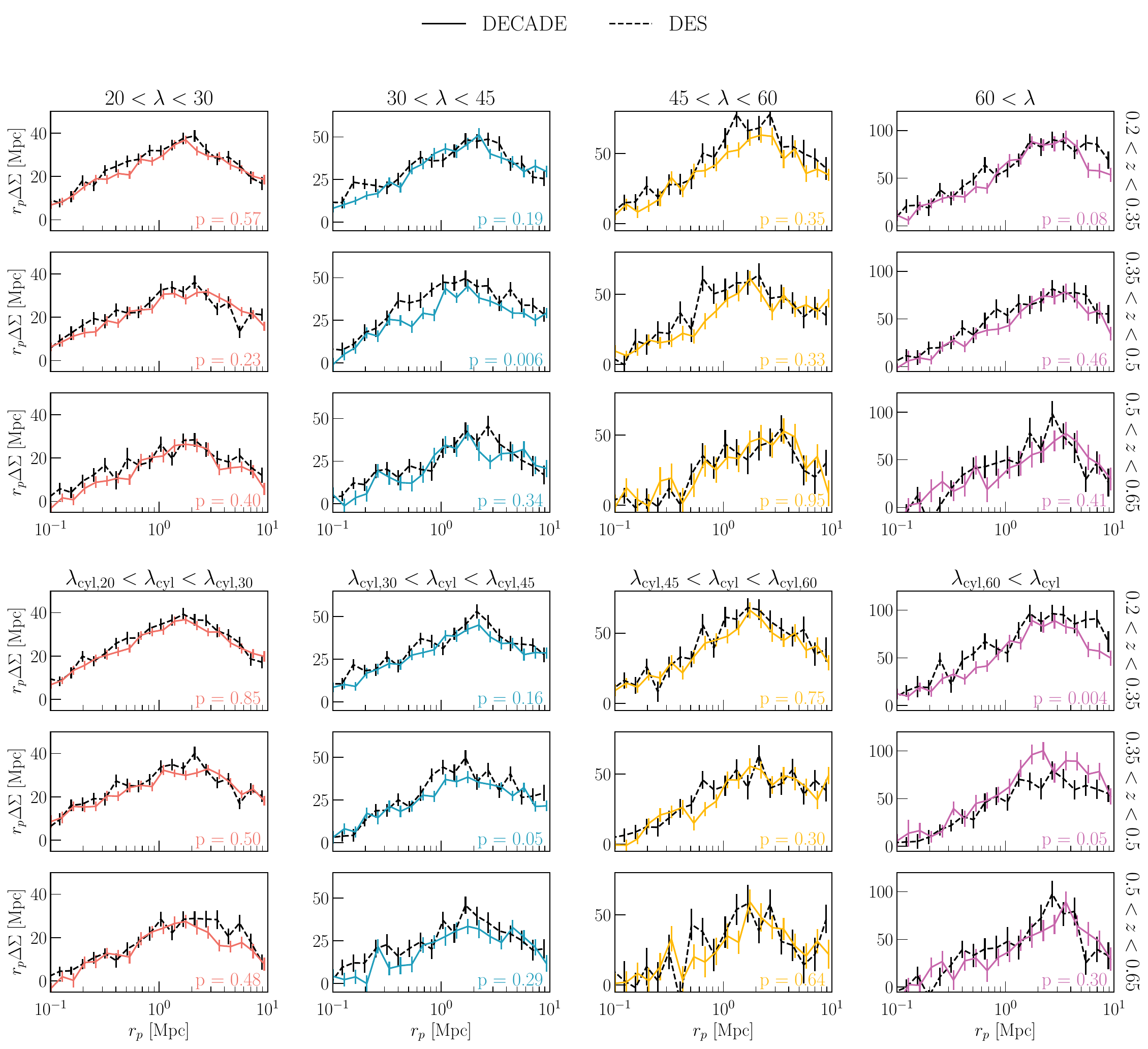}
    \caption{The weak lensing profiles of clusters in bins of $\lambda$ (or $\lambdacyl$) and $z$. Each panel shows the profiles for both DECADE and DES clusters, alongside a p-value corresponding to the $\chi^2$ estimated between the two profiles. The joint probability of consistency across all twelve bins is $p_{\rm comb} > 0.05$ (for both $\lambda$ and $\lambdacyl$ cases) which indicates that the richness--mass relations of the two datasets are consistent with each other. The joint estimate leverages the fact that the uncertainties are well-approximated as being independent across bins \citep{Wu:2019:ClusterWLCov, McClintock:2019:ClusterWL}, and then follows Appendix A of \citet{paper5} in using Fisher's method \citep{Fisher:1925} to combine the probabilities. Error bars are obtained using jackknife resampling over 100 spatial patches within each survey footprint. The bin edges for $\lambdacyl$ are obtained by abundance matching against $\lambda$; see Section \ref{sec:MassCalibration:WL} for details.}
    \label{fig:DeltaSigma}
\end{figure*}

Weak gravitational lensing \citep{Bartelmann2001, Schneider:2005:Lensing} provides a robust way to calibrate the mean masses of a given sample of galaxy clusters. As a result, it is the main mass calibration approach in cluster cosmology analyses across different surveys and wavelengths \citep[\eg][]{McClintock:2019:ClusterWL, bocquet2024sptclustersdeshst, Grandis:2024:eROSITAMass, Kleinebreil:2025:eROSITAMass}. In this work, we provide measurements of the weak lensing profiles (including the mean profiles in bins of richness and redshift) for the DECADE and DES clusters.

Our discussion below focuses solely on the binned measurements. We first divide the cluster sample into bins of richness and redshift, using $\lambda \in [20, 30, 45, 60, \infty]$ and $z \in [0.2, 0.35, 0.5, 0.65]$ as the bin edges. Note that the last redshift bin ($0.5 < z < 0.65$) for the DECADE sample incurs depth-related selection effects that are calibrated within \redmapper (Appendix \ref{appx:SurveyArea}). Some analyses may choose to discard this bin; for completeness, we still discuss this bin and perform the systematics validation tests. We also bin the clusters according to the $\lambdacyl$ observable described in Section \ref{sec:ClusterFinding:cylinder}. In this case, we determine the richness thresholds using abundance matching. Within a given redshift bin, we determine the percentiles corresponding to the bin-edges in $\lambda$ and find their corresponding values in $\lambdacyl$. We denote the latter thresholds as $\lambda_{\rm cyl, X}$, determined using the percentile where $\lambda = X$.

Having defined the binning, we measure the average weak lensing profile around the clusters in a given bin. This is done using the DECADE \citep{paper1, paper5} and DES \citep{y3-shapecatalog} source catalogs. For the latter, we use the DES Y3 catalog, which is the latest public DES dataset at the time of this work. The DES Y6 data has $1.3\times$ the number of weak lensing source galaxies, and will enable more precise mass calibration. Our main observable is the excess surface density, $\Delta \Sigma$. Following \cite{2025arXiv250920458T}, this is estimated from the galaxy shapes as,
\begin{eqnarray}\label{eqn:deltasigma}
    \Delta \Sigma (r) &=& \Delta \Sigma_L (r)-\Delta \Sigma_R(r) \nonumber, \\[10pt]
    \Delta \Sigma_{\alpha\in{L,R}} (r) &=&\frac{\sum_{\alpha, s} w_s \langle \Sigma^{-1}_{\rm{crit}}\rangle_{\alpha,s} \bm{e_{t,s}}}{\sum_{\alpha, s} w_s \langle \Sigma^{-1}_{\rm{crit}}\rangle_{\alpha,s}^2}\nonumber\\ 
    &&\times \frac{\sum_{\alpha, s} w_s \langle \Sigma^{-1}_{\rm{crit}}\rangle_{\alpha,s}^2}{\sum_{\alpha, s} w_s (\langle R_t \rangle_l + \langle R_s\rangle)\langle \Sigma^{-1}_{\rm{crit}} \rangle_{\alpha,s}^2}\nonumber\\
    &&\times \frac{1}{B(r)}, 
\end{eqnarray}
where $L$ and $R$ denote the cluster (lens) and random catalogs, respectively, $w_s$ are the source weights provided by the shape catalogs, and $\bm{e_t}$ are the tangential shapes of source galaxy $s$ around lens $L$ or random point $R$. The inverse critical surface density for a lens at redshift $z_l$ and source at redshift $z_s$ is \begin{equation}
\Sigma_{\rm crit}^{-1}(z_l,z_s)
=
\frac{4\pi G}{c^2}
\frac{D_{\rm A}(z_l)D_{\rm A}(z_l,z_s)}
     {D_{\rm A}(z_s)}
\Theta(z_s-z_l),
\end{equation}
where $D_{\rm A}(z_l)$ and $D_{\rm A}(z_s)$ are the angular-diameter
distances to the lens and source, respectively, and
$D_{\rm A}(z_l,z_s)$ is the angular-diameter distance between them.
We account for uncertainty in the source redshift by averaging over
its redshift probability distribution,
\begin{equation}
\left\langle\Sigma_{\rm crit}^{-1}\right\rangle_{\alpha,s}
=
\int {\rm d}z_s\,
p_s(z_s)\,
\Sigma_{\rm crit}^{-1}(z_\alpha,z_s).
\end{equation}

In Equation \eqref{eqn:deltasigma}, the first factor is the minimum variance estimator of $\Delta \Sigma$. The second factor is the weighted mean of the \textsc{Metacalibration} shear response. The third factor is the boost factor correction derived using the $p(z)$ decomposition method \citep{2019MNRAS.489.2511V,2021PhDT........16P, 2025arXiv250920458T}. This corrects for the dilution in lensing signal caused by some source galaxies being at the same redshift as (or being members of) the clusters. The redshift point estimates for this third factor are obtained using the Directional Neighborhood Fitting (DNF) algorithm \citep{DNF} run on the DECADE and DES SOF-based photometry. The full profile calculation is done using a modified pipeline based on \textsc{dsigma} v1.1.0 \citep{2022ascl.soft04006L}.\footnote{The original repo is in \url{https://github.com/johannesulf/dsigma}, and the modified variant is \url{https://github.com/chto/dsigma}}

Figure \ref{fig:DeltaSigma} presents our measurements for the different cluster bins. The amplitude of the signal increases with richness as expected (large $\lambda$ clusters have more mass, and therefore higher lensing amplitudes). The uncertainties are obtained from jackknife estimates across 100 spatial patches in a given survey. The covariance matrices are essentially diagonal in nature as the source-galaxy shape-noise contribution dominates the overall error budget \citep[\eg][]{McClintock:2019:ClusterWL}. Using these error estimates, we compare the $\Delta \Sigma$ profiles across DES and DECADE, focusing on one bin at a time. If the mass--richness relation is the same across the two datasets (as indicated by the results in Table \ref{tab:posterior_summary}), which in turn means the richness estimates are comparable across the datasets, then we expect the lensing profiles to be consistent. The bottom of each panel shows the $p$-values of this comparison, and we find all profiles are consistent within $3\sigma$ across the scales, $r_p > 0.1 \mpc/h$. The joint probability, which can be estimated using the same approach as Appendix A in \citet{paper5} --- given the profile uncertainties are essentially uncorrelated across cluster redshift/richness bins \citep{McClintock:2019:ClusterWL, Wu:2019:ClusterWLCov} --- also indicates consistency within $3\sigma$. This is strong evidence that the techniques and calibrations used for DES cluster mass calibration can be readily applied to the DECADE sample as well.

\begin{figure}
    \centering
    \includegraphics[width=\columnwidth]{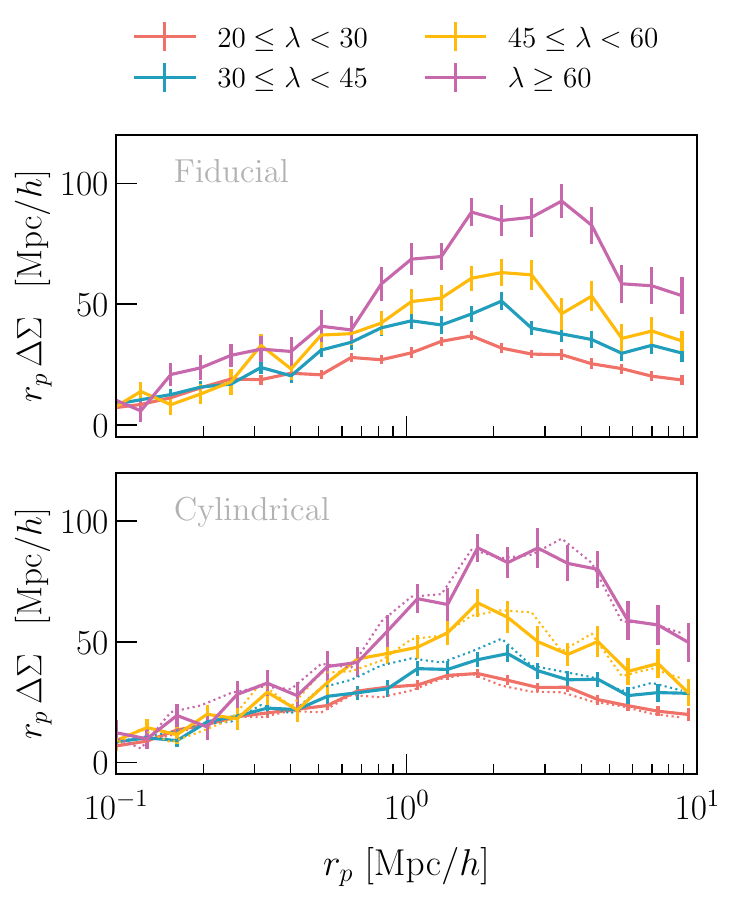}
    \caption{A comparison of the lensing profiles in the first redshift bin, $0.2 < z < 0.35$, when binned by the \redmapper richness (top), and by the cylindrical richness (bottom). The richness-based estimates are repeated in the bottom panel as dotted lines for easy comparison. The lensing measurements across the different bins are separated by similar factors regardless of the choice of binning variable, indicating that the cylindrical richness is a similarly precise observable as the fiducial \redmapper richness. The $\lambdacyl$ bins are defined following Figure \ref{fig:DeltaSigma}.}
    \label{fig:RM_vs_cyl}
\end{figure}

A key concern, particularly in DECADE, is the impact of inhomogeneous survey conditions (depth, seeing, etc.) on the richness estimates and the lensing-based mass calibrations. We quantify this impact extensively in Appendix \ref{appx:Inhomog} and show the effects are subdominant. In particular, Appendix \ref{appx:Inhomog} shows the consistency between cluster lensing profiles measured in sky regions where a given observing condition is higher/lower than the median value across the survey. In all cases, we find the two profiles are consistent within $3\sigma$. Thus, the mass--richness relation in DECADE does not exhibit a statistically significant bias from spatially varying systematics and shows no sign of such systematics being larger than that of DES. This is somewhat expected as the inhomogeneity in DECADE observing conditions affects objects near the detection threshold. \citet{paper3} noted that the source galaxy sample was significantly brighter than this threshold and already exhibited only minimal-to-no dependence on the observing conditions. The cluster sample is at lower redshifts, where the cluster member galaxies are observed at much higher SNR and therefore have far less sensitivity to observing conditions. Furthermore, as discussed in Section \ref{sec:ClusterFinding:Redmapper}, \redmapper accounts for incompleteness in the survey (either due to masking or noisy photometric data) in a data-driven way. As a result, our findings here are unsurprising.

Section \ref{sec:ClusterFinding:cylinder} discusses the benefits of $\lambdacyl$ as the primary cluster observable. In Figure \ref{fig:RM_vs_cyl}, we show that within a fixed redshift bin, the WL profiles from the $\lambda$-based binning exhibit similar separation to that from the $\lambdacyl$-based binning. Following the arguments in \citet{Huang:2022:OuterStellarMass}, this separation in $\Delta \Sigma$ provides key evidence that the cylinder-based richness is sufficiently precise to serve as an alternative observable to the traditional \redmapper richness.

As mentioned previously, we do not fit mass profiles to these lensing measurements. There are a variety of different methods for performing such an estimate, ranging from analytic to simulation-based approaches \citep[\eg][]{McClintock:2019:ClusterWL, 2024PhRvL.133v1002S, 2025arXiv251025706S, Andylee}. We release the lensing profiles and their error estimates to the community so anyone can perform the mass modeling of their choice.

\section{Summary} \label{sec:Summary}

Galaxy clusters provide a rich avenue to study both astrophysics and cosmology, and particularly to study the interaction of the two. We run the red-sequence cluster-finding algorithm, \redmapper, on 13,000 $\deg^2$ of photometric data from DECam, using coadded images from DELVE DR3 and DES DR2. We characterize the catalogs from these datasets through a variety of methods, and find that they show consistent behavior with existing \redmapper catalogs in the literature. We also test a modified optical cluster observable, which uses the counts of LRGs within cylinders, as a way of simplifying the cluster modeling. Our tests include this observable as well.

Our main findings can be summarized as follows:
\begin{itemize}
    \item The \redmapper catalogs deliver redshift estimates with a precision of $\sigma_z / (1 + z) \leq 0.008$ across $0.2 < z < 0.65$. The reported Gaussian uncertainties within \redmapper are well-calibrated, and the outlier rate is sub-percent (Figure \ref{fig:Catalog}).

    \item The $5\times$ larger spectroscopic calibration sample in DECADE relative to that of DES does not result in an improved redshift precision/accuracy in DECADE (Figure \ref{fig:Catalog}) as the redshift performance is predominantly set by the red-sequence relation and by the photometric uncertainty in our data.

    \item The X-ray--richness scaling relation is qualitatively consistent with results in the literature (Figure \ref{fig:Xrayscaling}), exhibiting a power-law scaling of roughly $L_{\rm 500c} \propto \lambda^{1.3}$.

    \item Our Bayesian population model for the optical--tSZ measurements provides parameter constraints that are consistent across all four possible pairs of DECADE/DES and SPT/ACT (Table \ref{tab:posterior_summary} and Figure \ref{fig:tSZscaling}). The derived mass--richness scaling relation is consistent with existing estimates (Figure \ref{fig:richness_relation}).

    \item When applying the same richness/redshift selection to DECADE and DES, the mean cluster WL profiles are consistent within $3\sigma$ (Figure \ref{fig:DeltaSigma}). Thus, the mass--richness relations of DES and DECADE are consistent with each other.

    \item Binning cluster by the traditional \redmapper richness $\lambda$, or the modified cylindrical richness, $\lambdacyl$ leads to bin-averaged cluster WL profiles with similar separations (Figure \ref{fig:RM_vs_cyl}). This provides strong evidence that $\lambdacyl$ is a sufficiently precise cluster observable for use in cosmology analyses.   

    \item Spatial inhomogeneity in survey conditions such as seeing, magnitude limit, stellar density, extinction, etc. has a negligible effect on the redshift and richness estimates (Figure \ref{fig:redshift_inhomog} and \ref{fig:richness_inhomog}). The cluster WL profiles are also statistically consistent when subdividing clusters into quantiles based on a survey systematic (Table \ref{tab:DsigmaSys}), and this is true when binning by \redmapper richness and by cylinder-based richness.
\end{itemize}

Our results showcase the reliability of \redmapper-based cluster finding, and subsequent cluster WL measurements, against variations in survey conditions such as seeing, stellar density, and imaging depth. This builds on previous validations of \redmapper cluster catalogs and their WL measurements \citep{Rykoff:2014:AlgorithmCatalog,McClintock:2019:ClusterWL}. Our results provide an empirical benchmark for cluster finding in next-generation imaging datasets from the Vera C. Rubin Observatory \citep{LSST2018SRD}, the Nancy Grace Roman Space Telescope \citep{Spergel:2015:Roman}, and the \textit{Euclid} space mission \citep{Euclid}. The combined DECADE and DES cluster catalog provides a wide-field, precision cluster sample to use in preparing cluster-analysis pipelines for those surveys. Furthermore, the wide-area coverage brings improvements to multi-wavelength cluster observations and enables a broad range of science applications, including measurements of cluster profiles \citep[\eg][]{Jimeno:2018:redMaPPerGas,Shin:2019:Splashback,Anbajagane:2024:Shocks}, constraints on baryonic feedback and the baryon content of halos \citep[\eg][]{Popesso:2026:GasFraction}, and cluster cosmology \citep[\eg][]{2025arXiv251025706S, DES:2025:6x2pt_Y6}. The catalogs and their supplementary products are made publicly available alongside this work to facilitate these and other science applications.

\section*{Acknowledgements}

DA is supported by the National Science Foundation (NSF) Graduate Research Fellowship under Grant No.\ DGE 1746045. CHT is supported by the Eric and Wendy Schmidt Fellowship.
CC is supported by the Department of Energy (DOE) grant DE-SC0021949. 
The DECADE project is supported by NSF AST-2108168 and AST-2108169. We acknowledge the use of ChatGPT 5.6 for literature searches, and of Claude Code for implementing fast numerical integrals used in both the HMC analysis and the miscentering analysis. No LLMs were used in writing the draft text, in producing figures, or in other aspects of the analysis.

The DELVE Survey gratefully acknowledges support from Fermilab LDRD (L2019.011), the NASA {\it Fermi} Guest Investigator Program Cycle 9 (No.\ 91201), and the NSF (AST-2108168, AST-2108169, AST-2307126,  AST-2407526, AST-2407527, AST-2407528). This work was completed in part with resources provided by the University of Chicago’s Research Computing Center. The project that gave rise to these results received the support of a fellowship from "la Caixa" Foundation (ID 100010434). The fellowship code is LCF/BQ/PI23/11970028. C.E.M.-V. is supported by the international Gemini Observatory, a program of NSF NOIRLab, which is managed by the Association of Universities for Research in Astronomy (AURA) under a cooperative agreement with the U.S. National Science Foundation, on behalf of the Gemini partnership of Argentina, Brazil, Canada, Chile, the Republic of Korea, and the United States of America.

Funding for the DES Projects has been provided by the U.S. Department of Energy, the U.S. National Science Foundation, the Ministry of Science and Education of Spain, 
the Science and Technology Facilities Council of the United Kingdom, the Higher Education Funding Council for England, the National Center for Supercomputing 
Applications at the University of Illinois at Urbana-Champaign, the Kavli Institute of Cosmological Physics at the University of Chicago, 
the Center for Cosmology and Astro-Particle Physics at the Ohio State University,
the Mitchell Institute for Fundamental Physics and Astronomy at Texas A\&M University, Financiadora de Estudos e Projetos, 
Funda{\c c}{\~a}o Carlos Chagas Filho de Amparo {\`a} Pesquisa do Estado do Rio de Janeiro, Conselho Nacional de Desenvolvimento Cient{\'i}fico e Tecnol{\'o}gico and 
the Minist{\'e}rio da Ci{\^e}ncia, Tecnologia e Inova{\c c}{\~a}o, the Deutsche Forschungsgemeinschaft and the Collaborating Institutions in the Dark Energy Survey. 

The Collaborating Institutions are Argonne National Laboratory, the University of California at Santa Cruz, the University of Cambridge, Centro de Investigaciones Energ{\'e}ticas, 
Medioambientales y Tecnol{\'o}gicas-Madrid, the University of Chicago, University College London, the DES-Brazil Consortium, the University of Edinburgh, 
the Eidgen{\"o}ssische Technische Hochschule (ETH) Z{\"u}rich, 
Fermi National Accelerator Laboratory, the University of Illinois at Urbana-Champaign, the Institut de Ci{\`e}ncies de l'Espai (IEEC/CSIC), 
the Institut de F{\'i}sica d'Altes Energies, Lawrence Berkeley National Laboratory, the Ludwig-Maximilians Universit{\"a}t M{\"u}nchen and the associated Excellence Cluster Universe, 
the University of Michigan, NSF's NOIRLab, the University of Nottingham, The Ohio State University, the University of Pennsylvania, the University of Portsmouth, 
SLAC National Accelerator Laboratory, Stanford University, the University of Sussex, Texas A\&M University, and the OzDES Membership Consortium.

The DES data management system is supported by the National Science Foundation under Grant Numbers AST-1138766 and AST-1536171.
The DES participants from Spanish institutions are partially supported by MICINN under grants ESP2017-89838, PGC2018-094773, PGC2018-102021, SEV-2016-0588, SEV-2016-0597, and MDM-2015-0509, some of which include ERDF funds from the European Union. IFAE is partially funded by the CERCA program of the Generalitat de Catalunya.
Research leading to these results has received funding from the European Research
Council under the European Union's Seventh Framework Program (FP7/2007-2013) including ERC grant agreements 240672, 291329, and 306478.
We  acknowledge support from the Brazilian Instituto Nacional de Ci\^encia
e Tecnologia (INCT) do e-Universo (CNPq grant 465376/2014-2).

%(NOIRLab Prop. ID 2012B-0001; PI: J. Frieman)
Based in part on observations at Cerro Tololo Inter-American Observatory at NSF's NOIRLab, which is managed by the Association of Universities for Research in Astronomy (AURA) under a cooperative agreement with the National Science Foundation.

This work has made use of data from the European Space Agency (ESA) mission {\it Gaia} (\url{https://www.cosmos.esa.int/gaia}), processed by the {\it Gaia} Data Processing and Analysis Consortium (DPAC, \url{https://www.cosmos.esa.int/web/gaia/dpac/consortium}).
Funding for the DPAC has been provided by national institutions, in particular the institutions participating in the {\it Gaia} Multilateral Agreement.

This paper is based on data collected at the Subaru Telescope and retrieved from the HSC data archive system, which is operated by the Subaru Telescope and Astronomy Data Center (ADC) at NAOJ. Data analysis was in part carried out with the cooperation of Center for Computational Astrophysics (CfCA), NAOJ. We are honored and grateful for the opportunity of observing the Universe from Maunakea, which has the cultural, historical and natural significance in Hawaii. 

This manuscript has been authored by Fermi Forward Discovery Group, LLC under Contract No.\ 89243024CSC000002 with the U.S. Department of Energy, Office of Science, Office of High Energy Physics.

All analysis in this work was enabled greatly by the following software: \textsc{Pandas} \citep{Mckinney2011pandas}, \textsc{NumPy} \citep{vanderWalt2011Numpy}, \textsc{SciPy} \citep{Virtanen2020Scipy}, \textsc{Matplotlib} \citep{Hunter2007Matplotlib}, and \textsc{Getdist} \citep{Lewis:2019:Getdist}. We have also used
the Astrophysics Data Service (\href{https://ui.adsabs.harvard.edu/}{ADS}) and \href{https://arxiv.org/}{\texttt{arXiv}} preprint repository extensively during this project and the writing of the paper.

\section*{Data Availability}

All cluster catalogs and measurements made in this work are publicly available at via Globus. Please visit \url{dhayaaanbajagane.github.io/data_release/decade_clusters} for a list of the available dataproducts and their corresponding data access. The details. The SOF photometry that underlies this work is publicly available as part of the DELVE DR3 release at \url{https://datalab.noirlab.edu/data/delve}. 

\bibliographystyle{mnras}
\bibliography{References}

@ARTICLE{paper1,
       author = {{Anbajagane}, D. and {Chang}, C. and {Zhang}, Z. and {Tan}, C.~Y. and {Adamow}, M. and {Secco}, L.~F. and {Becker}, M.~R. and {Ferguson}, P.~S. and {Drlica-Wagner}, A. and {Gruendl}, R.~A. and {Herron}, K. and {Tong}, A. and {Troxel}, M.~A. and {Sanchez-Cid}, D. and {Sevilla-Noarbe}, I. and {Chicoine}, N. and {Teixeira}, R. and {Alarcon}, A. and {Suson}, D. and {Alsina}, A.~N. and {Amon}, A. and {Bom}, C.~R. and {Carballo-Bello}, J.~A. and {Cerny}, W. and {Choi}, A. and {Choi}, Y. and {Doux}, C. and {Eckert}, K. and {Gatti}, M. and {Gruen}, D. and {James}, D.~J. and {Jarvis}, M. and {Kuropatkin}, N. and {Mart{\'\i}nez-V{\'a}zquez}, C.~E. and {Massana}, P. and {Mau}, S. and {McCullough}, J. and {Medina}, G.~E. and {Mutlu-Pakdil}, B. and {Navabi}, M. and {No{\"e}l}, N.~E.~D. and {Pace}, A.~B. and {Prat}, J. and {Raveri}, M. and {Riley}, A.~H. and {Rykoff}, E.~S. and {Sakowska}, J.~D. and {Sand}, D.~J. and {Santana-Silva}, L. and {Shin}, T. and {Soares-Santos}, M. and {Stringfellow}, G.~S. and {Vivas}, A.~K. and {Yamamoto}, M.},
        title = "{The DECADE cosmic shear project I: A new weak lensing shape catalog of 107 million galaxies}",
      journal = {arXiv e-prints},
         year = 2025,
        month = feb,
          eid = {arXiv:2502.17674},
        pages = {arXiv:2502.17674},
          doi = {10.48550/arXiv.2502.17674},
archivePrefix = {arXiv},
       eprint = {2502.17674},
 primaryClass = {astro-ph.CO},
       adsurl = {https://ui.adsabs.harvard.edu/abs/2025arXiv250217674A}
}

@ARTICLE{2009ApJ...703..601R,
       author = {{Rozo}, Eduardo and {Rykoff}, Eli S. and {Koester}, Benjamin P. and {McKay}, Timothy and {Hao}, Jiangang and {Evrard}, August and {Wechsler}, Risa H. and {Hansen}, Sarah and {Sheldon}, Erin and {Johnston}, David and {Becker}, Matthew and {Annis}, James and {Bleem}, Lindsey and {Scranton}, Ryan},
        title = "{Improvement of the Richness Estimates of maxBCG Clusters}",
      journal = {\apj},
         year = 2009,
        month = sep,
       volume = {703},
       number = {1},
        pages = {601-613},
          doi = {10.1088/0004-637X/703/1/601},
       adsurl = {https://ui.adsabs.harvard.edu/abs/2009ApJ...703..601R}
}

@ARTICLE{cardinal,
       author = {{To}, Chun-Hao and {DeRose}, Joseph and {Wechsler}, Risa H. and {Rykoff}, Eli and {Wu}, Hao-Yi and {Adhikari}, Susmita and {Krause}, Elisabeth and {Rozo}, Eduardo and {Weinberg}, David H.},
        title = "{Buzzard to Cardinal: Improved Mock Catalogs for Large Galaxy Surveys}",
      journal = {\apj},
         year = 2024,
        month = jan,
       volume = {961},
       number = {1},
          eid = {59},
        pages = {59},
          doi = {10.3847/1538-4357/ad0e61},
archivePrefix = {arXiv},
       eprint = {2303.12104},
 primaryClass = {astro-ph.CO},
       adsurl = {https://ui.adsabs.harvard.edu/abs/2024ApJ...961...59T}
}

@ARTICLE{Andylee,
       author = {{Lee}, Andy and {Wu}, Hao-Yi and {Salcedo}, Andr{\'e}s N. and {Sunayama}, Tomomi and {Costanzi}, Matteo and {Myles}, Justin and {Cao}, Shulei and {Rozo}, Eduardo and {To}, Chun-Hao and {Weinberg}, David H. and {Yang}, Lei and {Zhou}, Conghao},
        title = "{Optical galaxy cluster mock catalogs with realistic projection effects: Validations with the SDSS clusters}",
      journal = {\prd},
         year = 2025,
        month = mar,
       volume = {111},
       number = {6},
          eid = {063502},
        pages = {063502},
          doi = {10.1103/PhysRevD.111.063502},
archivePrefix = {arXiv},
       eprint = {2410.02497},
 primaryClass = {astro-ph.CO},
       adsurl = {https://ui.adsabs.harvard.edu/abs/2025PhRvD.111f3502L}
}

@ARTICLE{2025arXiv251025706S,
       author = {{Salcedo}, Andr{\'e}s N. and {Rozo}, Eduardo and {Wu}, Hao-Yi and {Weinberg}, David H. and {Chiploonkar}, Pranav and {To}, Chun-Hao and {Cao}, Shulei and {Rykoff}, Eli S. and {Marcelina Gountanis}, Nicole and {Zhou}, Conghao},
        title = "{Cosmological Constraints from Dark Energy Survey Year 1 Cluster Lensing and Abundances with Simulation-based Forward-Modeling}",
      journal = {arXiv e-prints},
         year = 2025,
        month = oct,
          eid = {arXiv:2510.25706},
        pages = {arXiv:2510.25706},
          doi = {10.48550/arXiv.2510.25706},
archivePrefix = {arXiv},
       eprint = {2510.25706},
 primaryClass = {astro-ph.CO},
       adsurl = {https://ui.adsabs.harvard.edu/abs/2025arXiv251025706S}
}

@ARTICLE{2020MNRAS.496.4468S,
       author = {{Sunayama}, Tomomi and {Park}, Youngsoo and {Takada}, Masahiro and {Kobayashi}, Yosuke and {Nishimichi}, Takahiro and {Kurita}, Toshiki and {More}, Surhud and {Oguri}, Masamune and {Osato}, Ken},
        title = "{The impact of projection effects on cluster observables: stacked lensing and projected clustering}",
      journal = {\mnras},
         year = 2020,
        month = aug,
       volume = {496},
       number = {4},
        pages = {4468-4487},
          doi = {10.1093/mnras/staa1646},
archivePrefix = {arXiv},
       eprint = {2002.03867},
 primaryClass = {astro-ph.CO},
       adsurl = {https://ui.adsabs.harvard.edu/abs/2020MNRAS.496.4468S}
}

@ARTICLE{2001AJ....122.2267E,
       author = {{Eisenstein}, Daniel J. and {Annis}, James and {Gunn}, James E. and {Szalay}, Alexander S. and {Connolly}, Andrew J. and {Nichol}, R.~C. and {Bahcall}, Neta A. and {Bernardi}, Mariangela and {Burles}, Scott and {Castander}, Francisco J. and {Fukugita}, Masataka and {Hogg}, David W. and {Ivezi{\'c}}, {\v{Z}}eljko and {Knapp}, G.~R. and {Lupton}, Robert H. and {Narayanan}, Vijay and {Postman}, Marc and {Reichart}, Daniel E. and {Richmond}, Michael and {Schneider}, Donald P. and {Schlegel}, David J. and {Strauss}, Michael A. and {SubbaRao}, Mark and {Tucker}, Douglas L. and {Vanden Berk}, Daniel and {Vogeley}, Michael S. and {Weinberg}, David H. and {Yanny}, Brian},
        title = "{Spectroscopic Target Selection for the Sloan Digital Sky Survey: The Luminous Red Galaxy Sample}",
      journal = {\aj},
         year = 2001,
        month = nov,
       volume = {122},
       number = {5},
        pages = {2267-2280},
          doi = {10.1086/323717},
archivePrefix = {arXiv},
       eprint = {astro-ph/0108153},
 primaryClass = {astro-ph},
       adsurl = {https://ui.adsabs.harvard.edu/abs/2001AJ....122.2267E}
}

@ARTICLE{2023AJ....165...58Z,
       author = {{Zhou}, Rongpu and {Dey}, Biprateep and {Newman}, Jeffrey A. and {Eisenstein}, Daniel J. and {Dawson}, K. and {Bailey}, S. and {Berti}, A. and {Guy}, J. and {Lan}, Ting-Wen and {Zou}, H. and {Aguilar}, J. and {Ahlen}, S. and {Alam}, Shadab and {Brooks}, D. and {de la Macorra}, A. and {Dey}, A. and {Dhungana}, G. and {Fanning}, K. and {Font-Ribera}, A. and {Gontcho}, S. Gontcho A. and {Honscheid}, K. and {Ishak}, Mustapha and {Kisner}, T. and {Kov{\'a}cs}, A. and {Kremin}, A. and {Landriau}, M. and {Levi}, Michael E. and {Magneville}, C. and {Manera}, Marc and {Martini}, P. and {Meisner}, Aaron M. and {Miquel}, R. and {Moustakas}, J. and {Myers}, Adam D. and {Nie}, Jundan and {Palanque-Delabrouille}, N. and {Percival}, W.~J. and {Poppett}, C. and {Prada}, F. and {Raichoor}, A. and {Ross}, A.~J. and {Schlafly}, E. and {Schlegel}, D. and {Schubnell}, M. and {Tarl{\'e}}, Gregory and {Weaver}, B.~A. and {Wechsler}, R.~H. and {Y{\'e}che}, Christophe and {Zhou}, Zhimin},
        title = "{Target Selection and Validation of DESI Luminous Red Galaxies}",
      journal = {\aj},
         year = 2023,
        month = feb,
       volume = {165},
       number = {2},
          eid = {58},
        pages = {58},
          doi = {10.3847/1538-3881/aca5fb},
archivePrefix = {arXiv},
       eprint = {2208.08515},
 primaryClass = {astro-ph.CO},
       adsurl = {https://ui.adsabs.harvard.edu/abs/2023AJ....165...58Z}
}

@ARTICLE{2024MNRAS.530..947Y,
       author = {{Yuan}, Sihan and {Zhang}, Hanyu and {Ross}, Ashley J. and {Donald-McCann}, Jamie and {Hadzhiyska}, Boryana and {Wechsler}, Risa H. and {Zheng}, Zheng and {Alam}, Shadab and {Gonzalez-Perez}, Violeta and {Aguilar}, Jessica Nicole and {Ahlen}, Steven and {Bianchi}, Davide and {Brooks}, David and {de la Macorra}, Axel and {Fanning}, Kevin and {Forero-Romero}, Jaime E. and {Honscheid}, Klaus and {Ishak}, Mustapha and {Kehoe}, Robert and {Lasker}, James and {Landriau}, Martin and {Manera}, Marc and {Martini}, Paul and {Meisner}, Aaron and {Miquel}, Ramon and {Moustakas}, John and {Nadathur}, Seshadri and {Newman}, Jeffrey A. and {Nie}, Jundan and {Percival}, Will and {Poppett}, Claire and {Rocher}, Antoine and {Rossi}, Graziano and {Sanchez}, Eusebio and {Samushia}, Lado and {Schubnell}, Michael and {Seo}, Hee-Jong and {Tarl{\'e}}, Gregory and {Weaver}, Benjamin Alan and {Yu}, Jiaxi and {Zhou}, Zhimin and {Zou}, Hu},
        title = "{The DESI one-per cent survey: exploring the halo occupation distribution of luminous red galaxies and quasi-stellar objects with ABACUSSUMMIT}",
      journal = {\mnras},
         year = 2024,
        month = may,
       volume = {530},
       number = {1},
        pages = {947-965},
          doi = {10.1093/mnras/stae359},
archivePrefix = {arXiv},
       eprint = {2306.06314},
 primaryClass = {astro-ph.CO},
       adsurl = {https://ui.adsabs.harvard.edu/abs/2024MNRAS.530..947Y}
}

@ARTICLE{2025arXiv250920458T,
       author = {{To}, Chun-Hao and {Chang}, Chihway and {Anbajagane}, Dhayaa and {Wechsler}, Risa H. and {Drlica-Wagner}, Alex and {Adam{\'o}w}, M. and {Alarcon}, A. and {Becker}, M.~R. and {Carballo-Bello}, J.~A. and {Cawthon}, R. and {Chicoine}, N. and {Doux}, C. and {Esteves}, J.~H. and {Ferguson}, P.~S. and {Gatti}, M. and {Gruen}, D. and {Gruendl}, R.~A. and {Herron}, K. and {James}, David J. and {Mart{\'\i}nez-V{\'a}zquez}, C.~E. and {Mau}, S. and {McCullough}, J. and {Medina}, G.~E. and {Mutlu-Pakdil}, B. and {Navarro-Alsina}, A. and {No{\"e}l}, N.~E.~D. and {Pace}, A.~B. and {Raveri}, M. and {Riley}, A.~H. and {Sand}, D.~J. and {Secco}, L.~F. and {Shin}, T. and {Stringfellow}, G.~S. and {Suson}, D. and {Tan}, C.~Y. and {Teixeira}, R. and {Zenteno}, A. and {Zhang}, Z.},
        title = "{A DECADE of dwarfs: first detection of weak lensing around spectroscopically confirmed low-mass galaxies}",
      journal = {arXiv e-prints},
         year = 2025,
        month = sep,
          eid = {arXiv:2509.20458},
        pages = {arXiv:2509.20458},
          doi = {10.48550/arXiv.2509.20458},
archivePrefix = {arXiv},
       eprint = {2509.20458},
 primaryClass = {astro-ph.GA},
       adsurl = {https://ui.adsabs.harvard.edu/abs/2025arXiv250920458T}
}

@PHDTHESIS{2021PhDT........16P,
       author = {{Paulus}, Maria},
        title = "{Toward accurate galaxy cluster masses with gravitational weak lensing}",
       school = {Ludwig-Maximilians University of Munich, Germany},
         year = 2021,
        month = jan,
       adsurl = {https://ui.adsabs.harvard.edu/abs/2021PhDT........16P}
}

@ARTICLE{DNF,
	title        = {{DNF - Galaxy photometric redshift by Directional Neighbourhood Fitting}},
	author       = {{De Vicente}, J. and {S{\'a}nchez}, E. and {Sevilla-Noarbe}, I.},
	year         = 2016,
	month        = jul,
	journal      = {\mnras},
	volume       = 459,
	number       = 3,
	pages        = {3078--3088},
	doi          = {10.1093/mnras/stw857},
	archivePrefix = {arXiv},
	eprint       = {1511.07623},
	primaryClass = {astro-ph.CO},
	adsurl       = {https://ui.adsabs.harvard.edu/abs/2016MNRAS.459.3078D}
}

@software{2022ascl.soft04006L,
       author = {{Lange}, Johannes and {Huang}, Song},
        title = "{dsigma: Galaxy-galaxy lensing Python package}",
 howpublished = {Astrophysics Source Code Library, record ascl:2204.006},
         year = 2022,
        month = apr,
          eid = {ascl:2204.006},
       adsurl = {https://ui.adsabs.harvard.edu/abs/2022ascl.soft04006L}
}

@ARTICLE{2019MNRAS.489.2511V,
	title        = {{Dark Energy Survey Year 1 results: validation of weak lensing cluster member contamination estimates from P(z) decomposition}},
	author       = {{Varga}, T.~N. and {DeRose}, J. and {Gruen}, D. and {McClintock}, T. and {Seitz}, S. and {Rozo}, E. and {Costanzi}, M. and {Hoyle}, B. and {MacCrann}, N. and {Plazas}, A.~A. and {Rykoff}, E.~S. and {Simet}, M. and {von der Linden}, A. and {Wechsler}, R.~H. and {Annis}, J. and {Avila}, S. and {Bertin}, E. and {Brooks}, D. and {Buckley-Geer}, E. and {Burke}, D.~L. and {Carnero Rosell}, A. and {Carrasco Kind}, M. and {Carretero}, J. and {Cunha}, C.~E. and {D'Andrea}, C.~B. and {da Costa}, L.~N. and {De Vicente}, J. and {Desai}, S. and {Diehl}, H.~T. and {Dietrich}, J.~P. and {Doel}, P. and {Evrard}, A.~E. and {Flaugher}, B. and {Fosalba}, P. and {Frieman}, J. and {Garc{\'\i}a-Bellido}, J. and {Gaztanaga}, E. and {Gerdes}, D.~W. and {Gruendl}, R.~A. and {Gschwend}, J. and {Gutierrez}, G. and {Hartley}, W.~G. and {Hollowood}, D.~L. and {Honscheid}, K. and {James}, D.~J. and {Jeltema}, T. and {Kuehn}, K. and {Kuropatkin}, N. and {Lima}, M. and {Maia}, M.~A.~G. and {March}, M. and {Marshall}, J.~L. and {Melchior}, P. and {Menanteau}, F. and {Miller}, C.~J. and {Miquel}, R. and {Ogando}, R.~L.~C. and {Romer}, A.~K. and {Sanchez}, E. and {Scarpine}, V. and {Schubnell}, M. and {Serrano}, S. and {Sevilla-Noarbe}, I. and {Smith}, M. and {Sobreira}, F. and {Suchyta}, E. and {Swanson}, M.~E.~C. and {Tarle}, G. and {Thomas}, D. and {Tucker}, D.~L. and {Zhang}, Y. and {DES Collaboration}},
	year         = 2019,
	month        = oct,
	journal      = {\mnras},
	volume       = 489,
	number       = 2,
	pages        = {2511--2524},
	doi          = {10.1093/mnras/stz2185},
	archivePrefix = {arXiv},
	eprint       = {1812.05116},
	primaryClass = {astro-ph.CO},
	adsurl       = {https://ui.adsabs.harvard.edu/abs/2019MNRAS.489.2511V}
}

@ARTICLE{2022PhRvD.106d3520P,
       author = {{Pandey}, S. and {Krause}, E. and {DeRose}, J. and {MacCrann}, N. and {Jain}, B. and {Crocce}, M. and {Blazek}, J. and {Choi}, A. and {Huang}, H. and {To}, C. and {Fang}, X. and {Elvin-Poole}, J. and {Prat}, J. and {Porredon}, A. and {Secco}, L.~F. and {Rodriguez-Monroy}, M. and {Weaverdyck}, N. and {Park}, Y. and {Raveri}, M. and {Rozo}, E. and {Rykoff}, E.~S. and {Bernstein}, G.~M. and {S{\'a}nchez}, C. and {Jarvis}, M. and {Troxel}, M.~A. and {Zacharegkas}, G. and {Chang}, C. and {Alarcon}, A. and {Alves}, O. and {Amon}, A. and {Andrade-Oliveira}, F. and {Baxter}, E. and {Bechtol}, K. and {Becker}, M.~R. and {Camacho}, H. and {Campos}, A. and {Carnero Rosell}, A. and {Carrasco Kind}, M. and {Cawthon}, R. and {Chen}, R. and {Chintalapati}, P. and {Davis}, C. and {Di Valentino}, E. and {Diehl}, H.~T. and {Dodelson}, S. and {Doux}, C. and {Drlica-Wagner}, A. and {Eckert}, K. and {Eifler}, T.~F. and {Elsner}, F. and {Everett}, S. and {Farahi}, A. and {Fert{\'e}}, A. and {Fosalba}, P. and {Friedrich}, O. and {Gatti}, M. and {Giannini}, G. and {Gruen}, D. and {Gruendl}, R.~A. and {Harrison}, I. and {Hartley}, W.~G. and {Huff}, E.~M. and {Huterer}, D. and {Kovacs}, A. and {Leget}, P.~F. and {McCullough}, J. and {Muir}, J. and {Myles}, J. and {Navarro-Alsina}, A. and {Omori}, Y. and {Rollins}, R.~P. and {Roodman}, A. and {Rosenfeld}, R. and {Sevilla-Noarbe}, I. and {Sheldon}, E. and {Shin}, T. and {Troja}, A. and {Tutusaus}, I. and {Varga}, T.~N. and {Wechsler}, R.~H. and {Yanny}, B. and {Yin}, B. and {Zhang}, Y. and {Zuntz}, J. and {Abbott}, T.~M.~C. and {Aguena}, M. and {Allam}, S. and {Annis}, J. and {Bacon}, D. and {Bertin}, E. and {Brooks}, D. and {Burke}, D.~L. and {Carretero}, J. and {Conselice}, C. and {Costanzi}, M. and {da Costa}, L.~N. and {Pereira}, M.~E.~S. and {De Vicente}, J. and {Dietrich}, J.~P. and {Doel}, P. and {Evrard}, A.~E. and {Ferrero}, I. and {Flaugher}, B. and {Frieman}, J. and {Garc{\'\i}a-Bellido}, J. and {Gaztanaga}, E. and {Gerdes}, D.~W. and {Giannantonio}, T. and {Gschwend}, J. and {Gutierrez}, G. and {Hinton}, S.~R. and {Hollowood}, D.~L. and {Honscheid}, K. and {James}, D.~J. and {Jeltema}, T. and {Kuehn}, K. and {Kuropatkin}, N. and {Lahav}, O. and {Lima}, M. and {Lin}, H. and {Maia}, M.~A.~G. and {Marshall}, J.~L. and {Melchior}, P. and {Menanteau}, F. and {Miller}, C.~J. and {Miquel}, R. and {Mohr}, J.~J. and {Morgan}, R. and {Palmese}, A. and {Paz-Chinch{\'o}n}, F. and {Petravick}, D. and {Pieres}, A. and {Plazas Malag{\'o}n}, A.~A. and {Sanchez}, E. and {Scarpine}, V. and {Serrano}, S. and {Smith}, M. and {Soares-Santos}, M. and {Suchyta}, E. and {Tarle}, G. and {Thomas}, D. and {Weller}, J. and {DES Collaboration}},
        title = "{Dark Energy Survey year 3 results: Constraints on cosmological parameters and galaxy-bias models from galaxy clustering and galaxy-galaxy lensing using the redMaGiC sample}",
      journal = {\prd},
         year = 2022,
        month = aug,
       volume = {106},
       number = {4},
          eid = {043520},
        pages = {043520},
          doi = {10.1103/PhysRevD.106.043520},
archivePrefix = {arXiv},
       eprint = {2105.13545},
 primaryClass = {astro-ph.CO},
       adsurl = {https://ui.adsabs.harvard.edu/abs/2022PhRvD.106d3520P}
}

@ARTICLE{2018PhRvD..98d2006E,
       author = {{Elvin-Poole}, J. and {Crocce}, M. and {Ross}, A.~J. and {Giannantonio}, T. and {Rozo}, E. and {Rykoff}, E.~S. and {Avila}, S. and {Banik}, N. and {Blazek}, J. and {Bridle}, S.~L. and {Cawthon}, R. and {Drlica-Wagner}, A. and {Friedrich}, O. and {Kokron}, N. and {Krause}, E. and {MacCrann}, N. and {Prat}, J. and {S{\'a}nchez}, C. and {Secco}, L.~F. and {Sevilla-Noarbe}, I. and {Troxel}, M.~A. and {Abbott}, T.~M.~C. and {Abdalla}, F.~B. and {Allam}, S. and {Annis}, J. and {Asorey}, J. and {Bechtol}, K. and {Becker}, M.~R. and {Benoit-L{\'e}vy}, A. and {Bernstein}, G.~M. and {Bertin}, E. and {Brooks}, D. and {Buckley-Geer}, E. and {Burke}, D.~L. and {Carnero Rosell}, A. and {Carollo}, D. and {Carrasco Kind}, M. and {Carretero}, J. and {Castander}, F.~J. and {Cunha}, C.~E. and {D'Andrea}, C.~B. and {da Costa}, L.~N. and {Davis}, T.~M. and {Davis}, C. and {Desai}, S. and {Diehl}, H.~T. and {Dietrich}, J.~P. and {Dodelson}, S. and {Doel}, P. and {Eifler}, T.~F. and {Evrard}, A.~E. and {Fernandez}, E. and {Flaugher}, B. and {Fosalba}, P. and {Frieman}, J. and {Garc{\'\i}a-Bellido}, J. and {Gaztanaga}, E. and {Gerdes}, D.~W. and {Glazebrook}, K. and {Gruen}, D. and {Gruendl}, R.~A. and {Gschwend}, J. and {Gutierrez}, G. and {Hartley}, W.~G. and {Hinton}, S.~R. and {Honscheid}, K. and {Hoormann}, J.~K. and {Jain}, B. and {James}, D.~J. and {Jarvis}, M. and {Jeltema}, T. and {Johnson}, M.~W.~G. and {Johnson}, M.~D. and {King}, A. and {Kuehn}, K. and {Kuhlmann}, S. and {Kuropatkin}, N. and {Lahav}, O. and {Lewis}, G. and {Li}, T.~S. and {Lidman}, C. and {Lima}, M. and {Lin}, H. and {Macaulay}, E. and {March}, M. and {Marshall}, J.~L. and {Martini}, P. and {Melchior}, P. and {Menanteau}, F. and {Miquel}, R. and {Mohr}, J.~J. and {M{\"o}ller}, A. and {Nichol}, R.~C. and {Nord}, B. and {O'Neill}, C.~R. and {Percival}, W.~J. and {Petravick}, D. and {Plazas}, A.~A. and {Romer}, A.~K. and {Sako}, M. and {Sanchez}, E. and {Scarpine}, V. and {Schindler}, R. and {Schubnell}, M. and {Sheldon}, E. and {Smith}, M. and {Smith}, R.~C. and {Soares-Santos}, M. and {Sobreira}, F. and {Sommer}, N.~E. and {Suchyta}, E. and {Swanson}, M.~E.~C. and {Tarle}, G. and {Thomas}, D. and {Tucker}, B.~E. and {Tucker}, D.~L. and {Uddin}, S.~A. and {Vikram}, V. and {Walker}, A.~R. and {Wechsler}, R.~H. and {Weller}, J. and {Wester}, W. and {Wolf}, R.~C. and {Yuan}, F. and {Zhang}, B. and {Zuntz}, J. and {DES Collaboration}},
        title = "{Dark Energy Survey year 1 results: Galaxy clustering for combined probes}",
      journal = {\prd},
         year = 2018,
        month = aug,
       volume = {98},
       number = {4},
          eid = {042006},
        pages = {042006},
          doi = {10.1103/PhysRevD.98.042006},
archivePrefix = {arXiv},
       eprint = {1708.01536},
 primaryClass = {astro-ph.CO},
       adsurl = {https://ui.adsabs.harvard.edu/abs/2018PhRvD..98d2006E}
}

@ARTICLE{2009ApJ...707..554Z,
       author = {{Zheng}, Zheng and {Zehavi}, Idit and {Eisenstein}, Daniel J. and {Weinberg}, David H. and {Jing}, Y.~P.},
        title = "{Halo Occupation Distribution Modeling of Clustering of Luminous Red Galaxies}",
      journal = {\apj},
         year = 2009,
        month = dec,
       volume = {707},
       number = {1},
        pages = {554-572},
          doi = {10.1088/0004-637X/707/1/554},
archivePrefix = {arXiv},
       eprint = {0809.1868},
 primaryClass = {astro-ph},
       adsurl = {https://ui.adsabs.harvard.edu/abs/2009ApJ...707..554Z}
}

@ARTICLE{2002AJ....123..485S,
       author = {{Stoughton}, Chris and {Lupton}, Robert H. and {Bernardi}, Mariangela and {Blanton}, Michael R. and {Burles}, Scott and {Castander}, Francisco J. and {Connolly}, A.~J. and {Eisenstein}, Daniel J. and {Frieman}, Joshua A. and {Hennessy}, G.~S. and {Hindsley}, Robert B. and {Ivezi{\'c}}, {\v{Z}}eljko and {Kent}, Stephen and {Kunszt}, Peter Z. and {Lee}, Brian C. and {Meiksin}, Avery and {Munn}, Jeffrey A. and {Newberg}, Heidi Jo and {Nichol}, R.~C. and {Nicinski}, Tom and {Pier}, Jeffrey R. and {Richards}, Gordon T. and {Richmond}, Michael W. and {Schlegel}, David J. and {Smith}, J. Allyn and {Strauss}, Michael A. and {SubbaRao}, Mark and {Szalay}, Alexander S. and {Thakar}, Aniruddha R. and {Tucker}, Douglas L. and {Vanden Berk}, Daniel E. and {Yanny}, Brian and {Adelman}, Jennifer K. and {Anderson}, Jr., John E. and {Anderson}, Scott F. and {Annis}, James and {Bahcall}, Neta A. and {Bakken}, J.~A. and {Bartelmann}, Matthias and {Bastian}, Steven and {Bauer}, Amanda and {Berman}, Eileen and {B{\"o}hringer}, Hans and {Boroski}, William N. and {Bracker}, Steve and {Briegel}, Charlie and {Briggs}, John W. and {Brinkmann}, J. and {Brunner}, Robert and {Carey}, Larry and {Carr}, Michael A. and {Chen}, Bing and {Christian}, Damian and {Colestock}, Patrick L. and {Crocker}, J.~H. and {Csabai}, Istv{\'a}n and {Czarapata}, Paul C. and {Dalcanton}, Julianne and {Davidsen}, Arthur F. and {Davis}, John Eric and {Dehnen}, Walter and {Dodelson}, Scott and {Doi}, Mamoru and {Dombeck}, Tom and {Donahue}, Megan and {Ellman}, Nancy and {Elms}, Brian R. and {Evans}, Michael L. and {Eyer}, Laurent and {Fan}, Xiaohui and {Federwitz}, Glenn R. and {Friedman}, Scott and {Fukugita}, Masataka and {Gal}, Roy and {Gillespie}, Bruce and {Glazebrook}, Karl and {Gray}, Jim and {Grebel}, Eva K. and {Greenawalt}, Bruce and {Greene}, Gretchen and {Gunn}, James E. and {de Haas}, Ernst and {Haiman}, Zolt{\'a}n and {Haldeman}, Merle and {Hall}, Patrick B. and {Hamabe}, Masaru and {Hansen}, Brad and {Harris}, Frederick H. and {Harris}, Hugh and {Harvanek}, Michael and {Hawley}, Suzanne L. and {Hayes}, J.~J.~E. and {Heckman}, Timothy M. and {Helmi}, Amina and {Henden}, Arne and {Hogan}, Craig J. and {Hogg}, David W. and {Holmgren}, Donald J. and {Holtzman}, Jon and {Huang}, Chih-Hao and {Hull}, Charles and {Ichikawa}, Shin-Ichi and {Ichikawa}, Takashi and {Johnston}, David E. and {Kauffmann}, Guinevere and {Kim}, Rita S.~J. and {Kimball}, Tim and {Kinney}, E. and {Klaene}, Mark and {Kleinman}, S.~J. and {Klypin}, Anatoly and {Knapp}, G.~R. and {Korienek}, John and {Krolik}, Julian and {Kron}, Richard G. and {Krzesi{\'n}ski}, Jurek and {Lamb}, D.~Q. and {Leger}, R. French and {Limmongkol}, Siriluk and {Lindenmeyer}, Carl and {Long}, Daniel C. and {Loomis}, Craig and {Loveday}, Jon and {MacKinnon}, Bryan and {Mannery}, Edward J. and {Mantsch}, P.~M. and {Margon}, Bruce and {McGehee}, Peregrine and {McKay}, Timothy A. and {McLean}, Brian and {Menou}, Kristen and {Merelli}, Aronne and {Mo}, H.~J. and {Monet}, David G. and {Nakamura}, Osamu and {Narayanan}, Vijay K. and {Nash}, Thomas and {Neilsen}, Jr., Eric H. and {Newman}, Peter R. and {Nitta}, Atsuko and {Odenkirchen}, Michael and {Okada}, Norio and {Okamura}, Sadanori and {Ostriker}, Jeremiah P. and {Owen}, Russell and {Pauls}, A. George and {Peoples}, John and {Peterson}, R.~S. and {Petravick}, Donald and {Pope}, Adrian and {Pordes}, Ruth and {Postman}, Marc and {Prosapio}, Angela and {Quinn}, Thomas R. and {Rechenmacher}, Ron and {Rivetta}, Claudio H. and {Rix}, Hans-Walter and {Rockosi}, Constance M. and {Rosner}, Robert and {Ruthmansdorfer}, Kurt and {Sandford}, Dale and {Schneider}, Donald P. and {Scranton}, Ryan and {Sekiguchi}, Maki and {Sergey}, Gary and {Sheth}, Ravi and {Shimasaku}, Kazuhiro and {Smee}, Stephen and {Snedden}, Stephanie A. and {Stebbins}, Albert and {Stubbs}, Christopher and {Szapudi}, Istv{\'a}n and {Szkody}, Paula and {Szokoly}, Gyula P. and {Tabachnik}, Serge and {Tsvetanov}, Zlatan and {Uomoto}, Alan and {Vogeley}, Michael S. and {Voges}, Wolfgang and {Waddell}, Patrick and {Walterbos}, Ren{\'e} and {Wang}, Shu-i. and {Watanabe}, Masaru and {Weinberg}, David H. and {White}, Richard L. and {White}, Simon D.~M. and {Wilhite}, Brian and {Wolfe}, David and {Yasuda}, Naoki and {York}, Donald G. and {Zehavi}, Idit and {Zheng}, Wei},
        title = "{Sloan Digital Sky Survey: Early Data Release}",
      journal = {\aj},
         year = 2002,
        month = jan,
       volume = {123},
       number = {1},
        pages = {485-548},
          doi = {10.1086/324741},
       adsurl = {https://ui.adsabs.harvard.edu/abs/2002AJ....123..485S}
}

@ARTICLE{2024PhRvL.133v1002S,
       author = {{Salcedo}, Andr{\'e}s N. and {Wu}, Hao-Yi and {Rozo}, Eduardo and {Weinberg}, David H. and {To}, Chun-Hao and {Sunayama}, Tomomi and {Lee}, Andy},
        title = "{Consistency of Dark Energy Survey Year 1 Galaxy Clusters with Planck}",
      journal = {\prl},
         year = 2024,
        month = nov,
       volume = {133},
       number = {22},
          eid = {221002},
        pages = {221002},
          doi = {10.1103/PhysRevLett.133.221002},
archivePrefix = {arXiv},
       eprint = {2310.03944},
 primaryClass = {astro-ph.CO},
       adsurl = {https://ui.adsabs.harvard.edu/abs/2024PhRvL.133v1002S}
}

@ARTICLE{paper2,
       author = {{Anbajagane}, D. and {Alarcon}, A. and {Teixeira}, R. and {Chang}, C. and {Secco}, L.~F. and {Tan}, C.~Y. and {Drlica-Wagner}, A. and {Adamow}, M. and {Gruendl}, R.~A. and {Giannini}, G. and {Becker}, M.~R. and {Ferguson}, P.~S. and {Chicoine}, N. and {Zhang}, Z. and {Herron}, K. and {Suson}, D. and {Alsina}, A.~N. and {Amon}, A. and {Bom}, C.~R. and {Carballo-Bello}, J.~A. and {Cerny}, W. and {Choi}, A. and {Choi}, Y. and {Doux}, C. and {Eckert}, K. and {Gatti}, M. and {Gruen}, D. and {Hartley}, W.~G. and {Herner}, K. and {Huff}, E.~M. and {James}, D.~J. and {Kuropatkin}, N. and {Mart{\'\i}nez-V{\'a}zquez}, C.~E. and {Massana}, P. and {Mau}, S. and {McCullough}, J. and {Medina}, G.~E. and {Mutlu-Pakdil}, B. and {Myles}, J. and {Navabi}, M. and {No{\"e}l}, N.~E.~D. and {Pace}, A.~B. and {Raveri}, M. and {Riley}, A.~H. and {Sakowska}, J.~D. and {Sanchez-Cid}, D. and {Sand}, D.~J. and {Santana-Silva}, L. and {Sevilla-Noarbe}, I. and {Soares-Santos}, M. and {Stringfellow}, G.~S. and {Vivas}, A.~K. and {Yamamoto}, M.},
        title = "{The DECADE cosmic shear project II: photometric redshift calibration of the source galaxy sample}",
      journal = {arXiv e-prints},
         year = 2025,
        month = feb,
          eid = {arXiv:2502.17675},
        pages = {arXiv:2502.17675},
          doi = {10.48550/arXiv.2502.17675},
archivePrefix = {arXiv},
       eprint = {2502.17675},
 primaryClass = {astro-ph.CO},
       adsurl = {https://ui.adsabs.harvard.edu/abs/2025arXiv250217675A}
}

@ARTICLE{paper3,
       author = {{Anbajagane}, D. and {Chang}, C. and {Chicoine}, N. and {Secco}, L.~F. and {Tan}, C.~Y. and {Ferguson}, P.~S. and {Drlica-Wagner}, A. and {Herron}, K. and {Adamow}, M. and {Gruendl}, R.~A. and {Becker}, M.~R. and {Teixeira}, R. and {Zhang}, Z. and {Alarcon}, A. and {Suson}, D. and {Alsina}, A.~N. and {Amon}, A. and {Andrade-Oliveira}, F. and {Blazek}, J. and {Camacho}, H. and {Carballo-Bello}, J.~A. and {Cerny}, W. and {Choi}, Y. and {Doux}, C. and {Gatti}, M. and {Gruen}, D. and {James}, D.~J. and {Krause}, E. and {Kuropatkin}, N. and {Mart{\'\i}nez-V{\'a}zquez}, C.~E. and {Massana}, P. and {Mau}, S. and {McCullough}, J. and {Medina}, G.~E. and {Mutlu-Pakdil}, B. and {Navabi}, M. and {No{\"e}l}, N.~E.~D. and {Pace}, A.~B. and {Porredon}, A. and {Raveri}, M. and {Riley}, A.~H. and {Sakowska}, J.~D. and {Samuroff}, S. and {Sanchez-Cid}, D. and {Sand}, D.~J. and {Santana-Silva}, L. and {Soares-Santos}, M. and {Stringfellow}, G.~S. and {To}, C. and {Vivas}, A.~K. and {Yamamoto}, M. and {Zenteno}, A. and {Zuntz}, J.},
        title = "{The DECADE cosmic shear project III: validation of analysis pipeline using spatially inhomogeneous data}",
      journal = {arXiv e-prints},
         year = 2025,
        month = feb,
          eid = {arXiv:2502.17676},
        pages = {arXiv:2502.17676},
          doi = {10.48550/arXiv.2502.17676},
archivePrefix = {arXiv},
       eprint = {2502.17676},
 primaryClass = {astro-ph.CO},
       adsurl = {https://ui.adsabs.harvard.edu/abs/2025arXiv250217676A}
}

@ARTICLE{paper4,
       author = {{Anbajagane}, D. and {Chang}, C. and {Drlica-Wagner}, A. and {Tan}, C.~Y. and {Adamow}, M. and {Gruendl}, R.~A. and {Secco}, L.~F. and {Zhang}, Z. and {Becker}, M.~R. and {Ferguson}, P.~S. and {Chicoine}, N. and {Herron}, K. and {Alarcon}, A. and {Teixeira}, R. and {Suson}, D. and {Alsina}, A.~N. and {Amon}, A. and {Andrade-Oliveira}, F. and {Blazek}, J. and {Bom}, C.~R. and {Camacho}, H. and {Carballo-Bello}, J.~A. and {Carnero Rosell}, A. and {Cawthon}, R. and {Cerny}, W. and {Choi}, A. and {Choi}, Y. and {Dodelson}, S. and {Doux}, C. and {Eckert}, K. and {Elvin-Poole}, J. and {Esteves}, J. and {Gatti}, M. and {Giannini}, G. and {Gruen}, D. and {Hartley}, W.~G. and {Herner}, K. and {Huff}, E.~M. and {James}, D.~J. and {Jarvis}, M. and {Krause}, E. and {Kuropatkin}, N. and {Mart{\'\i}nez-V{\'a}zquez}, C.~E. and {Massana}, P. and {Mau}, S. and {McCullough}, J. and {Medina}, G.~E. and {Mutlu-Pakdil}, B. and {Myles}, J. and {Navabi}, M. and {No{\"e}l}, N.~E.~D. and {Pace}, A.~B. and {Porredon}, A. and {Prat}, J. and {Raveri}, M. and {Riley}, A.~H. and {Rykoff}, E.~S. and {Sakowska}, J.~D. and {Samuroff}, S. and {Sanchez-Cid}, D. and {Sand}, D.~J. and {Santana-Silva}, L. and {Sevilla-Noarbe}, I. and {Shin}, T. and {Soares-Santos}, M. and {Stringfellow}, G.~S. and {To}, C. and {Tong}, A. and {Troxel}, M.~A. and {Vivas}, A.~K. and {Yamamoto}, M. and {Yanny}, B. and {Yin}, B. and {Zhang}, Y. and {Zuntz}, J.},
        title = "{The DECADE cosmic shear project IV: cosmological constraints from 107 million galaxies across 5,400 deg$^2$ of the sky}",
      journal = {arXiv e-prints},
         year = 2025,
        month = feb,
          eid = {arXiv:2502.17677},
        pages = {arXiv:2502.17677},
          doi = {10.48550/arXiv.2502.17677},
archivePrefix = {arXiv},
       eprint = {2502.17677},
 primaryClass = {astro-ph.CO},
       adsurl = {https://ui.adsabs.harvard.edu/abs/2025arXiv250217677A}
}

@ARTICLE{paper5,
       author = {{Anbajagane}, D. and {Chang}, C. and {Drlica-Wagner}, A. and {Tan}, C.~Y. and {Adamow}, M. and {Gruendl}, R.~A. and {Secco}, L.~F. and {Zhang}, Z. and {Becker}, M.~R. and {Ferguson}, P.~S. and {Chicoine}, N. and {Herron}, K. and {Alarcon}, A. and {Teixeira}, R. and {Suson}, D. and {Alsina}, A.~N. and {Amon}, A. and {Andrade-Oliveira}, F. and {Blazek}, J. and {Bom}, C.~R. and {Camacho}, H. and {Carballo-Bello}, J.~A. and {Carnero Rosell}, A. and {Cawthon}, R. and {Cerny}, W. and {Choi}, A. and {Choi}, Y. and {Dodelson}, S. and {Doux}, C. and {Eckert}, K. and {Elvin-Poole}, J. and {Esteves}, J. and {Gatti}, M. and {Giannini}, G. and {Gruen}, D. and {Hartley}, W.~G. and {Herner}, K. and {Huff}, E.~M. and {James}, D.~J. and {Jarvis}, M. and {Krause}, E. and {Kuropatkin}, N. and {Mart{\'\i}nez-V{\'a}zquez}, C.~E. and {Massana}, P. and {Mau}, S. and {McCullough}, J. and {Medina}, G.~E. and {Mutlu-Pakdil}, B. and {Myles}, J. and {Navabi}, M. and {No{\"e}l}, N.~E.~D. and {Pace}, A.~B. and {Porredon}, A. and {Prat}, J. and {Raveri}, M. and {Riley}, A.~H. and {Rykoff}, E.~S. and {Sakowska}, J.~D. and {Samuroff}, S. and {Sanchez-Cid}, D. and {Sand}, D.~J. and {Santana-Silva}, L. and {Sevilla-Noarbe}, I. and {Shin}, T. and {Soares-Santos}, M. and {Stringfellow}, G.~S. and {To}, C. and {Tong}, A. and {Troxel}, M.~A. and {Vivas}, A.~K. and {Yamamoto}, M. and {Yanny}, B. and {Yin}, B. and {Zhang}, Y. and {Zuntz}, J.},
        title = "{The DECADE cosmic shear project IV: cosmological constraints from 107 million galaxies across 5,400 deg$^2$ of the sky}",
      journal = {arXiv e-prints},
         year = 2025,
        month = feb,
          eid = {arXiv:2502.17677},
        pages = {arXiv:2502.17677},
          doi = {10.48550/arXiv.2502.17677},
archivePrefix = {arXiv},
       eprint = {2502.17677},
 primaryClass = {astro-ph.CO},
       adsurl = {https://ui.adsabs.harvard.edu/abs/2025arXiv250217677A}
}

@ARTICLE{Yamamoto2025,
       author = {{Yamamoto \& Becker} and {Becker}, M.~R. and {Sheldon}, E. and {Jarvis}, M. and {Gruendl}, R.~A. and {Menanteau}, F. and {Rykoff}, E.~S. and {Mau}, S. and {Schutt}, T. and {Gatti}, M. and {Troxel}, M.~A. and {Amon}, A. and {Anbajagane}, D. and {Bernstein}, G.~M. and {Gruen}, D. and {Huff}, E.~M. and {Tabbutt}, M. and {Tong}, A. and {Yanny}, B. and {Abbott}, T.~M.~C. and {Aguena}, M. and {Alarcon}, A. and {Andrade-Oliveira}, F. and {Bechtol}, K. and {Blazek}, J. and {Brooks}, D. and {Carnero Rosell}, A. and {Carretero}, J. and {Chang}, C. and {Choi}, A. and {Costanzi}, M. and {Crocce}, M. and {da Costa}, L.~N. and {Davis}, T.~M. and {De Vicente}, J. and {Desai}, S. and {Diehl}, H.~T. and {Dodelson}, S. and {Doel}, P. and {Doux}, C. and {Drlica-Wagner}, A. and {Fert{\'e}}, A. and {Flaugher}, B. and {Frieman}, J. and {Garc{\'\i}a-Bellido}, J. and {Gaztanaga}, E. and {Giannini}, G. and {Gutierrez}, G. and {Hartley}, W.~G. and {Herner}, K. and {Hinton}, S.~R. and {Hollowood}, D.~L. and {Honscheid}, K. and {Huterer}, D. and {Krause}, E. and {Kuehn}, K. and {Lahav}, O. and {Lima}, M. and {Marshall}, J.~L. and {Mena-Fern{\'a}ndez}, J. and {Miquel}, R. and {Mohr}, J.~J. and {Muir}, J. and {Myles}, J. and {Ogando}, R.~L.~C. and {Pieres}, A. and {Plazas Malag{\'o}n}, A.~A. and {Porredon}, A. and {Prat}, J. and {Raveri}, M. and {Rodriguez-Monroy}, M. and {Roodman}, A. and {Samuroff}, S. and {Sanchez}, E. and {Sanchez Cid}, D. and {Scarpine}, V. and {Sevilla-Noarbe}, I. and {Smith}, M. and {Suchyta}, E. and {Tarle}, G. and {Vikram}, V. and {Weaverdyck}, N. and {Wiseman}, P. and {Zhang}, Y.},
        title = "{Dark Energy Survey Year 6 Results: Cell-based Coadds and Metadetection Weak Lensing Shape Catalogue}",
      journal = {arXiv e-prints},
         year = 2025,
        month = jan,
          eid = {arXiv:2501.05665},
        pages = {arXiv:2501.05665},
archivePrefix = {arXiv},
       eprint = {2501.05665},
 primaryClass = {astro-ph.CO},
       adsurl = {https://ui.adsabs.harvard.edu/abs/2025arXiv250105665Y}
}

@ARTICLE{Gatti:2022:WzY3,
       author = {{Gatti \& Giannini} and {Giannini}, G. and {Bernstein}, G.~M. and {Alarcon}, A. and {Myles}, J. and {Amon}, A. and {Cawthon}, R. and {Troxel}, M. and {DeRose}, J. and {Everett}, S. and {Ross}, A.~J. and {Rykoff}, E.~S. and {Elvin-Poole}, J. and {Cordero}, J. and {Harrison}, I. and {Sanchez}, C. and {Prat}, J. and {Gruen}, D. and {Lin}, H. and {Crocce}, M. and {Rozo}, E. and {Abbott}, T.~M.~C. and {Aguena}, M. and {Allam}, S. and {Annis}, J. and {Avila}, S. and {Bacon}, D. and {Bertin}, E. and {Brooks}, D. and {Burke}, D.~L. and {Rosell}, A. Carnero and {Kind}, M. Carrasco and {Carretero}, J. and {Castander}, F.~J. and {Choi}, A. and {Conselice}, C. and {Costanzi}, M. and {Crocce}, M. and {da Costa}, L.~N. and {Pereira}, M.~E.~S. and {Dawson}, K. and {Desai}, S. and {Diehl}, H.~T. and {Eckert}, K. and {Eifler}, T.~F. and {Evrard}, A.~E. and {Ferrero}, I. and {Flaugher}, B. and {Fosalba}, P. and {Frieman}, J. and {Garc{\'\i}a-Bellido}, J. and {Gaztanaga}, E. and {Giannantonio}, T. and {Gruendl}, R.~A. and {Gschwend}, J. and {Hinton}, S.~R. and {Hollowood}, D.~L. and {Honscheid}, K. and {Hoyle}, B. and {Huterer}, D. and {James}, D.~J. and {Kuehn}, K. and {Kuropatkin}, N. and {Lahav}, O. and {Lima}, M. and {MacCrann}, N. and {Maia}, M.~A.~G. and {March}, M. and {Marshall}, J.~L. and {Melchior}, P. and {Menanteau}, F. and {Miquel}, R. and {Mohr}, J.~J. and {Morgan}, R. and {Ogando}, R.~L.~C. and {Palmese}, A. and {Paz-Chinch{\'o}n}, F. and {Percival}, W.~J. and {Plazas}, A.~A. and {Rodriguez-Monroy}, M. and {Roodman}, A. and {Rossi}, G. and {Samuroff}, S. and {Sanchez}, E. and {Scarpine}, V. and {Secco}, L.~F. and {Serrano}, S. and {Sevilla-Noarbe}, I. and {Smith}, M. and {Soares-Santos}, M. and {Suchyta}, E. and {Swanson}, M.~E.~C. and {Tarle}, G. and {Thomas}, D. and {To}, C. and {Varga}, T.~N. and {Weller}, J. and {Wilkinson}, R.~D. and {Wilkinson}, R.~D. and {DES Collaboration}},
        title = "{Dark Energy Survey Year 3 Results: clustering redshifts - calibration of the weak lensing source redshift distributions with redMaGiC and BOSS/eBOSS}",
      journal = {\mnras},
         year = 2022,
        month = feb,
       volume = {510},
       number = {1},
        pages = {1223-1247},
          doi = {10.1093/mnras/stab3311},
archivePrefix = {arXiv},
       eprint = {2012.08569},
 primaryClass = {astro-ph.CO},
       adsurl = {https://ui.adsabs.harvard.edu/abs/2022MNRAS.510.1223G}
}

@article{Flaugher2005,
	Adsurl = {http://adsabs.harvard.edu/abs/2005IJMPA..20.3121F},
	Author = {{Flaugher}, B.},
	Doi = {10.1142/S0217751X05025917},
	Journal = {International Journal of Modern Physics A},
	Pages = {3121-3123},
	Title = {{The Dark Energy Survey}},
	Volume = 20,
	Year = 2005}

@ARTICLE{Dey:2019,
       author = {{Dey}, Arjun and {Schlegel}, David J. and {Lang}, Dustin and
         {Blum}, Robert and {Burleigh}, Kaylan and {Fan}, Xiaohui and
         {Findlay}, Joseph R. and {Finkbeiner}, Doug and {Herrera}, David and
         {Juneau}, St{\'e}phanie and {Landriau}, Martin and {Levi}, Michael and
         {McGreer}, Ian and {Meisner}, Aaron and {Myers}, Adam D. and
         {Moustakas}, John and {Nugent}, Peter and {Patej}, Anna and
         {Schlafly}, Edward F. and {Walker}, Alistair R. and
         {Valdes}, Francisco and {Weaver}, Benjamin A. and
         {Y{\`e}che}, Christophe and {Zou}, Hu and {Zhou}, Xu and
         {Abareshi}, Behzad and {Abbott}, T.~M.~C. and {Abolfathi}, Bela and
         {Aguilera}, C. and {Alam}, Shadab and {Allen}, Lori and {Alvarez}, A. and
         {Annis}, James and {Ansarinejad}, Behzad and {Aubert}, Marie and
         {Beechert}, Jacqueline and {Bell}, Eric F. and {BenZvi}, Segev Y. and
         {Beutler}, Florian and {Bielby}, Richard M. and {Bolton}, Adam S. and
         {Brice{\~n}o}, C{\'e}sar and {Buckley-Geer}, Elizabeth J. and
         {Butler}, Karen and {Calamida}, Annalisa and {Carlberg}, Raymond G. and
         {Carter}, Paul and {Casas}, Ricard and {Castander}, Francisco J. and
         {Choi}, Yumi and {Comparat}, Johan and {Cukanovaite}, Elena and
         {Delubac}, Timoth{\'e}e and {DeVries}, Kaitlin and {Dey}, Sharmila and
         {Dhungana}, Govinda and {Dickinson}, Mark and {Ding}, Zhejie and
         {Donaldson}, John B. and {Duan}, Yutong and
         {Duckworth}, Christopher J. and {Eftekharzadeh}, Sarah and
         {Eisenstein}, Daniel J. and {Etourneau}, Thomas and
         {Fagrelius}, Parker A. and {Farihi}, Jay and {Fitzpatrick}, Mike and
         {Font-Ribera}, Andreu and {Fulmer}, Leah and {G{\"a}nsicke}, Boris T. and
         {Gaztanaga}, Enrique and {George}, Koshy and {Gerdes}, David W. and
         {Gontcho}, Satya Gontcho A. and {Gorgoni}, Claudio and
         {Green}, Gregory and {Guy}, Julien and {Harmer}, Diane and {Hernand
        ez}, M. and {Honscheid}, Klaus and {Huang}, Lijuan Wendy and
         {James}, David J. and {Jannuzi}, Buell T. and {Jiang}, Linhua and
         {Joyce}, Richard and {Karcher}, Armin and {Karkar}, Sonia and
         {Kehoe}, Robert and {Kneib}, Jean-Paul and {Kueter-Young}, Andrea and
         {Lan}, Ting-Wen and {Lauer}, Tod R. and {Le Guillou}, Laurent and
         {Le Van Suu}, Auguste and {Lee}, Jae Hyeon and {Lesser}, Michael and
         {Perreault Levasseur}, Laurence and {Li}, Ting S. and
         {Mann}, Justin L. and {Marshall}, Robert and
         {Mart{\'\i}nez-V{\'a}zquez}, C.~E. and {Martini}, Paul and
         {du Mas des Bourboux}, H{\'e}lion and {McManus}, Sean and
         {Meier}, Tobias Gabriel and {M{\'e}nard}, Brice and {Metcalfe}, Nigel and
         {Mu{\~n}oz-Guti{\'e}rrez}, Andrea and {Najita}, Joan and
         {Napier}, Kevin and {Narayan}, Gautham and {Newman}, Jeffrey A. and
         {Nie}, Jundan and {Nord}, Brian and {Norman}, Dara J. and
         {Olsen}, Knut A.~G. and {Paat}, Anthony and
         {Palanque-Delabrouille}, Nathalie and {Peng}, Xiyan and
         {Poppett}, Claire L. and {Poremba}, Megan R. and {Prakash}, Abhishek and
         {Rabinowitz}, David and {Raichoor}, Anand and {Rezaie}, Mehdi and
         {Robertson}, A.~N. and {Roe}, Natalie A. and {Ross}, Ashley J. and
         {Ross}, Nicholas P. and {Rudnick}, Gregory and {Safonova}, Sasha and
         {Saha}, Abhijit and {S{\'a}nchez}, F. Javier and {Savary}, Elodie and
         {Schweiker}, Heidi and {Scott}, Adam and {Seo}, Hee-Jong and
         {Shan}, Huanyuan and {Silva}, David R. and {Slepian}, Zachary and
         {Soto}, Christian and {Sprayberry}, David and {Staten}, Ryan and
         {Stillman}, Coley M. and {Stupak}, Robert J. and {Summers}, David L. and
         {Sien Tie}, Suk and {Tirado}, H. and {Vargas-Maga{\~n}a}, Mariana and
         {Vivas}, A. Katherina and {Wechsler}, Risa H. and {Williams}, Doug and
         {Yang}, Jinyi and {Yang}, Qian and {Yapici}, Tolga and
         {Zaritsky}, Dennis and {Zenteno}, A. and {Zhang}, Kai and
         {Zhang}, Tianmeng and {Zhou}, Rongpu and {Zhou}, Zhimin},
        title = "{Overview of the DESI Legacy Imaging Surveys}",
      journal = {\aj},
         year = "2019",
        month = "May",
       volume = {157},
       number = {5},
          eid = {168},
        pages = {168},
          doi = {10.3847/1538-3881/ab089d},
archivePrefix = {arXiv},
       eprint = {1804.08657},
 primaryClass = {astro-ph.IM},
       adsurl = {https://ui.adsabs.harvard.edu/abs/2019AJ....157..168D}
}

@ARTICLE{Morganson:2018,
       author = {{Morganson}, E. and {Gruendl}, R.~A. and {Menanteau}, F. and
         {Carrasco Kind}, M. and {Chen}, Y. -C. and {Daues}, G. and
         {Drlica-Wagner}, A. and {Friedel}, D.~N. and {Gower}, M. and
         {Johnson}, M.~W.~G. and {Johnson}, M.~D. and {Kessler}, R. and
         {Paz-Chinch{\'o}n}, F. and {Petravick}, D. and {Pond}, C. and
         {Yanny}, B. and {Allam}, S. and {Armstrong}, R. and {Barkhouse}, W. and
         {Bechtol}, K. and {Benoit-L{\'e}vy}, A. and {Bernstein}, G.~M. and
         {Bertin}, E. and {Buckley-Geer}, E. and {Covarrubias}, R. and
         {Desai}, S. and {Diehl}, H.~T. and {Goldstein}, D.~A. and {Gruen}, D. and
         {Li}, T.~S. and {Lin}, H. and {Marriner}, J. and {Mohr}, J.~J. and
         {Neilsen}, E. and {Ngeow}, C. -C. and {Paech}, K. and {Rykoff}, E.~S. and
         {Sako}, M. and {Sevilla-Noarbe}, I. and {Sheldon}, E. and
         {Sobreira}, F. and {Tucker}, D.~L. and {Wester}, W. and
         {DES Collaboration}},
        title = "{The Dark Energy Survey Image Processing Pipeline}",
      journal = {\pasp},
         year = "2018",
        month = "Jul",
       volume = {130},
       number = {989},
        pages = {074501},
          doi = {10.1088/1538-3873/aab4ef},
archivePrefix = {arXiv},
       eprint = {1801.03177},
 primaryClass = {astro-ph.IM},
       adsurl = {https://ui.adsabs.harvard.edu/abs/2018PASP..130g4501M}
}

@ARTICLE{vanderWalt2011Numpy,
       author = {{Van der Walt}, St{\'e}fan and {Colbert}, S. Chris and
         {Varoquaux}, Ga{\"e}l},
        title = "{The NumPy Array: A Structure for Efficient Numerical Computation}",
      journal = {Computing in Science and Engineering},
         year = 2011,
        month = mar,
       volume = {13},
       number = {2},
        pages = {22-30},
          doi = {10.1109/MCSE.2011.37},
archivePrefix = {arXiv},
       eprint = {1102.1523},
 primaryClass = {cs.MS},
       adsurl = {https://ui.adsabs.harvard.edu/abs/2011CSE....13b..22V}
}

@ARTICLE{2025PhRvD.112f3537T,
       author = {{To}, Chun-Hao and {Krause}, Elisabeth and {Chang}, Chihway and {Wu}, Hao-Yi and {Wechsler}, Risa H. and {Rozo}, Eduardo and {Weinberg}, David H. and {Anbajagane}, D. and {Avila}, S. and {Blazek}, J. and {Bocquet}, S. and {Costanzi}, M. and {De Vicente}, J. and {Elvin-Poole}, J. and {Fert{\'e}}, A. and {Grandis}, S. and {Muir}, J. and {Porredon}, A. and {Samuroff}, S. and {Sanchez}, E. and {Cid}, D. Sanchez and {Sevilla-Noarbe}, I. and {Weaverdyck}, N. and {Abbott}, T.~M.~C. and {Aguena}, M. and {Andrade-Oliveira}, F. and {Bacon}, D. and {Becker}, M.~R. and {Brooks}, D. and {Rosell}, A. Carnero and {Carretero}, J. and {Choi}, A. and {da Costa}, L.~N. and {Pereira}, M.~E.~S. and {Davis}, T.~M. and {Desai}, S. and {Doel}, P. and {Doux}, C. and {Everett}, S. and {Frieman}, J. and {Garc{\'\i}a-Bellido}, J. and {Gatti}, M. and {Gaztanaga}, E. and {Giannini}, G. and {Gruen}, D. and {Gutierrez}, G. and {Hinton}, S.~R. and {Hollowood}, D.~L. and {Honscheid}, K. and {Jeltema}, T. and {Kuehn}, K. and {Lee}, S. and {Marshall}, J.~L. and {Mena-Fern{\'a}ndez}, J. and {Miquel}, R. and {Mohr}, J.~J. and {Myles}, J. and {Palmese}, A. and {Malag{\'o}n}, A.~A. Plazas and {Romer}, A.~K. and {Shin}, T. and {Smith}, M. and {Suchyta}, E. and {Tarle}, G. and {Vikram}, V. and {Walker}, A.~R. and {Weller}, J. and {DES Collaboration}},
        title = "{Dark energy survey: Modeling strategy for multiprobe cluster cosmology and validation for the full six-year dataset}",
      journal = {\prd},
         year = 2025,
        month = sep,
       volume = {112},
       number = {6},
          eid = {063537},
        pages = {063537},
          doi = {10.1103/ynqj-6hsb},
archivePrefix = {arXiv},
       eprint = {2503.13631},
 primaryClass = {astro-ph.CO},
       adsurl = {https://ui.adsabs.harvard.edu/abs/2025PhRvD.112f3537T}
}

@ARTICLE{2021PhRvL.126n1301T,
       author = {{To}, C. and {Krause}, E. and {Rozo}, E. and {Wu}, H. and {Gruen}, D. and {Wechsler}, R.~H. and {Eifler}, T.~F. and {Rykoff}, E.~S. and {Costanzi}, M. and {Becker}, M.~R. and {Bernstein}, G.~M. and {Blazek}, J. and {Bocquet}, S. and {Bridle}, S.~L. and {Cawthon}, R. and {Choi}, A. and {Crocce}, M. and {Davis}, C. and {DeRose}, J. and {Drlica-Wagner}, A. and {Elvin-Poole}, J. and {Fang}, X. and {Farahi}, A. and {Friedrich}, O. and {Gatti}, M. and {Gaztanaga}, E. and {Giannantonio}, T. and {Hartley}, W.~G. and {Hoyle}, B. and {Jarvis}, M. and {MacCrann}, N. and {McClintock}, T. and {Miranda}, V. and {Pereira}, M.~E.~S. and {Park}, Y. and {Porredon}, A. and {Prat}, J. and {Rau}, M.~M. and {Ross}, A.~J. and {Samuroff}, S. and {S{\'a}nchez}, C. and {Sevilla-Noarbe}, I. and {Sheldon}, E. and {Troxel}, M.~A. and {Varga}, T.~N. and {Vielzeuf}, P. and {Zhang}, Y. and {Zuntz}, J. and {Abbott}, T.~M.~C. and {Aguena}, M. and {Amon}, A. and {Annis}, J. and {Avila}, S. and {Bertin}, E. and {Bhargava}, S. and {Brooks}, D. and {Burke}, D.~L. and {Carnero Rosell}, A. and {Carrasco Kind}, M. and {Carretero}, J. and {Chang}, C. and {Conselice}, C. and {da Costa}, L.~N. and {Davis}, T.~M. and {Desai}, S. and {Diehl}, H.~T. and {Dietrich}, J.~P. and {Everett}, S. and {Evrard}, A.~E. and {Ferrero}, I. and {Flaugher}, B. and {Fosalba}, P. and {Frieman}, J. and {Garc{\'\i}a-Bellido}, J. and {Gruendl}, R.~A. and {Gutierrez}, G. and {Hinton}, S.~R. and {Hollowood}, D.~L. and {Honscheid}, K. and {Huterer}, D. and {James}, D.~J. and {Jeltema}, T. and {Kron}, R. and {Kuehn}, K. and {Kuropatkin}, N. and {Lima}, M. and {Maia}, M.~A.~G. and {Marshall}, J.~L. and {Menanteau}, F. and {Miquel}, R. and {Morgan}, R. and {Muir}, J. and {Myles}, J. and {Palmese}, A. and {Paz-Chinch{\'o}n}, F. and {Plazas}, A.~A. and {Romer}, A.~K. and {Roodman}, A. and {Sanchez}, E. and {Santiago}, B. and {Scarpine}, V. and {Serrano}, S. and {Smith}, M. and {Suchyta}, E. and {Swanson}, M.~E.~C. and {Tarle}, G. and {Thomas}, D. and {Tucker}, D.~L. and {Weller}, J. and {Wester}, W. and {Wilkinson}, R.~D. and {DES Collaboration}},
        title = "{Dark Energy Survey Year 1 Results: Cosmological Constraints from Cluster Abundances, Weak Lensing, and Galaxy Correlations}",
      journal = {\prl},
         year = 2021,
        month = apr,
       volume = {126},
       number = {14},
          eid = {141301},
        pages = {141301},
          doi = {10.1103/PhysRevLett.126.141301},
archivePrefix = {arXiv},
       eprint = {2010.01138},
 primaryClass = {astro-ph.CO},
       adsurl = {https://ui.adsabs.harvard.edu/abs/2021PhRvL.126n1301T}
}

@ARTICLE{Virtanen2020Scipy,
       author = {{Virtanen}, Pauli and {Gommers}, Ralf and {Oliphant},
         Travis E. and {Haberland},      Matt and {Reddy}, Tyler and
         {Cournapeau}, David and {Burovski}, Evgeni and {Peterson}, Pearu
         and {Weckesser}, Warren and {Bright}, Jonathan and {van der Walt},
         St{\'e}fan J.  and {Brett}, Matthew and {Wilson}, Joshua and
         {Jarrod Millman}, K.  and {Mayorov}, Nikolay and {Nelson}, Andrew
         R.~J. and {Jones}, Eric and {Kern}, Robert and {Larson}, Eric and
         {Carey}, CJ and {Polat}, {\.I}lhan and {Feng}, Yu and {Moore},
         Eric W. and {Vand erPlas}, Jake and {Laxalde}, Denis and
         {Perktold}, Josef and {Cimrman}, Robert and {Henriksen}, Ian and
         {Quintero}, E.~A. and {Harris}, Charles R and {Archibald}, Anne M.
         and {Ribeiro}, Ant{\^o}nio H. and {Pedregosa}, Fabian and
         {van Mulbregt}, Paul and {Contributors}, SciPy 1. 0},
        title = "{SciPy 1.0: Fundamental Algorithms for Scientific
                  Computing in Python}",
      journal = {Nature Methods},
      year = "2020",
      volume={17},
      pages={261--272},
      adsurl = {https://rdcu.be/b08Wh},
      doi = {https://doi.org/10.1038/s41592-019-0686-2},
}

@ARTICLE{Hunter2007Matplotlib,
       author = {{Hunter}, John D.},
        title = "{Matplotlib: A 2D Graphics Environment}",
      journal = {Computing in Science and Engineering},
         year = 2007,
        month = may,
       volume = {9},
       number = {3},
        pages = {90-95},
          doi = {10.1109/MCSE.2007.55},
       adsurl = {https://ui.adsabs.harvard.edu/abs/2007CSE.....9...90H}
}

@article{Mckinney2011pandas,
    title={pandas: a foundational Python library for data analysis and statistics},
    author={Wes McKinney},
    journal={Python for High Performance and Scientific Computing},
    volume={14},
    year={2011}
}

@article{y3-shapecatalog,
       author = {{Gatti \& Sheldon} and {Sheldon}, E. and {Amon}, A. and {Becker}, M. and {Troxel}, M. and {Choi}, A. and {Doux}, C. and {MacCrann}, N. and {Navarro-Alsina}, A. and {Harrison}, I. and {Gruen}, D. and {Bernstein}, G. and {Jarvis}, M. and {Secco}, L.~F. and {Fert{\'e}}, A. and {Shin}, T. and {McCullough}, J. and {Rollins}, R.~P. and {Chen}, R. and {Chang}, C. and {Pandey}, S. and {Tutusaus}, I. and {Prat}, J. and {Elvin-Poole}, J. and {Sanchez}, C. and {Plazas}, A.~A. and {Roodman}, A. and {Zuntz}, J. and {Abbott}, T.~M.~C. and {Aguena}, M. and {Allam}, S. and {Annis}, J. and {Avila}, S. and {Bacon}, D. and {Bertin}, E. and {Bhargava}, S. and {Brooks}, D. and {Burke}, D.~L. and {Carnero Rosell}, A. and {Carrasco Kind}, M. and {Carretero}, J. and {Castander}, F.~J. and {Conselice}, C. and {Costanzi}, M. and {Crocce}, M. and {da Costa}, L.~N. and {Davis}, T.~M. and {De Vicente}, J. and {Desai}, S. and {Diehl}, H.~T. and {Dietrich}, J.~P. and {Doel}, P. and {Drlica-Wagner}, A. and {Eckert}, K. and {Everett}, S. and {Ferrero}, I. and {Frieman}, J. and {Garc{\'\i}a-Bellido}, J. and {Gerdes}, D.~W. and {Giannantonio}, T. and {Gruendl}, R.~A. and {Gschwend}, J. and {Gutierrez}, G. and {Hartley}, W.~G. and {Hinton}, S.~R. and {Hollowood}, D.~L. and {Honscheid}, K. and {Hoyle}, B. and {Huff}, E.~M. and {Huterer}, D. and {Jain}, B. and {James}, D.~J. and {Jeltema}, T. and {Krause}, E. and {Kron}, R. and {Kuropatkin}, N. and {Lima}, M. and {Maia}, M.~A.~G. and {Marshall}, J.~L. and {Miquel}, R. and {Morgan}, R. and {Myles}, J. and {Palmese}, A. and {Paz-Chinch{\'o}n}, F. and {Rykoff}, E.~S. and {Samuroff}, S. and {Sanchez}, E. and {Scarpine}, V. and {Schubnell}, M. and {Serrano}, S. and {Sevilla-Noarbe}, I. and {Smith}, M. and {Soares-Santos}, M. and {Suchyta}, E. and {Swanson}, M.~E.~C. and {Tarle}, G. and {Thomas}, D. and {To}, C. and {Tucker}, D.~L. and {Varga}, T.~N. and {Wechsler}, R.~H. and {Weller}, J. and {Wester}, W. and {Wilkinson}, R.~D.},
        title = "{Dark energy survey year 3 results: weak lensing shape catalogue}",
      journal = {\mnras},
         year = 2021,
        month = jul,
       volume = {504},
       number = {3},
        pages = {4312-4336},
          doi = {10.1093/mnras/stab918},
archivePrefix = {arXiv},
       eprint = {2011.03408},
 primaryClass = {astro-ph.CO},
       adsurl = {https://ui.adsabs.harvard.edu/abs/2021MNRAS.504.4312G}
}

@article{Bartelmann2001,
	Adsurl = {http://adsabs.harvard.edu/abs/2001PhR...340..291B},
	Author = {{Bartelmann}, M. and {Schneider}, P.},
	Doi = {10.1016/S0370-1573(00)00082-X},
	Eprint = {astro-ph/9912508},
	Journal = {\physrep},
	Month = jan,
	Pages = {291-472},
	Title = {{Weak gravitational lensing}},
	Volume = 340,
	Year = 2001}

@ARTICLE{Huff2017,
       author = {{Huff}, Eric and {Mandelbaum}, Rachel},
        title = "{Metacalibration: Direct Self-Calibration of Biases in Shear Measurement}",
      journal = {arXiv e-prints},
         year = 2017,
        month = feb,
          eid = {arXiv:1702.02600},
        pages = {arXiv:1702.02600},
          doi = {10.48550/arXiv.1702.02600},
archivePrefix = {arXiv},
       eprint = {1702.02600},
 primaryClass = {astro-ph.CO},
       adsurl = {https://ui.adsabs.harvard.edu/abs/2017arXiv170202600H}
}

@article{York_2000,
   author = {{York}, D.~G. and {Adelman}, J. and {Anderson}, Jr., J.~E. and {Anderson}, S.~F. and {Annis}, J. and {Bahcall}, N.~A. and {Bakken}, J.~A. and {Barkhouser}, R. and {Bastian}, S. and {Berman}, E. and {Boroski}, W.~N. and {Bracker}, S. and {Briegel}, C. and {Briggs}, J.~W. and {Brinkmann}, J. and {Brunner}, R. and {Burles}, S. and {Carey}, L. and {Carr}, M.~A. and {Castander}, F.~J. and {Chen}, B. and {Colestock}, P.~L. and {Connolly}, A.~J. and {Crocker}, J.~H. and {Csabai}, I. and {Czarapata}, P.~C. and {Davis}, J.~E. and {Doi}, M. and {Dombeck}, T. and {Eisenstein}, D. and {Ellman}, N. and {Elms}, B.~R. and {Evans}, M.~L. and {Fan}, X. and {Federwitz}, G.~R. and {Fiscelli}, L. and {Friedman}, S. and {Frieman}, J.~A. and {Fukugita}, M. and {Gillespie}, B. and {Gunn}, J.~E. and {Gyuk}, G. and {Haldeman}, M. and {Harris}, F.~H. and {Hayes}, J. and {Heckman}, T.~M. and {Hennessy}, G.~S. and {Hindsley}, R.~B. and {Holm}, S. and {Holmgren}, D.~J. and {Huang}, C. and {Hull}, C. and {Husby}, D. and {Ichikawa}, S.-I. and {Ichikawa}, T. and {Ivezi{\'c}}, Z. and {Kent}, S. and {Kim}, R.~S.~J. and {Kinney}, E. and {Klaene}, M. and {Kleinman}, A.~N. and {Kron}, R.~G. and {Kunszt}, P.~Z. and {Lamb}, D.~Q. and {Lee}, B. and {Leger}, R.~F. and {Limmongkol}, S. and {Lindenmeyer}, C. and {Long}, D.~C. and {Lupton}, R.~H. and {MacKinnon}, B. and {Mannery}, E.~J. and {Mantsch}, P.~M. and {Margon}, B. and {McGehee}, P. and {McKay}, T.~A. and {Meiksin}, A. and {Merelli}, A. and {Monet}, D.~G. and {Munn}, J.~A. and {Narayanan}, V.~K. and {Nash}, T. and {Neilsen}, Jr., E.~H. and {Neswold}, R. and {Newberg}, H.~J. and {Nichol}, R.~C. and {Nicinski}, T. and {Nonino}, M. and {Okada}, N. and {Okamura}, S. and {Ostriker}, J.~P. and {Owen}, R. and {Pauls}, A.~G. and {Peoples}, Jr., J. and {Peterson}, R.~L. and {Petravick}, D. and {Pier}, J.~R. and {Pope}, A. and {Pordes}, R. and {Prosapio}, A. and {Rechenmacher}, R. and {Rivetta}, C.~H. and {Rivetta}, S. and {Rockosi}, C.~M. and {Schlegel}, D.~J. and {Siegmund}, W.~A. and {Smee}, S. and {Smith}, J.~A. and {Snedden}, S. and {Stone}, R. and {Stoughton}, C. and {Szalay}, A.~S. and {Szapudi}, I. and {Szokoly}, G.~P. and {Thakar}, A.~R. and {Tremonti}, C. and {Tucker}, D.~L. and {Uomoto}, A. and {Vanden Berk}, D. and {Vogeley}, M.~S. and {Waddell}, P. and {Wang}, S.-i. and {Watanabe}, M. and {Weinberg}, D.~H. and {Yanny}, B. and {Yasuda}, N.},
   title = "{The Sloan Digital Sky Survey: Technical Summary}",
   journal = {The Astronomical Journal},
   volume = {120},
   pages = {1579-1587},
   year = 2000,
   month = sep,
   doi = {10.1086/301513},
   adsurl = {https://ui.adsabs.harvard.edu/abs/2000AJ....120.1579Y}
}

@ARTICLE{DESI:2016:Part1,
       author = {{DESI Collaboration} and {Aghamousa}, Amir and {Aguilar}, Jessica and {Ahlen}, Steve and {Alam}, Shadab and {Allen}, Lori E. and {Allende Prieto}, Carlos and {Annis}, James and {Bailey}, Stephen and {Balland}, Christophe and {Ballester}, Otger and {Baltay}, Charles and {Beaufore}, Lucas and {Bebek}, Chris and {Beers}, Timothy C. and {Bell}, Eric F. and {Bernal}, Jos{\'e} Luis and {Besuner}, Robert and {Beutler}, Florian and {Blake}, Chris and {Bleuler}, Hannes and {Blomqvist}, Michael and {Blum}, Robert and {Bolton}, Adam S. and {Briceno}, Cesar and {Brooks}, David and {Brownstein}, Joel R. and {Buckley-Geer}, Elizabeth and {Burden}, Angela and {Burtin}, Etienne and {Busca}, Nicolas G. and {Cahn}, Robert N. and {Cai}, Yan-Chuan and {Cardiel-Sas}, Laia and {Carlberg}, Raymond G. and {Carton}, Pierre-Henri and {Casas}, Ricard and {Castander}, Francisco J. and {Cervantes-Cota}, Jorge L. and {Claybaugh}, Todd M. and {Close}, Madeline and {Coker}, Carl T. and {Cole}, Shaun and {Comparat}, Johan and {Cooper}, Andrew P. and {Cousinou}, M. -C. and {Crocce}, Martin and {Cuby}, Jean-Gabriel and {Cunningham}, Daniel P. and {Davis}, Tamara M. and {Dawson}, Kyle S. and {de la Macorra}, Axel and {De Vicente}, Juan and {Delubac}, Timoth{\'e}e and {Derwent}, Mark and {Dey}, Arjun and {Dhungana}, Govinda and {Ding}, Zhejie and {Doel}, Peter and {Duan}, Yutong T. and {Ealet}, Anne and {Edelstein}, Jerry and {Eftekharzadeh}, Sarah and {Eisenstein}, Daniel J. and {Elliott}, Ann and {Escoffier}, St{\'e}phanie and {Evatt}, Matthew and {Fagrelius}, Parker and {Fan}, Xiaohui and {Fanning}, Kevin and {Farahi}, Arya and {Farihi}, Jay and {Favole}, Ginevra and {Feng}, Yu and {Fernandez}, Enrique and {Findlay}, Joseph R. and {Finkbeiner}, Douglas P. and {Fitzpatrick}, Michael J. and {Flaugher}, Brenna and {Flender}, Samuel and {Font-Ribera}, Andreu and {Forero-Romero}, Jaime E. and {Fosalba}, Pablo and {Frenk}, Carlos S. and {Fumagalli}, Michele and {Gaensicke}, Boris T. and {Gallo}, Giuseppe and {Garcia-Bellido}, Juan and {Gaztanaga}, Enrique and {Pietro Gentile Fusillo}, Nicola and {Gerard}, Terry and {Gershkovich}, Irena and {Giannantonio}, Tommaso and {Gillet}, Denis and {Gonzalez-de-Rivera}, Guillermo and {Gonzalez-Perez}, Violeta and {Gott}, Shelby and {Graur}, Or and {Gutierrez}, Gaston and {Guy}, Julien and {Habib}, Salman and {Heetderks}, Henry and {Heetderks}, Ian and {Heitmann}, Katrin and {Hellwing}, Wojciech A. and {Herrera}, David A. and {Ho}, Shirley and {Holland}, Stephen and {Honscheid}, Klaus and {Huff}, Eric and {Hutchinson}, Timothy A. and {Huterer}, Dragan and {Hwang}, Ho Seong and {Illa Laguna}, Joseph Maria and {Ishikawa}, Yuzo and {Jacobs}, Dianna and {Jeffrey}, Niall and {Jelinsky}, Patrick and {Jennings}, Elise and {Jiang}, Linhua and {Jimenez}, Jorge and {Johnson}, Jennifer and {Joyce}, Richard and {Jullo}, Eric and {Juneau}, St{\'e}phanie and {Kama}, Sami and {Karcher}, Armin and {Karkar}, Sonia and {Kehoe}, Robert and {Kennamer}, Noble and {Kent}, Stephen and {Kilbinger}, Martin and {Kim}, Alex G. and {Kirkby}, David and {Kisner}, Theodore and {Kitanidis}, Ellie and {Kneib}, Jean-Paul and {Koposov}, Sergey and {Kovacs}, Eve and {Koyama}, Kazuya and {Kremin}, Anthony and {Kron}, Richard and {Kronig}, Luzius and {Kueter-Young}, Andrea and {Lacey}, Cedric G. and {Lafever}, Robin and {Lahav}, Ofer and {Lambert}, Andrew and {Lampton}, Michael and {Landriau}, Martin and {Lang}, Dustin and {Lauer}, Tod R. and {Le Goff}, Jean-Marc and {Le Guillou}, Laurent and {Le Van Suu}, Auguste and {Lee}, Jae Hyeon and {Lee}, Su-Jeong and {Leitner}, Daniela and {Lesser}, Michael and {Levi}, Michael E. and {L'Huillier}, Benjamin and {Li}, Baojiu and {Liang}, Ming and {Lin}, Huan and {Linder}, Eric and {Loebman}, Sarah R. and {Luki{\'c}}, Zarija and {Ma}, Jun and {MacCrann}, Niall and {Magneville}, Christophe and {Makarem}, Laleh and {Manera}, Marc and {Manser}, Christopher J. and {Marshall}, Robert and {Martini}, Paul and {Massey}, Richard and {Matheson}, Thomas and {McCauley}, Jeremy and {McDonald}, Patrick and {McGreer}, Ian D. and {Meisner}, Aaron and {Metcalfe}, Nigel and {Miller}, Timothy N. and {Miquel}, Ramon and {Moustakas}, John and {Myers}, Adam and {Naik}, Milind and {Newman}, Jeffrey A. and {Nichol}, Robert C. and {Nicola}, Andrina and {Nicolati da Costa}, Luiz and {Nie}, Jundan and {Niz}, Gustavo and {Norberg}, Peder and {Nord}, Brian and {Norman}, Dara and {Nugent}, Peter and {O'Brien}, Thomas and {Oh}, Minji and {Olsen}, Knut A.~G. and {Padilla}, Cristobal and {Padmanabhan}, Hamsa and {Padmanabhan}, Nikhil and {Palanque-Delabrouille}, Nathalie and {Palmese}, Antonella and {Pappalardo}, Daniel and {P{\^a}ris}, Isabelle and {Park}, Changbom and {Patej}, Anna and {Peacock}, John A. and {Peiris}, Hiranya V. and {Peng}, Xiyan and {Percival}, Will J. and {Perruchot}, Sandrine and {Pieri}, Matthew M. and {Pogge}, Richard and {Pollack}, Jennifer E. and {Poppett}, Claire and {Prada}, Francisco and {Prakash}, Abhishek and {Probst}, Ronald G. and {Rabinowitz}, David and {Raichoor}, Anand and {Ree}, Chang Hee and {Refregier}, Alexandre and {Regal}, Xavier and {Reid}, Beth and {Reil}, Kevin and {Rezaie}, Mehdi and {Rockosi}, Constance M. and {Roe}, Natalie and {Ronayette}, Samuel and {Roodman}, Aaron and {Ross}, Ashley J. and {Ross}, Nicholas P. and {Rossi}, Graziano and {Rozo}, Eduardo and {Ruhlmann-Kleider}, Vanina and {Rykoff}, Eli S. and {Sabiu}, Cristiano and {Samushia}, Lado and {Sanchez}, Eusebio and {Sanchez}, Javier and {Schlegel}, David J. and {Schneider}, Michael and {Schubnell}, Michael and {Secroun}, Aur{\'e}lia and {Seljak}, Uros and {Seo}, Hee-Jong and {Serrano}, Santiago and {Shafieloo}, Arman and {Shan}, Huanyuan and {Sharples}, Ray and {Sholl}, Michael J. and {Shourt}, William V. and {Silber}, Joseph H. and {Silva}, David R. and {Sirk}, Martin M. and {Slosar}, Anze and {Smith}, Alex and {Smoot}, George F. and {Som}, Debopam and {Song}, Yong-Seon and {Sprayberry}, David and {Staten}, Ryan and {Stefanik}, Andy and {Tarle}, Gregory and {Sien Tie}, Suk and {Tinker}, Jeremy L. and {Tojeiro}, Rita and {Valdes}, Francisco and {Valenzuela}, Octavio and {Valluri}, Monica and {Vargas-Magana}, Mariana and {Verde}, Licia and {Walker}, Alistair R. and {Wang}, Jiali and {Wang}, Yuting and {Weaver}, Benjamin A. and {Weaverdyck}, Curtis and {Wechsler}, Risa H. and {Weinberg}, David H. and {White}, Martin and {Yang}, Qian and {Yeche}, Christophe and {Zhang}, Tianmeng and {Zhao}, Gong-Bo and {Zheng}, Yi and {Zhou}, Xu and {Zhou}, Zhimin and {Zhu}, Yaling and {Zou}, Hu and {Zu}, Ying},
        title = "{The DESI Experiment Part I: Science,Targeting, and Survey Design}",
      journal = {arXiv e-prints},
         year = 2016,
        month = oct,
          eid = {arXiv:1611.00036},
        pages = {arXiv:1611.00036},
          doi = {10.48550/arXiv.1611.00036},
archivePrefix = {arXiv},
       eprint = {1611.00036},
 primaryClass = {astro-ph.IM},
       adsurl = {https://ui.adsabs.harvard.edu/abs/2016arXiv161100036D}
}

@ARTICLE{LSST2018SRD,
       author = {{The LSST Dark Energy Science Collaboration} and {Mandelbaum}, Rachel and {Eifler}, Tim and {Hlo{\v{z}}ek}, Ren{\'e}e and {Collett}, Thomas and {Gawiser}, Eric and {Scolnic}, Daniel and {Alonso}, David and {Awan}, Humna and {Biswas}, Rahul and {Blazek}, Jonathan and {Burchat}, Patricia and {Chisari}, Nora Elisa and {Dell'Antonio}, Ian and {Digel}, Seth and {Frieman}, Josh and {Goldstein}, Daniel A. and {Hook}, Isobel and {Ivezi{\'c}}, {\v{Z}}eljko and {Kahn}, Steven M. and {Kamath}, Sowmya and {Kirkby}, David and {Kitching}, Thomas and {Krause}, Elisabeth and {Leget}, Pierre-Fran{\c{c}}ois and {Marshall}, Philip J. and {Meyers}, Joshua and {Miyatake}, Hironao and {Newman}, Jeffrey A. and {Nichol}, Robert and {Rykoff}, Eli and {Sanchez}, F. Javier and {Slosar}, An{\v{z}}e and {Sullivan}, Mark and {Troxel}, M.~A.},
        title = "{The LSST Dark Energy Science Collaboration (DESC) Science Requirements Document}",
      journal = {arXiv e-prints},
         year = 2018,
        month = sep,
          eid = {arXiv:1809.01669},
        pages = {arXiv:1809.01669},
          doi = {10.48550/arXiv.1809.01669},
archivePrefix = {arXiv},
       eprint = {1809.01669},
 primaryClass = {astro-ph.CO},
       adsurl = {https://ui.adsabs.harvard.edu/abs/2018arXiv180901669T}
}

@INPROCEEDINGS{Euclid,
       author = {{Racca}, Giuseppe D. and {Laureijs}, Ren{\'e} and {Stagnaro}, Luca and {Salvignol}, Jean-Christophe and {Lorenzo Alvarez}, Jos{\'e} and {Saavedra Criado}, Gonzalo and {Gaspar Venancio}, Luis and {Short}, Alex and {Strada}, Paolo and {B{\"o}nke}, Tobias and {Colombo}, Cyril and {Calvi}, Adriano and {Maiorano}, Elena and {Piersanti}, Osvaldo and {Prezelus}, Sylvain and {Rosato}, Pierluigi and {Pinel}, Jacques and {Rozemeijer}, Hans and {Lesna}, Valentina and {Musi}, Paolo and {Sias}, Marco and {Anselmi}, Alberto and {Cazaubiel}, Vincent and {Vaillon}, Ludovic and {Mellier}, Yannick and {Amiaux}, J{\'e}r{\^o}me and {Berth{\'e}}, Michel and {Sauvage}, Marc and {Azzollini}, Ruyman and {Cropper}, Mark and {Pottinger}, Sabrina and {Jahnke}, Knud and {Ealet}, Anne and {Maciaszek}, Thierry and {Pasian}, Fabio and {Zacchei}, Andrea and {Scaramella}, Roberto and {Hoar}, John and {Kohley}, Ralf and {Vavrek}, Roland and {Rudolph}, Andreas and {Schmidt}, Micha},
        title = "{The Euclid mission design}",
    booktitle = {Space Telescopes and Instrumentation 2016: Optical, Infrared, and Millimeter Wave},
         year = 2016,
       editor = {{MacEwen}, Howard A. and {Fazio}, Giovanni G. and {Lystrup}, Makenzie and {Batalha}, Natalie and {Siegler}, Nicholas and {Tong}, Edward C.},
       series = {Society of Photo-Optical Instrumentation Engineers (SPIE) Conference Series},
       volume = {9904},
        month = jul,
          eid = {99040O},
        pages = {99040O},
          doi = {10.1117/12.2230762},
archivePrefix = {arXiv},
       eprint = {1610.05508},
 primaryClass = {astro-ph.IM},
       adsurl = {https://ui.adsabs.harvard.edu/abs/2016SPIE.9904E..0OR}
}

@ARTICLE{Spergel:2015:Roman,
       author = {{Spergel}, D. and {Gehrels}, N. and {Baltay}, C. and {Bennett}, D. and {Breckinridge}, J. and {Donahue}, M. and {Dressler}, A. and {Gaudi}, B.~S. and {Greene}, T. and {Guyon}, O. and {Hirata}, C. and {Kalirai}, J. and {Kasdin}, N.~J. and {Macintosh}, B. and {Moos}, W. and {Perlmutter}, S. and {Postman}, M. and {Rauscher}, B. and {Rhodes}, J. and {Wang}, Y. and {Weinberg}, D. and {Benford}, D. and {Hudson}, M. and {Jeong}, W. -S. and {Mellier}, Y. and {Traub}, W. and {Yamada}, T. and {Capak}, P. and {Colbert}, J. and {Masters}, D. and {Penny}, M. and {Savransky}, D. and {Stern}, D. and {Zimmerman}, N. and {Barry}, R. and {Bartusek}, L. and {Carpenter}, K. and {Cheng}, E. and {Content}, D. and {Dekens}, F. and {Demers}, R. and {Grady}, K. and {Jackson}, C. and {Kuan}, G. and {Kruk}, J. and {Melton}, M. and {Nemati}, B. and {Parvin}, B. and {Poberezhskiy}, I. and {Peddie}, C. and {Ruffa}, J. and {Wallace}, J.~K. and {Whipple}, A. and {Wollack}, E. and {Zhao}, F.},
        title = "{Wide-Field InfrarRed Survey Telescope-Astrophysics Focused Telescope Assets WFIRST-AFTA 2015 Report}",
      journal = {arXiv e-prints},
         year = 2015,
        month = mar,
          eid = {arXiv:1503.03757},
        pages = {arXiv:1503.03757},
          doi = {10.48550/arXiv.1503.03757},
archivePrefix = {arXiv},
       eprint = {1503.03757},
 primaryClass = {astro-ph.IM},
       adsurl = {https://ui.adsabs.harvard.edu/abs/2015arXiv150303757S}
}

@ARTICLE{Planck:2020:Cosmo,
       author = {{Planck Collaboration} and {Aghanim}, N. and {Akrami}, Y. and {Ashdown}, M. and {Aumont}, J. and {Baccigalupi}, C. and {Ballardini}, M. and {Banday}, A.~J. and {Barreiro}, R.~B. and {Bartolo}, N. and {Basak}, S. and {Battye}, R. and {Benabed}, K. and {Bernard}, J. -P. and {Bersanelli}, M. and {Bielewicz}, P. and {Bock}, J.~J. and {Bond}, J.~R. and {Borrill}, J. and {Bouchet}, F.~R. and {Boulanger}, F. and {Bucher}, M. and {Burigana}, C. and {Butler}, R.~C. and {Calabrese}, E. and {Cardoso}, J. -F. and {Carron}, J. and {Challinor}, A. and {Chiang}, H.~C. and {Chluba}, J. and {Colombo}, L.~P.~L. and {Combet}, C. and {Contreras}, D. and {Crill}, B.~P. and {Cuttaia}, F. and {de Bernardis}, P. and {de Zotti}, G. and {Delabrouille}, J. and {Delouis}, J. -M. and {Di Valentino}, E. and {Diego}, J.~M. and {Dor{\'e}}, O. and {Douspis}, M. and {Ducout}, A. and {Dupac}, X. and {Dusini}, S. and {Efstathiou}, G. and {Elsner}, F. and {En{\ss}lin}, T.~A. and {Eriksen}, H.~K. and {Fantaye}, Y. and {Farhang}, M. and {Fergusson}, J. and {Fernandez-Cobos}, R. and {Finelli}, F. and {Forastieri}, F. and {Frailis}, M. and {Fraisse}, A.~A. and {Franceschi}, E. and {Frolov}, A. and {Galeotta}, S. and {Galli}, S. and {Ganga}, K. and {G{\'e}nova-Santos}, R.~T. and {Gerbino}, M. and {Ghosh}, T. and {Gonz{\'a}lez-Nuevo}, J. and {G{\'o}rski}, K.~M. and {Gratton}, S. and {Gruppuso}, A. and {Gudmundsson}, J.~E. and {Hamann}, J. and {Handley}, W. and {Hansen}, F.~K. and {Herranz}, D. and {Hildebrandt}, S.~R. and {Hivon}, E. and {Huang}, Z. and {Jaffe}, A.~H. and {Jones}, W.~C. and {Karakci}, A. and {Keih{\"a}nen}, E. and {Keskitalo}, R. and {Kiiveri}, K. and {Kim}, J. and {Kisner}, T.~S. and {Knox}, L. and {Krachmalnicoff}, N. and {Kunz}, M. and {Kurki-Suonio}, H. and {Lagache}, G. and {Lamarre}, J. -M. and {Lasenby}, A. and {Lattanzi}, M. and {Lawrence}, C.~R. and {Le Jeune}, M. and {Lemos}, P. and {Lesgourgues}, J. and {Levrier}, F. and {Lewis}, A. and {Liguori}, M. and {Lilje}, P.~B. and {Lilley}, M. and {Lindholm}, V. and {L{\'o}pez-Caniego}, M. and {Lubin}, P.~M. and {Ma}, Y. -Z. and {Mac{\'\i}as-P{\'e}rez}, J.~F. and {Maggio}, G. and {Maino}, D. and {Mandolesi}, N. and {Mangilli}, A. and {Marcos-Caballero}, A. and {Maris}, M. and {Martin}, P.~G. and {Martinelli}, M. and {Mart{\'\i}nez-Gonz{\'a}lez}, E. and {Matarrese}, S. and {Mauri}, N. and {McEwen}, J.~D. and {Meinhold}, P.~R. and {Melchiorri}, A. and {Mennella}, A. and {Migliaccio}, M. and {Millea}, M. and {Mitra}, S. and {Miville-Desch{\^e}nes}, M. -A. and {Molinari}, D. and {Montier}, L. and {Morgante}, G. and {Moss}, A. and {Natoli}, P. and {N{\o}rgaard-Nielsen}, H.~U. and {Pagano}, L. and {Paoletti}, D. and {Partridge}, B. and {Patanchon}, G. and {Peiris}, H.~V. and {Perrotta}, F. and {Pettorino}, V. and {Piacentini}, F. and {Polastri}, L. and {Polenta}, G. and {Puget}, J. -L. and {Rachen}, J.~P. and {Reinecke}, M. and {Remazeilles}, M. and {Renzi}, A. and {Rocha}, G. and {Rosset}, C. and {Roudier}, G. and {Rubi{\~n}o-Mart{\'\i}n}, J.~A. and {Ruiz-Granados}, B. and {Salvati}, L. and {Sandri}, M. and {Savelainen}, M. and {Scott}, D. and {Shellard}, E.~P.~S. and {Sirignano}, C. and {Sirri}, G. and {Spencer}, L.~D. and {Sunyaev}, R. and {Suur-Uski}, A. -S. and {Tauber}, J.~A. and {Tavagnacco}, D. and {Tenti}, M. and {Toffolatti}, L. and {Tomasi}, M. and {Trombetti}, T. and {Valenziano}, L. and {Valiviita}, J. and {Van Tent}, B. and {Vibert}, L. and {Vielva}, P. and {Villa}, F. and {Vittorio}, N. and {Wandelt}, B.~D. and {Wehus}, I.~K. and {White}, M. and {White}, S.~D.~M. and {Zacchei}, A. and {Zonca}, A.},
        title = "{Planck 2018 results. VI. Cosmological parameters}",
      journal = {\aap},
         year = 2020,
        month = sep,
       volume = {641},
          eid = {A6},
        pages = {A6},
          doi = {10.1051/0004-6361/201833910},
archivePrefix = {arXiv},
       eprint = {1807.06209},
 primaryClass = {astro-ph.CO},
       adsurl = {https://ui.adsabs.harvard.edu/abs/2020A&A...641A...6P}
}

@ARTICLE{Allen:2011:Clusters,
       author = {{Allen}, Steven W. and {Evrard}, August E. and {Mantz}, Adam B.},
        title = "{Cosmological Parameters from Observations of Galaxy Clusters}",
      journal = {\araa},
         year = 2011,
        month = sep,
       volume = {49},
       number = {1},
        pages = {409-470},
          doi = {10.1146/annurev-astro-081710-102514},
archivePrefix = {arXiv},
       eprint = {1103.4829},
 primaryClass = {astro-ph.CO},
       adsurl = {https://ui.adsabs.harvard.edu/abs/2011ARA&A..49..409A}
}

@ARTICLE{Carlstrom2011,
       author = {{Carlstrom}, J.~E. and {Ade}, P.~A.~R. and {Aird}, K.~A. and {Benson}, B.~A. and {Bleem}, L.~E. and {Busetti}, S. and {Chang}, C.~L. and {Chauvin}, E. and {Cho}, H. -M. and {Crawford}, T.~M. and {Crites}, A.~T. and {Dobbs}, M.~A. and {Halverson}, N.~W. and {Heimsath}, S. and {Holzapfel}, W.~L. and {Hrubes}, J.~D. and {Joy}, M. and {Keisler}, R. and {Lanting}, T.~M. and {Lee}, A.~T. and {Leitch}, E.~M. and {Leong}, J. and {Lu}, W. and {Lueker}, M. and {Luong-Van}, D. and {McMahon}, J.~J. and {Mehl}, J. and {Meyer}, S.~S. and {Mohr}, J.~J. and {Montroy}, T.~E. and {Padin}, S. and {Plagge}, T. and {Pryke}, C. and {Ruhl}, J.~E. and {Schaffer}, K.~K. and {Schwan}, D. and {Shirokoff}, E. and {Spieler}, H.~G. and {Staniszewski}, Z. and {Stark}, A.~A. and {Tucker}, C. and {Vanderlinde}, K. and {Vieira}, J.~D. and {Williamson}, R.},
        title = "{The 10 Meter South Pole Telescope}",
      journal = {\pasp},
         year = 2011,
        month = may,
       volume = {123},
       number = {903},
        pages = {568},
          doi = {10.1086/659879},
archivePrefix = {arXiv},
       eprint = {0907.4445},
 primaryClass = {astro-ph.IM},
       adsurl = {https://ui.adsabs.harvard.edu/abs/2011PASP..123..568C}
}

@ARTICLE{ACT:2016,
       author = {{Thornton}, R.~J. and {Ade}, P.~A.~R. and {Aiola}, S. and {Angil{\`e}}, F.~E. and {Amiri}, M. and {Beall}, J.~A. and {Becker}, D.~T. and {Cho}, H. -M. and {Choi}, S.~K. and {Corlies}, P. and {Coughlin}, K.~P. and {Datta}, R. and {Devlin}, M.~J. and {Dicker}, S.~R. and {D{\"u}nner}, R. and {Fowler}, J.~W. and {Fox}, A.~E. and {Gallardo}, P.~A. and {Gao}, J. and {Grace}, E. and {Halpern}, M. and {Hasselfield}, M. and {Henderson}, S.~W. and {Hilton}, G.~C. and {Hincks}, A.~D. and {Ho}, S.~P. and {Hubmayr}, J. and {Irwin}, K.~D. and {Klein}, J. and {Koopman}, B. and {Li}, Dale and {Louis}, T. and {Lungu}, M. and {Maurin}, L. and {McMahon}, J. and {Munson}, C.~D. and {Naess}, S. and {Nati}, F. and {Newburgh}, L. and {Nibarger}, J. and {Niemack}, M.~D. and {Niraula}, P. and {Nolta}, M.~R. and {Page}, L.~A. and {Pappas}, C.~G. and {Schillaci}, A. and {Schmitt}, B.~L. and {Sehgal}, N. and {Sievers}, J.~L. and {Simon}, S.~M. and {Staggs}, S.~T. and {Tucker}, C. and {Uehara}, M. and {van Lanen}, J. and {Ward}, J.~T. and {Wollack}, E.~J.},
        title = "{The Atacama Cosmology Telescope: The Polarization-sensitive ACTPol Instrument}",
      journal = {\apjs},
         year = 2016,
        month = dec,
       volume = {227},
       number = {2},
          eid = {21},
        pages = {21},
          doi = {10.3847/1538-4365/227/2/21},
archivePrefix = {arXiv},
       eprint = {1605.06569},
 primaryClass = {astro-ph.IM},
       adsurl = {https://ui.adsabs.harvard.edu/abs/2016ApJS..227...21T}
}

@ARTICLE{Merloni:2012:Erosita,
       author = {{Merloni}, A. and {Predehl}, P. and {Becker}, W. and {B{\"o}hringer}, H. and {Boller}, T. and {Brunner}, H. and {Brusa}, M. and {Dennerl}, K. and {Freyberg}, M. and {Friedrich}, P. and {Georgakakis}, A. and {Haberl}, F. and {Hasinger}, G. and {Meidinger}, N. and {Mohr}, J. and {Nandra}, K. and {Rau}, A. and {Reiprich}, T.~H. and {Robrade}, J. and {Salvato}, M. and {Santangelo}, A. and {Sasaki}, M. and {Schwope}, A. and {Wilms}, J. and {German eROSITA Consortium}, the},
        title = "{eROSITA Science Book: Mapping the Structure of the Energetic Universe}",
      journal = {arXiv e-prints},
         year = 2012,
        month = sep,
          eid = {arXiv:1209.3114},
        pages = {arXiv:1209.3114},
          doi = {10.48550/arXiv.1209.3114},
archivePrefix = {arXiv},
       eprint = {1209.3114},
 primaryClass = {astro-ph.HE},
       adsurl = {https://ui.adsabs.harvard.edu/abs/2012arXiv1209.3114M}
}

@ARTICLE{Schneider:2005:Lensing,
       author = {{Schneider}, Peter},
        title = "{Weak Gravitational Lensing}",
      journal = {arXiv e-prints},
         year = 2005,
        month = sep,
          eid = {astro-ph/0509252},
        pages = {astro-ph/0509252},
          doi = {10.48550/arXiv.astro-ph/0509252},
archivePrefix = {arXiv},
       eprint = {astro-ph/0509252},
 primaryClass = {astro-ph},
       adsurl = {https://ui.adsabs.harvard.edu/abs/2005astro.ph..9252S}
}

@ARTICLE{Anbajagane2020StellarProp,
       author = {{Anbajagane}, Dhayaa and {Evrard}, August E. and {Farahi}, Arya and {Barnes}, David J. and {Dolag}, Klaus and {McCarthy}, Ian G. and {Nelson}, Dylan and {Pillepich}, Annalisa},
        title = "{Stellar property statistics of massive haloes from cosmological hydrodynamics simulations: common kernel shapes}",
      journal = {\mnras},
         year = 2020,
        month = jun,
       volume = {495},
       number = {1},
        pages = {686-704},
          doi = {10.1093/mnras/staa1147},
archivePrefix = {arXiv},
       eprint = {2001.02283},
 primaryClass = {astro-ph.GA},
       adsurl = {https://ui.adsabs.harvard.edu/abs/2020MNRAS.495..686A}
}

@ARTICLE{Hartley:2020:DESY3Deep,
       author = {{Hartley}, W.~G. and {Choi}, A. and {Amon}, A. and {Gruendl}, R.~A. and {Sheldon}, E. and {Harrison}, I. and {Bernstein}, G.~M. and {Sevilla-Noarbe}, I. and {Yanny}, B. and {Eckert}, K. and {Diehl}, H.~T. and {Alarcon}, A. and {Banerji}, M. and {Bechtol}, K. and {Buchs}, R. and {Cantu}, S. and {Conselice}, C. and {Cordero}, J. and {Davis}, C. and {Davis}, T.~M. and {Dodelson}, S. and {Drlica-Wagner}, A. and {Everett}, S. and {Fert{\'e}}, A. and {Gruen}, D. and {Honscheid}, K. and {Jarvis}, M. and {Johnson}, M.~D. and {Kokron}, N. and {MacCrann}, N. and {Myles}, J. and {Pace}, A.~B. and {Palmese}, A. and {Paz-Chinch{\'o}n}, F. and {Pereira}, M.~E.~S. and {Plazas}, A.~A. and {Prat}, J. and {Rodriguez-Monroy}, M. and {Rykoff}, E.~S. and {Samuroff}, S. and {S{\'a}nchez}, C. and {Secco}, L.~F. and {Tarsitano}, F. and {Tong}, A. and {Troxel}, M.~A. and {Vasquez}, Z. and {Wang}, K. and {Zhou}, C. and {Abbott}, T.~M.~C. and {Aguena}, M. and {Allam}, S. and {Annis}, J. and {Bacon}, D. and {Bertin}, E. and {Bhargava}, S. and {Brooks}, D. and {Burke}, D.~L. and {Carnero Rosell}, A. and {Carrasco Kind}, M. and {Carretero}, J. and {Castander}, F.~J. and {Costanzi}, M. and {Crocce}, M. and {da Costa}, L.~N. and {De Vicente}, J. and {DeRose}, J. and {Desai}, S. and {Dietrich}, J.~P. and {Eifler}, T.~F. and {Elvin-Poole}, J. and {Ferrero}, I. and {Flaugher}, B. and {Fosalba}, P. and {Garc{\'\i}a-Bellido}, J. and {Gaztanaga}, E. and {Gerdes}, D.~W. and {Gschwend}, J. and {Gutierrez}, G. and {Hinton}, S.~R. and {Hollowood}, D.~L. and {Huterer}, D. and {James}, D.~J. and {Kent}, S. and {Krause}, E. and {Kuehn}, K. and {Kuropatkin}, N. and {Lahav}, O. and {Lin}, H. and {Maia}, M.~A.~G. and {March}, M. and {Marshall}, J.~L. and {Martini}, P. and {Melchior}, P. and {Menanteau}, F. and {Miquel}, R. and {Mohr}, J.~J. and {Morgan}, R. and {Neilsen}, E. and {Ogando}, R.~L.~C. and {Pandey}, S. and {Romer}, A.~K. and {Roodman}, A. and {Sako}, M. and {Sanchez}, E. and {Scarpine}, V. and {Serrano}, S. and {Smith}, M. and {Soares-Santos}, M. and {Suchyta}, E. and {Swanson}, M.~E.~C. and {Tarle}, G. and {Thomas}, D. and {To}, C. and {Varga}, T.~N. and {Walker}, A.~R. and {Wester}, W. and {Wilkinson}, R.~D. and {Zuntz}, J. and {Zuntz}, J. and {DES Collaboration}},
        title = "{Dark Energy Survey Year 3 Results: Deep Field optical + near-infrared images and catalogue}",
      journal = {\mnras},
         year = 2022,
        month = jan,
       volume = {509},
       number = {3},
        pages = {3547-3579},
          doi = {10.1093/mnras/stab3055},
archivePrefix = {arXiv},
       eprint = {2012.12824},
 primaryClass = {astro-ph.CO},
       adsurl = {https://ui.adsabs.harvard.edu/abs/2022MNRAS.509.3547H}
}

@ARTICLE{Sevilla-Noarbe2021,
       author = {{Sevilla-Noarbe}, I. and {Bechtol}, K. and {Carrasco Kind}, M. and {Carnero Rosell}, A. and {Becker}, M.~R. and {Drlica-Wagner}, A. and {Gruendl}, R.~A. and {Rykoff}, E.~S. and {Sheldon}, E. and {Yanny}, B. and {Alarcon}, A. and {Allam}, S. and {Amon}, A. and {Benoit-L{\'e}vy}, A. and {Bernstein}, G.~M. and {Bertin}, E. and {Burke}, D.~L. and {Carretero}, J. and {Choi}, A. and {Diehl}, H.~T. and {Everett}, S. and {Flaugher}, B. and {Gaztanaga}, E. and {Gschwend}, J. and {Harrison}, I. and {Hartley}, W.~G. and {Hoyle}, B. and {Jarvis}, M. and {Johnson}, M.~D. and {Kessler}, R. and {Kron}, R. and {Kuropatkin}, N. and {Leistedt}, B. and {Li}, T.~S. and {Menanteau}, F. and {Morganson}, E. and {Ogando}, R.~L.~C. and {Palmese}, A. and {Paz-Chinch{\'o}n}, F. and {Pieres}, A. and {Pond}, C. and {Rodriguez-Monroy}, M. and {Smith}, J. Allyn and {Stringer}, K.~M. and {Troxel}, M.~A. and {Tucker}, D.~L. and {de Vicente}, J. and {Wester}, W. and {Zhang}, Y. and {Abbott}, T.~M.~C. and {Aguena}, M. and {Annis}, J. and {Avila}, S. and {Bhargava}, S. and {Bridle}, S.~L. and {Brooks}, D. and {Brout}, D. and {Castander}, F.~J. and {Cawthon}, R. and {Chang}, C. and {Conselice}, C. and {Costanzi}, M. and {Crocce}, M. and {da Costa}, L.~N. and {Pereira}, M.~E.~S. and {Davis}, T.~M. and {Desai}, S. and {Dietrich}, J.~P. and {Doel}, P. and {Eckert}, K. and {Evrard}, A.~E. and {Ferrero}, I. and {Fosalba}, P. and {Garc{\'\i}a-Bellido}, J. and {Gerdes}, D.~W. and {Giannantonio}, T. and {Gruen}, D. and {Gutierrez}, G. and {Hinton}, S.~R. and {Hollowood}, D.~L. and {Honscheid}, K. and {Huff}, E.~M. and {Huterer}, D. and {James}, D.~J. and {Jeltema}, T. and {Kuehn}, K. and {Lahav}, O. and {Lidman}, C. and {Lima}, M. and {Lin}, H. and {Maia}, M.~A.~G. and {Marshall}, J.~L. and {Martini}, P. and {Melchior}, P. and {Miquel}, R. and {Mohr}, J.~J. and {Morgan}, R. and {Neilsen}, E. and {Plazas}, A.~A. and {Romer}, A.~K. and {Roodman}, A. and {Sanchez}, E. and {Scarpine}, V. and {Schubnell}, M. and {Serrano}, S. and {Smith}, M. and {Suchyta}, E. and {Tarle}, G. and {Thomas}, D. and {To}, C. and {Varga}, T.~N. and {Wechsler}, R.~H. and {Weller}, J. and {Wilkinson}, R.~D. and {DES Collaboration}},
        title = "{Dark Energy Survey Year 3 Results: Photometric Data Set for Cosmology}",
      journal = {\apjs},
         year = 2021,
        month = jun,
       volume = {254},
       number = {2},
          eid = {24},
        pages = {24},
          doi = {10.3847/1538-4365/abeb66},
archivePrefix = {arXiv},
       eprint = {2011.03407},
 primaryClass = {astro-ph.CO},
       adsurl = {https://ui.adsabs.harvard.edu/abs/2021ApJS..254...24S}
}

@ARTICLE{Bechtol:2025:Y6GOLD,
       author = {{Bechtol}, K. and {Sevilla-Noarbe}, I. and {Drlica-Wagner}, A. and {Yanny}, B. and {Gruendl}, R.~A. and {Sheldon}, E. and {Rykoff}, E.~S. and {De Vicente}, J. and {Adamow}, M. and {Anbajagane}, D. and {Becker}, M.~R. and {Bernstein}, G.~M. and {Carnero Rosell}, A. and {Gschwend}, J. and {Gorsuch}, M. and {Hartley}, W.~G. and {Jarvis}, M. and {Jeltema}, T. and {Kron}, R. and {Manning}, T.~A. and {O'Donnell}, J. and {Pieres}, A. and {Rodr{\'\i}guez-Monroy}, M. and {Sanchez Cid}, D. and {Tabbutt}, M. and {Toribio San Cipriano}, L. and {Tucker}, D.~L. and {Weaverdyck}, N. and {Yamamoto}, M. and {Abbott}, T.~M.~C. and {Aguena}, M. and {Alarc{\'o}n}, A. and {Allam}, S. and {Amon}, A. and {Andrade-Oliveira}, F. and {Avila}, S. and {Bernardinelli}, P.~H. and {Bertin}, E. and {Blazek}, J. and {Brooks}, D. and {Burke}, D.~L. and {Carretero}, J. and {Castander}, F.~J. and {Cawthon}, R. and {Chang}, C. and {Choi}, A. and {Conselice}, C. and {Costanzi}, M. and {Crocce}, M. and {da Costa}, L.~N. and {Davis}, T.~M. and {Desai}, S. and {Diehl}, H.~T. and {Dodelson}, S. and {Doel}, P. and {Doux}, C. and {Fert{\'e}}, A. and {Flaugher}, B. and {Fosalba}, P. and {Frieman}, J. and {Garc{\'\i}a-Bellido}, J. and {Gatti}, M. and {Gaztanaga}, E. and {Giannini}, G. and {Gruen}, D. and {Gutierrez}, G. and {Herner}, K. and {Hinton}, S.~R. and {Hollowood}, D.~L. and {Honscheid}, K. and {Huterer}, D. and {Jeffrey}, N. and {Krause}, E. and {Kuehn}, K. and {Lahav}, O. and {Lee}, S. and {Lidman}, C. and {Lima}, M. and {Lin}, H. and {Marshall}, J.~L. and {Mena-Fern{\'a}ndez}, J. and {Miquel}, R. and {Mohr}, J.~J. and {Muir}, J. and {Myles}, J. and {Ogando}, R.~L.~C. and {Palmese}, A. and {Plazas Malag{\'o}n}, A.~A. and {Porredon}, A. and {Prat}, J. and {Raveri}, M. and {Romer}, A.~K. and {Roodman}, A. and {Samuroff}, S. and {Sanchez}, E. and {Scarpine}, V. and {Smith}, M. and {Soares-Santos}, M. and {Suchyta}, E. and {Tarle}, G. and {Troxel}, M.~A. and {Vikram}, V. and {Walker}, A.~R. and {Weller}, J. and {Wiseman}, P. and {Zhang}, Y.},
        title = "{Dark Energy Survey Year 6 Results: Photometric Data Set for Cosmology}",
      journal = {arXiv e-prints},
         year = 2025,
        month = jan,
          eid = {arXiv:2501.05739},
        pages = {arXiv:2501.05739},
          doi = {10.48550/arXiv.2501.05739},
archivePrefix = {arXiv},
       eprint = {2501.05739},
 primaryClass = {astro-ph.CO},
       adsurl = {https://ui.adsabs.harvard.edu/abs/2025arXiv250105739B}
}

@ARTICLE{Anbajagane:2024:Shocks,
       author = {{Anbajagane}, D. and {Chang}, C. and {Baxter}, E.~J. and {Charney}, S. and {Lokken}, M. and {Aguena}, M. and {Allam}, S. and {Alves}, O. and {Amon}, A. and {An}, R. and {Andrade-Oliveira}, F. and {Bacon}, D. and {Battaglia}, N. and {Bechtol}, K. and {Becker}, M.~R. and {Benson}, B.~A. and {Bernstein}, G.~M. and {Bleem}, L. and {Bocquet}, S. and {Bond}, J.~R. and {Brooks}, D. and {Carnero Rosell}, A. and {Carrasco Kind}, M. and {Chen}, R. and {Choi}, A. and {Costanzi}, M. and {Crawford}, T.~M. and {Crocce}, M. and {da Costa}, L.~N. and {Pereira}, M.~E.~S. and {Davis}, T.~M. and {De Vicente}, J. and {Desai}, S. and {Devlin}, M.~J. and {Diehl}, H.~T. and {Doel}, P. and {Doux}, C. and {Drlica-Wagner}, A. and {Elvin-Poole}, J. and {Ferrero}, I. and {Fert{\'e}}, A. and {Flaugher}, B. and {Fosalba}, P. and {Friedel}, D. and {Frieman}, J. and {Garc{\'\i}a-Bellido}, J. and {Gatti}, M. and {Giannini}, G. and {Grandis}, S. and {Gruen}, D. and {Gruendl}, R.~A. and {Gutierrez}, G. and {Harrison}, I. and {Hill}, J.~C. and {Hilton}, M. and {Hinton}, S.~R. and {Hollowood}, D.~L. and {Honscheid}, K. and {Jain}, B. and {James}, D.~J. and {Jarvis}, M. and {Kuehn}, K. and {Lin}, M. and {MacCrann}, N. and {Marshall}, J.~L. and {McCullough}, J. and {McMahon}, J.~J. and {Mena-Fern{\'a}ndez}, J. and {Menanteau}, F. and {Miquel}, R. and {Moodley}, K. and {Mroczkowski}, T. and {Myles}, J. and {Naess}, S. and {Navarro-Alsina}, A. and {Ogando}, R.~L.~C. and {Page}, L.~A. and {Palmese}, A. and {Pandey}, S. and {Patridge}, B. and {Pieres}, A. and {Plazas Malag{\'o}n}, A.~A. and {Porredon}, A. and {Prat}, J. and {Reichardt}, C. and {Reil}, K. and {Rodriguez-Monroy}, M. and {Rollins}, R.~P. and {Romer}, A.~K. and {Rykoff}, E.~S. and {Sanchez}, E. and {S{\'a}nchez}, C. and {Sanchez Cid}, D. and {Schaan}, E. and {Schubnell}, M. and {Secco}, L.~F. and {Sevilla-Noarbe}, I. and {Sheldon}, E. and {Shin}, T. and {Sif{\'o}n}, C. and {Smith}, M. and {Staggs}, S.~T. and {Suchyta}, E. and {Swanson}, M.~E.~C. and {Tarle}, G. and {To}, C. and {Troxel}, M.~A. and {Tutusaus}, I. and {Vavagiakis}, E.~M. and {Weaverdyck}, N. and {Weller}, J. and {Wiseman}, P. and {Wollack}, E.~J. and {Yanny}, B.},
        title = "{Cosmological shocks around galaxy clusters: a coherent investigation with DES, SPT, and ACT}",
      journal = {\mnras},
         year = 2024,
        month = jan,
       volume = {527},
       number = {3},
        pages = {9378-9404},
          doi = {10.1093/mnras/stad3726},
archivePrefix = {arXiv},
       eprint = {2310.00059},
 primaryClass = {astro-ph.GA},
       adsurl = {https://ui.adsabs.harvard.edu/abs/2024MNRAS.527.9378A}
}

@ARTICLE{Shin:2019:Splashback,
       author = {{Shin}, T. and {Adhikari}, S. and {Baxter}, E.~J. and {Chang}, C. and {Jain}, B. and {Battaglia}, N. and {Bleem}, L. and {Bocquet}, S. and {DeRose}, J. and {Gruen}, D. and {Hilton}, M. and {Kravtsov}, A. and {McClintock}, T. and {Rozo}, E. and {Rykoff}, E.~S. and {Varga}, T.~N. and {Wechsler}, R.~H. and {Wu}, H. and {Zhang}, Z. and {Aiola}, S. and {Allam}, S. and {Bechtol}, K. and {Benson}, B.~A. and {Bertin}, E. and {Bond}, J.~R. and {Brodwin}, M. and {Brooks}, D. and {Buckley-Geer}, E. and {Burke}, D.~L. and {Carlstrom}, J.~E. and {Carnero Rosell}, A. and {Carrasco Kind}, M. and {Carretero}, J. and {Castander}, F.~J. and {Choi}, S.~K. and {Cunha}, C.~E. and {Crawford}, T.~M. and {da Costa}, L.~N. and {De Vicente}, J. and {Desai}, S. and {Devlin}, M.~J. and {Dietrich}, J.~P. and {Doel}, P. and {Dunkley}, J. and {Eifler}, T.~F. and {Evrard}, A.~E. and {Flaugher}, B. and {Fosalba}, P. and {Gallardo}, P.~A. and {Garc{\'\i}a-Bellido}, J. and {Gaztanaga}, E. and {Gerdes}, D.~W. and {Gralla}, M. and {Gruendl}, R.~A. and {Gschwend}, J. and {Gupta}, N. and {Gutierrez}, G. and {Hartley}, W.~G. and {Hill}, J.~C. and {Ho}, S.~P. and {Hollowood}, D.~L. and {Honscheid}, K. and {Hoyle}, B. and {Huffenberger}, K. and {Hughes}, J.~P. and {James}, D.~J. and {Jeltema}, T. and {Kim}, A.~G. and {Krause}, E. and {Kuehn}, K. and {Lahav}, O. and {Lima}, M. and {Madhavacheril}, M.~S. and {Maia}, M.~A.~G. and {Marshall}, J.~L. and {Maurin}, L. and {McMahon}, J. and {Menanteau}, F. and {Miller}, C.~J. and {Miquel}, R. and {Mohr}, J.~J. and {Naess}, S. and {Nati}, F. and {Newburgh}, L. and {Niemack}, M.~D. and {Ogando}, R.~L.~C. and {Page}, L.~A. and {Partridge}, B. and {Patil}, S. and {Plazas}, A.~A. and {Rapetti}, D. and {Reichardt}, C.~L. and {Romer}, A.~K. and {Sanchez}, E. and {Scarpine}, V. and {Schindler}, R. and {Serrano}, S. and {Smith}, M. and {Smith}, R.~C. and {Soares-Santos}, M. and {Sobreira}, F. and {Staggs}, S.~T. and {Stark}, A. and {Stein}, G. and {Suchyta}, E. and {Swanson}, M.~E.~C. and {Tarle}, G. and {Thomas}, D. and {van Engelen}, A. and {Wollack}, E.~J. and {Xu}, Z.},
        title = "{Measurement of the splashback feature around SZ-selected Galaxy clusters with DES, SPT, and ACT}",
      journal = {\mnras},
         year = 2019,
        month = aug,
       volume = {487},
       number = {2},
        pages = {2900-2918},
          doi = {10.1093/mnras/stz1434},
archivePrefix = {arXiv},
       eprint = {1811.06081},
 primaryClass = {astro-ph.CO},
       adsurl = {https://ui.adsabs.harvard.edu/abs/2019MNRAS.487.2900S}
}

@ARTICLE{Carlstrom2002SZReview,
       author = {{Carlstrom}, John E. and {Holder}, Gilbert P. and {Reese}, Erik D.},
        title = "{Cosmology with the Sunyaev-Zel'dovich Effect}",
      journal = {\araa},
         year = 2002,
        month = jan,
       volume = {40},
        pages = {643-680},
          doi = {10.1146/annurev.astro.40.060401.093803},
archivePrefix = {arXiv},
       eprint = {astro-ph/0208192},
 primaryClass = {astro-ph},
       adsurl = {https://ui.adsabs.harvard.edu/abs/2002ARA&A..40..643C}
}

@ARTICLE{Sunyaev1972SZEffect,
       author = {{Sunyaev}, R.~A. and {Zeldovich}, Ya. B.},
        title = "{The Observations of Relic Radiation as a Test of the Nature of X-Ray Radiation from the Clusters of Galaxies}",
      journal = {Comments on Astrophysics and Space Physics},
         year = 1972,
        month = nov,
       volume = {4},
        pages = {173},
       adsurl = {https://ui.adsabs.harvard.edu/abs/1972CoASP...4..173S}
}

@ARTICLE{Lewis:2019:Getdist,
       author = {{Lewis}, Antony},
        title = "{GetDist: a Python package for analysing Monte Carlo samples}",
      journal = {arXiv e-prints},
         year = 2019,
        month = oct,
          eid = {arXiv:1910.13970},
        pages = {arXiv:1910.13970},
          doi = {10.48550/arXiv.1910.13970},
archivePrefix = {arXiv},
       eprint = {1910.13970},
 primaryClass = {astro-ph.IM},
       adsurl = {https://ui.adsabs.harvard.edu/abs/2019arXiv191013970L}
}

@ARTICLE{Tinker:2008:HMF,
       author = {{Tinker}, Jeremy and {Kravtsov}, Andrey V. and {Klypin}, Anatoly and {Abazajian}, Kevork and {Warren}, Michael and {Yepes}, Gustavo and {Gottl{\"o}ber}, Stefan and {Holz}, Daniel E.},
        title = "{Toward a Halo Mass Function for Precision Cosmology: The Limits of Universality}",
      journal = {\apj},
         year = 2008,
        month = dec,
       volume = {688},
       number = {2},
        pages = {709-728},
          doi = {10.1086/591439},
archivePrefix = {arXiv},
       eprint = {0803.2706},
 primaryClass = {astro-ph},
       adsurl = {https://ui.adsabs.harvard.edu/abs/2008ApJ...688..709T}
}

@ARTICLE{Dawson:2013:BOSS,
       author = {{Dawson}, Kyle S. and {Schlegel}, David J. and {Ahn}, Christopher P. and {Anderson}, Scott F. and {Aubourg}, {\'E}ric and {Bailey}, Stephen and {Barkhouser}, Robert H. and {Bautista}, Julian E. and {Beifiori}, Alessandra and {Berlind}, Andreas A. and {Bhardwaj}, Vaishali and {Bizyaev}, Dmitry and {Blake}, Cullen H. and {Blanton}, Michael R. and {Blomqvist}, Michael and {Bolton}, Adam S. and {Borde}, Arnaud and {Bovy}, Jo and {Brandt}, W.~N. and {Brewington}, Howard and {Brinkmann}, Jon and {Brown}, Peter J. and {Brownstein}, Joel R. and {Bundy}, Kevin and {Busca}, N.~G. and {Carithers}, William and {Carnero}, Aurelio R. and {Carr}, Michael A. and {Chen}, Yanmei and {Comparat}, Johan and {Connolly}, Natalia and {Cope}, Frances and {Croft}, Rupert A.~C. and {Cuesta}, Antonio J. and {da Costa}, Luiz N. and {Davenport}, James R.~A. and {Delubac}, Timoth{\'e}e and {de Putter}, Roland and {Dhital}, Saurav and {Ealet}, Anne and {Ebelke}, Garrett L. and {Eisenstein}, Daniel J. and {Escoffier}, S. and {Fan}, Xiaohui and {Filiz Ak}, N. and {Finley}, Hayley and {Font-Ribera}, Andreu and {G{\'e}nova-Santos}, R. and {Gunn}, James E. and {Guo}, Hong and {Haggard}, Daryl and {Hall}, Patrick B. and {Hamilton}, Jean-Christophe and {Harris}, Ben and {Harris}, David W. and {Ho}, Shirley and {Hogg}, David W. and {Holder}, Diana and {Honscheid}, Klaus and {Huehnerhoff}, Joe and {Jordan}, Beatrice and {Jordan}, Wendell P. and {Kauffmann}, Guinevere and {Kazin}, Eyal A. and {Kirkby}, David and {Klaene}, Mark A. and {Kneib}, Jean-Paul and {Le Goff}, Jean-Marc and {Lee}, Khee-Gan and {Long}, Daniel C. and {Loomis}, Craig P. and {Lundgren}, Britt and {Lupton}, Robert H. and {Maia}, Marcio A.~G. and {Makler}, Martin and {Malanushenko}, Elena and {Malanushenko}, Viktor and {Mandelbaum}, Rachel and {Manera}, Marc and {Maraston}, Claudia and {Margala}, Daniel and {Masters}, Karen L. and {McBride}, Cameron K. and {McDonald}, Patrick and {McGreer}, Ian D. and {McMahon}, Richard G. and {Mena}, Olga and {Miralda-Escud{\'e}}, Jordi and {Montero-Dorta}, Antonio D. and {Montesano}, Francesco and {Muna}, Demitri and {Myers}, Adam D. and {Naugle}, Tracy and {Nichol}, Robert C. and {Noterdaeme}, Pasquier and {Nuza}, Sebasti{\'a}n E. and {Olmstead}, Matthew D. and {Oravetz}, Audrey and {Oravetz}, Daniel J. and {Owen}, Russell and {Padmanabhan}, Nikhil and {Palanque-Delabrouille}, Nathalie and {Pan}, Kaike and {Parejko}, John K. and {P{\^a}ris}, Isabelle and {Percival}, Will J. and {P{\'e}rez-Fournon}, Ismael and {P{\'e}rez-R{\`a}fols}, Ignasi and {Petitjean}, Patrick and {Pfaffenberger}, Robert and {Pforr}, Janine and {Pieri}, Matthew M. and {Prada}, Francisco and {Price-Whelan}, Adrian M. and {Raddick}, M. Jordan and {Rebolo}, Rafael and {Rich}, James and {Richards}, Gordon T. and {Rockosi}, Constance M. and {Roe}, Natalie A. and {Ross}, Ashley J. and {Ross}, Nicholas P. and {Rossi}, Graziano and {Rubi{\~n}o-Martin}, J.~A. and {Samushia}, Lado and {S{\'a}nchez}, Ariel G. and {Sayres}, Conor and {Schmidt}, Sarah J. and {Schneider}, Donald P. and {Sc{\'o}ccola}, C.~G. and {Seo}, Hee-Jong and {Shelden}, Alaina and {Sheldon}, Erin and {Shen}, Yue and {Shu}, Yiping and {Slosar}, An{\v{z}}e and {Smee}, Stephen A. and {Snedden}, Stephanie A. and {Stauffer}, Fritz and {Steele}, Oliver and {Strauss}, Michael A. and {Streblyanska}, Alina and {Suzuki}, Nao and {Swanson}, Molly E.~C. and {Tal}, Tomer and {Tanaka}, Masayuki and {Thomas}, Daniel and {Tinker}, Jeremy L. and {Tojeiro}, Rita and {Tremonti}, Christy A. and {Vargas Maga{\~n}a}, M. and {Verde}, Licia and {Viel}, Matteo and {Wake}, David A. and {Watson}, Mike and {Weaver}, Benjamin A. and {Weinberg}, David H. and {Weiner}, Benjamin J. and {West}, Andrew A. and {White}, Martin and {Wood-Vasey}, W.~M. and {Yeche}, Christophe and {Zehavi}, Idit and {Zhao}, Gong-Bo and {Zheng}, Zheng},
        title = "{The Baryon Oscillation Spectroscopic Survey of SDSS-III}",
      journal = {\aj},
         year = 2013,
        month = jan,
       volume = {145},
       number = {1},
          eid = {10},
        pages = {10},
          doi = {10.1088/0004-6256/145/1/10},
archivePrefix = {arXiv},
       eprint = {1208.0022},
 primaryClass = {astro-ph.CO},
       adsurl = {https://ui.adsabs.harvard.edu/abs/2013AJ....145...10D}
}

@ARTICLE{Dawson:2016:eBOSS,
       author = {{Dawson}, Kyle S. and {Kneib}, Jean-Paul and {Percival}, Will J. and {Alam}, Shadab and {Albareti}, Franco D. and {Anderson}, Scott F. and {Armengaud}, Eric and {Aubourg}, {\'E}ric and {Bailey}, Stephen and {Bautista}, Julian E. and {Berlind}, Andreas A. and {Bershady}, Matthew A. and {Beutler}, Florian and {Bizyaev}, Dmitry and {Blanton}, Michael R. and {Blomqvist}, Michael and {Bolton}, Adam S. and {Bovy}, Jo and {Brandt}, W.~N. and {Brinkmann}, Jon and {Brownstein}, Joel R. and {Burtin}, Etienne and {Busca}, N.~G. and {Cai}, Zheng and {Chuang}, Chia-Hsun and {Clerc}, Nicolas and {Comparat}, Johan and {Cope}, Frances and {Croft}, Rupert A.~C. and {Cruz-Gonzalez}, Irene and {da Costa}, Luiz N. and {Cousinou}, Marie-Claude and {Darling}, Jeremy and {de la Macorra}, Axel and {de la Torre}, Sylvain and {Delubac}, Timoth{\'e}e and {du Mas des Bourboux}, H{\'e}lion and {Dwelly}, Tom and {Ealet}, Anne and {Eisenstein}, Daniel J. and {Eracleous}, Michael and {Escoffier}, S. and {Fan}, Xiaohui and {Finoguenov}, Alexis and {Font-Ribera}, Andreu and {Frinchaboy}, Peter and {Gaulme}, Patrick and {Georgakakis}, Antonis and {Green}, Paul and {Guo}, Hong and {Guy}, Julien and {Ho}, Shirley and {Holder}, Diana and {Huehnerhoff}, Joe and {Hutchinson}, Timothy and {Jing}, Yipeng and {Jullo}, Eric and {Kamble}, Vikrant and {Kinemuchi}, Karen and {Kirkby}, David and {Kitaura}, Francisco-Shu and {Klaene}, Mark A. and {Laher}, Russ R. and {Lang}, Dustin and {Laurent}, Pierre and {Le Goff}, Jean-Marc and {Li}, Cheng and {Liang}, Yu and {Lima}, Marcos and {Lin}, Qiufan and {Lin}, Weipeng and {Lin}, Yen-Ting and {Long}, Daniel C. and {Lundgren}, Britt and {MacDonald}, Nicholas and {Geimba Maia}, Marcio Antonio and {Malanushenko}, Elena and {Malanushenko}, Viktor and {Mariappan}, Vivek and {McBride}, Cameron K. and {McGreer}, Ian D. and {M{\'e}nard}, Brice and {Merloni}, Andrea and {Meza}, Andres and {Montero-Dorta}, Antonio D. and {Muna}, Demitri and {Myers}, Adam D. and {Nandra}, Kirpal and {Naugle}, Tracy and {Newman}, Jeffrey A. and {Noterdaeme}, Pasquier and {Nugent}, Peter and {Ogando}, Ricardo and {Olmstead}, Matthew D. and {Oravetz}, Audrey and {Oravetz}, Daniel J. and {Padmanabhan}, Nikhil and {Palanque-Delabrouille}, Nathalie and {Pan}, Kaike and {Parejko}, John K. and {P{\^a}ris}, Isabelle and {Peacock}, John A. and {Petitjean}, Patrick and {Pieri}, Matthew M. and {Pisani}, Alice and {Prada}, Francisco and {Prakash}, Abhishek and {Raichoor}, Anand and {Reid}, Beth and {Rich}, James and {Ridl}, Jethro and {Rodriguez-Torres}, Sergio and {Carnero Rosell}, Aurelio and {Ross}, Ashley J. and {Rossi}, Graziano and {Ruan}, John and {Salvato}, Mara and {Sayres}, Conor and {Schneider}, Donald P. and {Schlegel}, David J. and {Seljak}, Uros and {Seo}, Hee-Jong and {Sesar}, Branimir and {Shandera}, Sarah and {Shu}, Yiping and {Slosar}, An{\v{z}}e and {Sobreira}, Flavia and {Streblyanska}, Alina and {Suzuki}, Nao and {Taylor}, Donna and {Tao}, Charling and {Tinker}, Jeremy L. and {Tojeiro}, Rita and {Vargas-Maga{\~n}a}, Mariana and {Wang}, Yuting and {Weaver}, Benjamin A. and {Weinberg}, David H. and {White}, Martin and {Wood-Vasey}, W.~M. and {Yeche}, Christophe and {Zhai}, Zhongxu and {Zhao}, Cheng and {Zhao}, Gong-bo and {Zheng}, Zheng and {Ben Zhu}, Guangtun and {Zou}, Hu},
        title = "{The SDSS-IV Extended Baryon Oscillation Spectroscopic Survey: Overview and Early Data}",
      journal = {\aj},
         year = 2016,
        month = feb,
       volume = {151},
       number = {2},
          eid = {44},
        pages = {44},
          doi = {10.3847/0004-6256/151/2/44},
archivePrefix = {arXiv},
       eprint = {1508.04473},
 primaryClass = {astro-ph.CO},
       adsurl = {https://ui.adsabs.harvard.edu/abs/2016AJ....151...44D}
}

@ARTICLE{Gaia:2021:eDR3,
       author = {{Gaia Collaboration} and {Brown}, A.~G.~A. and {Vallenari}, A. and {Prusti}, T. and {de Bruijne}, J.~H.~J. and {Babusiaux}, C. and {Biermann}, M. and {Creevey}, O.~L. and {Evans}, D.~W. and {Eyer}, L. and {Hutton}, A. and {Jansen}, F. and {Jordi}, C. and {Klioner}, S.~A. and {Lammers}, U. and {Lindegren}, L. and {Luri}, X. and {Mignard}, F. and {Panem}, C. and {Pourbaix}, D. and {Randich}, S. and {Sartoretti}, P. and {Soubiran}, C. and {Walton}, N.~A. and {Arenou}, F. and {Bailer-Jones}, C.~A.~L. and {Bastian}, U. and {Cropper}, M. and {Drimmel}, R. and {Katz}, D. and {Lattanzi}, M.~G. and {van Leeuwen}, F. and {Bakker}, J. and {Cacciari}, C. and {Casta{\~n}eda}, J. and {De Angeli}, F. and {Ducourant}, C. and {Fabricius}, C. and {Fouesneau}, M. and {Fr{\'e}mat}, Y. and {Guerra}, R. and {Guerrier}, A. and {Guiraud}, J. and {Jean-Antoine Piccolo}, A. and {Masana}, E. and {Messineo}, R. and {Mowlavi}, N. and {Nicolas}, C. and {Nienartowicz}, K. and {Pailler}, F. and {Panuzzo}, P. and {Riclet}, F. and {Roux}, W. and {Seabroke}, G.~M. and {Sordo}, R. and {Tanga}, P. and {Th{\'e}venin}, F. and {Gracia-Abril}, G. and {Portell}, J. and {Teyssier}, D. and {Altmann}, M. and {Andrae}, R. and {Bellas-Velidis}, I. and {Benson}, K. and {Berthier}, J. and {Blomme}, R. and {Brugaletta}, E. and {Burgess}, P.~W. and {Busso}, G. and {Carry}, B. and {Cellino}, A. and {Cheek}, N. and {Clementini}, G. and {Damerdji}, Y. and {Davidson}, M. and {Delchambre}, L. and {Dell'Oro}, A. and {Fern{\'a}ndez-Hern{\'a}ndez}, J. and {Galluccio}, L. and {Garc{\'\i}a-Lario}, P. and {Garcia-Reinaldos}, M. and {Gonz{\'a}lez-N{\'u}{\~n}ez}, J. and {Gosset}, E. and {Haigron}, R. and {Halbwachs}, J. -L. and {Hambly}, N.~C. and {Harrison}, D.~L. and {Hatzidimitriou}, D. and {Heiter}, U. and {Hern{\'a}ndez}, J. and {Hestroffer}, D. and {Hodgkin}, S.~T. and {Holl}, B. and {Jan{\ss}en}, K. and {Jevardat de Fombelle}, G. and {Jordan}, S. and {Krone-Martins}, A. and {Lanzafame}, A.~C. and {L{\"o}ffler}, W. and {Lorca}, A. and {Manteiga}, M. and {Marchal}, O. and {Marrese}, P.~M. and {Moitinho}, A. and {Mora}, A. and {Muinonen}, K. and {Osborne}, P. and {Pancino}, E. and {Pauwels}, T. and {Petit}, J. -M. and {Recio-Blanco}, A. and {Richards}, P.~J. and {Riello}, M. and {Rimoldini}, L. and {Robin}, A.~C. and {Roegiers}, T. and {Rybizki}, J. and {Sarro}, L.~M. and {Siopis}, C. and {Smith}, M. and {Sozzetti}, A. and {Ulla}, A. and {Utrilla}, E. and {van Leeuwen}, M. and {van Reeven}, W. and {Abbas}, U. and {Abreu Aramburu}, A. and {Accart}, S. and {Aerts}, C. and {Aguado}, J.~J. and {Ajaj}, M. and {Altavilla}, G. and {{\'A}lvarez}, M.~A. and {{\'A}lvarez Cid-Fuentes}, J. and {Alves}, J. and {Anderson}, R.~I. and {Anglada Varela}, E. and {Antoja}, T. and {Audard}, M. and {Baines}, D. and {Baker}, S.~G. and {Balaguer-N{\'u}{\~n}ez}, L. and {Balbinot}, E. and {Balog}, Z. and {Barache}, C. and {Barbato}, D. and {Barros}, M. and {Barstow}, M.~A. and {Bartolom{\'e}}, S. and {Bassilana}, J. -L. and {Bauchet}, N. and {Baudesson-Stella}, A. and {Becciani}, U. and {Bellazzini}, M. and {Bernet}, M. and {Bertone}, S. and {Bianchi}, L. and {Blanco-Cuaresma}, S. and {Boch}, T. and {Bombrun}, A. and {Bossini}, D. and {Bouquillon}, S. and {Bragaglia}, A. and {Bramante}, L. and {Breedt}, E. and {Bressan}, A. and {Brouillet}, N. and {Bucciarelli}, B. and {Burlacu}, A. and {Busonero}, D. and {Butkevich}, A.~G. and {Buzzi}, R. and {Caffau}, E. and {Cancelliere}, R. and {C{\'a}novas}, H. and {Cantat-Gaudin}, T. and {Carballo}, R. and {Carlucci}, T. and {Carnerero}, M.~I. and {Carrasco}, J.~M. and {Casamiquela}, L. and {Castellani}, M. and {Castro-Ginard}, A. and {Castro Sampol}, P. and {Chaoul}, L. and {Charlot}, P. and {Chemin}, L. and {Chiavassa}, A. and {Cioni}, M. -R.~L. and {Comoretto}, G. and {Cooper}, W.~J. and {Cornez}, T. and {Cowell}, S. and {Crifo}, F. and {Crosta}, M. and {Crowley}, C. and {Dafonte}, C. and {Dapergolas}, A. and {David}, M. and {David}, P. and {de Laverny}, P. and {De Luise}, F. and {De March}, R. and {De Ridder}, J. and {de Souza}, R. and {de Teodoro}, P. and {de Torres}, A. and {del Peloso}, E.~F. and {del Pozo}, E. and {Delbo}, M. and {Delgado}, A. and {Delgado}, H.~E. and {Delisle}, J. -B. and {Di Matteo}, P. and {Diakite}, S. and {Diener}, C. and {Distefano}, E. and {Dolding}, C. and {Eappachen}, D. and {Edvardsson}, B. and {Enke}, H. and {Esquej}, P. and {Fabre}, C. and {Fabrizio}, M. and {Faigler}, S. and {Fedorets}, G. and {Fernique}, P. and {Fienga}, A. and {Figueras}, F. and {Fouron}, C. and {Fragkoudi}, F. and {Fraile}, E. and {Franke}, F. and {Gai}, M. and {Garabato}, D. and {Garcia-Gutierrez}, A. and {Garc{\'\i}a-Torres}, M. and {Garofalo}, A. and {Gavras}, P. and {Gerlach}, E. and {Geyer}, R. and {Giacobbe}, P. and {Gilmore}, G. and {Girona}, S. and {Giuffrida}, G. and {Gomel}, R. and {Gomez}, A. and {Gonzalez-Santamaria}, I. and {Gonz{\'a}lez-Vidal}, J.~J. and {Granvik}, M. and {Guti{\'e}rrez-S{\'a}nchez}, R. and {Guy}, L.~P. and {Hauser}, M. and {Haywood}, M. and {Helmi}, A. and {Hidalgo}, S.~L. and {Hilger}, T. and {H{\l}adczuk}, N. and {Hobbs}, D. and {Holland}, G. and {Huckle}, H.~E. and {Jasniewicz}, G. and {Jonker}, P.~G. and {Juaristi Campillo}, J. and {Julbe}, F. and {Karbevska}, L. and {Kervella}, P. and {Khanna}, S. and {Kochoska}, A. and {Kontizas}, M. and {Kordopatis}, G. and {Korn}, A.~J. and {Kostrzewa-Rutkowska}, Z. and {Kruszy{\'n}ska}, K. and {Lambert}, S. and {Lanza}, A.~F. and {Lasne}, Y. and {Le Campion}, J. -F. and {Le Fustec}, Y. and {Lebreton}, Y. and {Lebzelter}, T. and {Leccia}, S. and {Leclerc}, N. and {Lecoeur-Taibi}, I. and {Liao}, S. and {Licata}, E. and {Lindstr{\o}m}, E.~P. and {Lister}, T.~A. and {Livanou}, E. and {Lobel}, A. and {Madrero Pardo}, P. and {Managau}, S. and {Mann}, R.~G. and {Marchant}, J.~M. and {Marconi}, M. and {Marcos Santos}, M.~M.~S. and {Marinoni}, S. and {Marocco}, F. and {Marshall}, D.~J. and {Martin Polo}, L. and {Mart{\'\i}n-Fleitas}, J.~M. and {Masip}, A. and {Massari}, D. and {Mastrobuono-Battisti}, A. and {Mazeh}, T. and {McMillan}, P.~J. and {Messina}, S. and {Michalik}, D. and {Millar}, N.~R. and {Mints}, A. and {Molina}, D. and {Molinaro}, R. and {Moln{\'a}r}, L. and {Montegriffo}, P. and {Mor}, R. and {Morbidelli}, R. and {Morel}, T. and {Morris}, D. and {Mulone}, A.~F. and {Munoz}, D. and {Muraveva}, T. and {Murphy}, C.~P. and {Musella}, I. and {Noval}, L. and {Ord{\'e}novic}, C. and {Orr{\`u}}, G. and {Osinde}, J. and {Pagani}, C. and {Pagano}, I. and {Palaversa}, L. and {Palicio}, P.~A. and {Panahi}, A. and {Pawlak}, M. and {Pe{\~n}alosa Esteller}, X. and {Penttil{\"a}}, A. and {Piersimoni}, A.~M. and {Pineau}, F. -X. and {Plachy}, E. and {Plum}, G. and {Poggio}, E. and {Poretti}, E. and {Poujoulet}, E. and {Pr{\v{s}}a}, A. and {Pulone}, L. and {Racero}, E. and {Ragaini}, S. and {Rainer}, M. and {Raiteri}, C.~M. and {Rambaux}, N. and {Ramos}, P. and {Ramos-Lerate}, M. and {Re Fiorentin}, P. and {Regibo}, S. and {Reyl{\'e}}, C. and {Ripepi}, V. and {Riva}, A. and {Rixon}, G. and {Robichon}, N. and {Robin}, C. and {Roelens}, M. and {Rohrbasser}, L. and {Romero-G{\'o}mez}, M. and {Rowell}, N. and {Royer}, F. and {Rybicki}, K.~A. and {Sadowski}, G. and {Sagrist{\`a} Sell{\'e}s}, A. and {Sahlmann}, J. and {Salgado}, J. and {Salguero}, E. and {Samaras}, N. and {Sanchez Gimenez}, V. and {Sanna}, N. and {Santove{\~n}a}, R. and {Sarasso}, M. and {Schultheis}, M. and {Sciacca}, E. and {Segol}, M. and {Segovia}, J.~C. and {S{\'e}gransan}, D. and {Semeux}, D. and {Shahaf}, S. and {Siddiqui}, H.~I. and {Siebert}, A. and {Siltala}, L. and {Slezak}, E. and {Smart}, R.~L. and {Solano}, E. and {Solitro}, F. and {Souami}, D. and {Souchay}, J. and {Spagna}, A. and {Spoto}, F. and {Steele}, I.~A. and {Steidelm{\"u}ller}, H. and {Stephenson}, C.~A. and {S{\"u}veges}, M. and {Szabados}, L. and {Szegedi-Elek}, E. and {Taris}, F. and {Tauran}, G. and {Taylor}, M.~B. and {Teixeira}, R. and {Thuillot}, W. and {Tonello}, N. and {Torra}, F. and {Torra}, J. and {Turon}, C. and {Unger}, N. and {Vaillant}, M. and {van Dillen}, E. and {Vanel}, O. and {Vecchiato}, A. and {Viala}, Y. and {Vicente}, D. and {Voutsinas}, S. and {Weiler}, M. and {Wevers}, T. and {Wyrzykowski}, {\L}. and {Yoldas}, A. and {Yvard}, P. and {Zhao}, H. and {Zorec}, J. and {Zucker}, S. and {Zurbach}, C. and {Zwitter}, T.},
        title = "{Gaia Early Data Release 3. Summary of the contents and survey properties}",
      journal = {\aap},
         year = 2021,
        month = may,
       volume = {649},
          eid = {A1},
        pages = {A1},
          doi = {10.1051/0004-6361/202039657},
archivePrefix = {arXiv},
       eprint = {2012.01533},
 primaryClass = {astro-ph.GA},
       adsurl = {https://ui.adsabs.harvard.edu/abs/2021A&A...649A...1G}
}

@ARTICLE{Skrutskie:2006:2Mass,
       author = {{Skrutskie}, M.~F. and {Cutri}, R.~M. and {Stiening}, R. and {Weinberg}, M.~D. and {Schneider}, S. and {Carpenter}, J.~M. and {Beichman}, C. and {Capps}, R. and {Chester}, T. and {Elias}, J. and {Huchra}, J. and {Liebert}, J. and {Lonsdale}, C. and {Monet}, D.~G. and {Price}, S. and {Seitzer}, P. and {Jarrett}, T. and {Kirkpatrick}, J.~D. and {Gizis}, J.~E. and {Howard}, E. and {Evans}, T. and {Fowler}, J. and {Fullmer}, L. and {Hurt}, R. and {Light}, R. and {Kopan}, E.~L. and {Marsh}, K.~A. and {McCallon}, H.~L. and {Tam}, R. and {Van Dyk}, S. and {Wheelock}, S.},
        title = "{The Two Micron All Sky Survey (2MASS)}",
      journal = {\aj},
         year = 2006,
        month = feb,
       volume = {131},
       number = {2},
        pages = {1163-1183},
          doi = {10.1086/498708},
       adsurl = {https://ui.adsabs.harvard.edu/abs/2006AJ....131.1163S}
}

@ARTICLE{DES:2025:6x2pt_Y6,
       author = {{DES Collaboration} and {Abbott}, T.~M.~C. and {Aguena}, M. and {Alarcon}, A. and {Amon}, A. and {Anbajagane}, D. and {Andrade-Oliveira}, F. and {Avila}, S. and {Allam}, S.~S. and {Bacon}, D. and {Becker}, M.~R. and {Bhargava}, S. and {Blazek}, J. and {Bocquet}, S. and {Brooks}, D. and {Rosell}, A. Carnero and {Carretero}, J. and {Castander}, F.~J. and {Chang}, C. and {Choi}, A. and {Conselice}, C. and {Costanzi}, M. and {Crocce}, M. and {da Costa}, L.~N. and {Pereira}, M.~E.~S. and {Doux}, C. and {Davis}, T.~M. and {Desai}, S. and {Diehl}, H.~T. and {Dodelson}, S. and {Doel}, P. and {Elvin-Poole}, J. and {Esteves}, J. and {Everett}, S. and {Farahi}, A. and {Fert{\'e}}, A. and {Flaugher}, B. and {Garc{\'\i}a-Bellido}, J. and {Gatti}, M. and {Giannini}, G. and {Giles}, P. and {Grandis}, S. and {Gruen}, D. and {Gruendl}, R.~A. and {Gutierrez}, G. and {Harrison}, I. and {Hinton}, S.~R. and {Hollowood}, D.~L. and {Honscheid}, K. and {Jeffrey}, N. and {Jeltema}, T. and {Krause}, E. and {Lahav}, O. and {Lee}, S. and {Lidman}, C. and {Lima}, M. and {Lin}, H. and {Mohr}, J.~J. and {Marshall}, J.~L. and {McCullough}, J. and {Mena-Fern'andez}, J. and {Miquel}, R. and {Muir}, J. and {Myles}, J. and {Ogando}, R.~L.~C. and {Palmese}, A. and {Paterno}, M. and {Malag{\'o}n}, A.~A. Plazas and {Porredon}, A. and {Prat}, J. and {Romer}, A.~K. and {Roodman}, A. and {Rozo}, E. and {Rykoff}, E.~S. and {Rosenfeld}, R. and {Sanchez}, E. and {Cid}, D. Sanchez and {Sevilla-Noarbe}, I. and {Smith}, M. and {Suchyta}, E. and {Shin}, Tae-hyeon and {Tarle}, G. and {Thomas}, D. and {To}, Chun-Hao and {Troxel}, M.~A. and {Vikram}, V. and {Walker}, A.~R. and {Weinberg}, David H. and {Weaverdyck}, N. and {Wechsler}, R.~H. and {Weller}, J. and {Wu}, H.-Y. and {Yamamoto}, M. and {Yanny}, B. and {Zhang}, Y. and {Zhou}, C.},
        title = "{Dark energy survey year 3 results: Cosmological constraints from cluster abundances, weak lensing, and galaxy clustering}",
      journal = {\prd},
         year = 2025,
        month = oct,
       volume = {112},
       number = {8},
          eid = {083535},
        pages = {083535},
          doi = {10.1103/3dzh-d8f5},
archivePrefix = {arXiv},
       eprint = {2503.13632},
 primaryClass = {astro-ph.CO},
       adsurl = {https://ui.adsabs.harvard.edu/abs/2025PhRvD.112h3535A}
}

@ARTICLE{Costanzi:2021:DESxSPT,
       author = {{Costanzi}, M. and {Saro}, A. and {Bocquet}, S. and {Abbott}, T.~M.~C. and {Aguena}, M. and {Allam}, S. and {Amara}, A. and {Annis}, J. and {Avila}, S. and {Bacon}, D. and {Benson}, B.~A. and {Bhargava}, S. and {Brooks}, D. and {Buckley-Geer}, E. and {Burke}, D.~L. and {Carnero Rosell}, A. and {Carrasco Kind}, M. and {Carretero}, J. and {Choi}, A. and {da Costa}, L.~N. and {Pereira}, M.~E.~S. and {De Vicente}, J. and {Desai}, S. and {Diehl}, H.~T. and {Dietrich}, J.~P. and {Doel}, P. and {Eifler}, T.~F. and {Everett}, S. and {Ferrero}, I. and {Fert{\'e}}, A. and {Flaugher}, B. and {Fosalba}, P. and {Frieman}, J. and {Garc{\'\i}a-Bellido}, J. and {Gaztanaga}, E. and {Gerdes}, D.~W. and {Giannantonio}, T. and {Giles}, P. and {Grandis}, S. and {Gruen}, D. and {Gruendl}, R.~A. and {Gupta}, N. and {Gutierrez}, G. and {Hartley}, W.~G. and {Hinton}, S.~R. and {Hollowood}, D.~L. and {Honscheid}, K. and {James}, D.~J. and {Jeltema}, T. and {Krause}, E. and {Kuehn}, K. and {Kuropatkin}, N. and {Lahav}, O. and {Lima}, M. and {MacCrann}, N. and {Maia}, M.~A.~G. and {Marshall}, J.~L. and {Menanteau}, F. and {Miquel}, R. and {Mohr}, J.~J. and {Morgan}, R. and {Myles}, J. and {Ogando}, R.~L.~C. and {Palmese}, A. and {Paz-Chinch{\'o}n}, F. and {Plazas}, A.~A. and {Rapetti}, D. and {Reichardt}, C.~L. and {Romer}, A.~K. and {Roodman}, A. and {Ruppin}, F. and {Salvati}, L. and {Samuroff}, S. and {Sanchez}, E. and {Scarpine}, V. and {Serrano}, S. and {Sevilla-Noarbe}, I. and {Singh}, P. and {Smith}, M. and {Soares-Santos}, M. and {Stark}, A.~A. and {Suchyta}, E. and {Swanson}, M.~E.~C. and {Tarle}, G. and {Thomas}, D. and {To}, C. and {Tucker}, D.~L. and {Varga}, T.~N. and {Wechsler}, R.~H. and {Zhang}, Z. and {DES} and {SPT Collaborations}},
        title = "{Cosmological constraints from DES Y1 cluster abundances and SPT multiwavelength data}",
      journal = {\prd},
         year = 2021,
        month = feb,
       volume = {103},
       number = {4},
          eid = {043522},
        pages = {043522},
          doi = {10.1103/PhysRevD.103.043522},
archivePrefix = {arXiv},
       eprint = {2010.13800},
 primaryClass = {astro-ph.CO},
       adsurl = {https://ui.adsabs.harvard.edu/abs/2021PhRvD.103d3522C}
}

@ARTICLE{Hilton:2026:ACTDR6,
       author = {{Aguena}, M. and {Aiola}, S. and {Allam}, S. and {Andrade-Oliveira}, F. and {Bacon}, D. and {Bahcall}, N. and {Battaglia}, N. and {Battistelli}, E.~S. and {Bocquet}, S. and {Bolliet}, B. and {Bond}, J.~R. and {Brooks}, D. and {Calabrese}, E. and {Carretero}, J. and {Choi}, S.~K. and {da Costa}, L.~N. and {Costanzi}, M. and {Coulton}, W. and {Davis}, T.~M. and {Desai}, S. and {Devlin}, M.~J. and {Dicker}, S. and {Doel}, P. and {Duivenvoorden}, A.~J. and {Dunkley}, J. and {Ferraro}, S. and {Flaugher}, B. and {Frieman}, J. and {Gallardo}, P.~A. and {Gatti}, M. and {Gaztanaga}, E. and {Gill}, A.~S. and {Golec}, J.~E. and {Gruen}, D. and {Gruendl}, R.~A. and {Halpern}, M. and {Hasselfield}, M. and {Hill}, J.~C. and {Hilton}, M. and {Hincks}, A.~D. and {Hinton}, S.~R. and {Hollowood}, D.~L. and {Honscheid}, K. and {Hubmayr}, J. and {Huffenberger}, K.~M. and {Hughes}, J.~P. and {James}, D.~J. and {Klein}, M. and {Knowles}, K. and {Koopman}, B.~J. and {Kosowsky}, A. and {Lahav}, O. and {Lee}, E. and {Lin}, Y. and {Lokken}, M. and {Madhavacheril}, M.~S. and {Malag{\'o}n}, A.~A. Plazas and {Marrewijk}, J. v. and {Marshall}, J.~L. and {McMahon}, J. and {Mena-Fern{\'a}ndez}, J. and {Miquel}, R. and {Miyatake}, H. and {Mohr}, J.~J. and {Moodley}, K. and {Mroczkowski}, T. and {Naess}, S. and {Nati}, F. and {Nicola}, A. and {Niemack}, M.~D. and {Ogando}, R.~L.~C. and {Oguri}, M. and {Orlowski-Scherer}, J. and {Page}, L.~A. and {Partridge}, B. and {da Silva Pereira}, M.~E. and {Porredon}, A. and {Qu}, F.~J. and {Ragavan}, D.~C. and {Guachalla}, B. Ried and {Romer}, A.~K. and {Rosell}, A. Carnero and {Rykoff}, E.~S. and {Samuroff}, S. and {Sanchez}, E. and {Sevilla-Noarbe}, I. and {Sierra}, C. and {Sif{\'o}n}, C. and {Smith}, M. and {Staggs}, S.~T. and {Suchyta}, E. and {Swanson}, M.~E.~C. and {Tucker}, D.~L. and {Vargas}, C. and {Vavagiakis}, E.~M. and {De Vicente}, J. and {Weaverdyck}, N. and {Weller}, J. and {Wollack}, E.~J. and {Zubeldia}, I.},
        title = "{The Atacama Cosmology Telescope: DR6 Sunyaev-Zel'dovich Selected Galaxy Clusters Catalog}",
      journal = {The Open Journal of Astrophysics},
         year = 2026,
        month = jan,
       volume = {9},
        pages = {55863},
          doi = {10.33232/001c.155863},
archivePrefix = {arXiv},
       eprint = {2507.21459},
 primaryClass = {astro-ph.CO},
       adsurl = {https://ui.adsabs.harvard.edu/abs/2026OJAp....955863A}
}

@ARTICLE{Bleem:2026:Clusters,
       author = {{Bleem}, L.~E. and {Klein}, M. and {Kornoelje}, K. and {Bocquet}, S. and {Sobrin}, J.~A. and {Aguena}, M. and {Anderes}, E. and {Anderson}, A.~J. and {Andrade-Oliveira}, F. and {Ansarinejad}, B. and {Archipley}, M. and {Balkenhol}, L. and {Barron}, D.~R. and {Barry}, P.~S. and {Bayliss}, M. and {Benabed}, K. and {Bender}, A.~N. and {Benson}, B.~A. and {Bianchini}, F. and {Bouchet}, F.~R. and {Brooks}, D. and {Burke}, D.~L. and {Calzadilla}, M. and {Camilleri}, R. and {Camphuis}, E. and {Campitiello}, M.~G. and {Carlstrom}, J.~E. and {Carnero Rosell}, A. and {Carretero}, J. and {Carron}, J. and {Chang}, C.~L. and {Chichura}, P.~M. and {Chokshi}, A. and {Chou}, T.-L. and {Coerver}, A. and {Costanzi}, M. and {Crawford}, T.~M. and {da Costa}, L.~N. and {Daley}, C. and {Davis}, T.~M. and {de Haan}, T. and {De Vicente}, J. and {Desai}, S. and {Dibert}, K.~R. and {Diehl}, H.~T. and {Dobbs}, M.~A. and {Doohan}, M. and {Dutcher}, D. and {Everett}, S. and {Evrard}, G. and {Feng}, C. and {Ferguson}, K.~R. and {Ferree}, N.~C. and {Fichman}, K. and {Flaugher}, B. and {Floyd}, B. and {Fosdick}, K. and {Foster}, A. and {Galli}, S. and {Gambrel}, A.~E. and {Gao}, A.~K. and {Garc{\'\i}a-Bellido}, J. and {Ge}, F. and {Gladders}, M.~D. and {Grandis}, S. and {Guidi}, F. and {Guns}, S. and {Gutierrez}, G. and {Halverson}, N.~W. and {Hinton}, S.~R. and {Hivon}, E. and {Holder}, G.~P. and {Hollowood}, D.~L. and {Holzapfel}, W.~L. and {Hood}, J.~C. and {Hryciuk}, A. and {Huang}, N. and {Jhaveri}, T. and {Kent}, S. and {K{\'e}ruzor{\'e}}, F. and {Khalife}, A.~R. and {Khullar}, G. and {Knox}, L. and {Kuehn}, K. and {Kuo}, C.-L. and {Lahav}, O. and {Levy}, K. and {Li}, Y. and {Lowitz}, A.~E. and {Lu}, C. and {Lynch}, G.~P. and {Maccarone}, T.~J. and {Mahler}, G. and {Maniyar}, A.~S. and {Marshall}, J.~L. and {Martsen}, E.~S. and {McDonald}, M. and {Mena-Fern{\'a}ndez}, J. and {Menanteau}, F. and {Millea}, M. and {Miquel}, R. and {Mohr}, J.~J. and {Montgomery}, J. and {Myles}, J. and {Nakato}, Y. and {Natoli}, T. and {Nichol}, R.~C. and {Novosad}, V. and {Ogando}, R.~L.~C. and {Omori}, Y. and {Ouellette}, A. and {Pan}, Z. and {Phadke}, K.~A. and {Plazas Malag{\'o}n}, A.~A. and {Pollak}, A.~W. and {Porredon}, A. and {Prabhu}, K. and {Prat}, J. and {Quan}, W. and {Raghunathan}, S. and {Rahimi}, M. and {Rahlin}, A. and {Reichardt}, C.~L. and {Romer}, A.~K. and {Rouble}, M. and {Ruhl}, J.~E. and {Sanchez}, E. and {Sanchez Cid}, D. and {Schrabback}, T. and {Sevilla-Noarbe}, I. and {Silva Oliveira}, A.~C. and {Simpson}, A. and {Somboonpanyakul}, T. and {Stark}, A.~A. and {Suchyta}, E. and {Swanson}, M.~E.~C. and {Tandoi}, C. and {To}, C. and {Trendafilova}, C. and {Vieira}, J.~D. and {Vieregg}, A.~G. and {Vikram}, V. and {Vitrier}, A. and {Wan}, Y. and {Weaverdyck}, N. and {Weller}, J. and {Whitehorn}, N. and {Wu}, W.~L.~K. and {Young}, M.~R. and {Zebrowski}, J.~A. and {Zhan}, J.},
        title = "{Galaxy Clusters Selected via the Sunyaev-Zel'dovich Effect in 5 year data from the SPT-3G Main Survey}",
      journal = {arXiv e-prints},
         year = 2026,
        month = jul,
          eid = {arXiv:2607.01175},
        pages = {arXiv:2607.01175},
          doi = {10.48550/arXiv.2607.01175},
archivePrefix = {arXiv},
       eprint = {2607.01175},
 primaryClass = {astro-ph.CO},
       adsurl = {https://ui.adsabs.harvard.edu/abs/2026arXiv260701175B}
}

@ARTICLE{Bleem:2015:SPT,
       author = {{Bleem}, L.~E. and {Stalder}, B. and {de Haan}, T. and {Aird}, K.~A. and {Allen}, S.~W. and {Applegate}, D.~E. and {Ashby}, M.~L.~N. and {Bautz}, M. and {Bayliss}, M. and {Benson}, B.~A. and {Bocquet}, S. and {Brodwin}, M. and {Carlstrom}, J.~E. and {Chang}, C.~L. and {Chiu}, I. and {Cho}, H.~M. and {Clocchiatti}, A. and {Crawford}, T.~M. and {Crites}, A.~T. and {Desai}, S. and {Dietrich}, J.~P. and {Dobbs}, M.~A. and {Foley}, R.~J. and {Forman}, W.~R. and {George}, E.~M. and {Gladders}, M.~D. and {Gonzalez}, A.~H. and {Halverson}, N.~W. and {Hennig}, C. and {Hoekstra}, H. and {Holder}, G.~P. and {Holzapfel}, W.~L. and {Hrubes}, J.~D. and {Jones}, C. and {Keisler}, R. and {Knox}, L. and {Lee}, A.~T. and {Leitch}, E.~M. and {Liu}, J. and {Lueker}, M. and {Luong-Van}, D. and {Mantz}, A. and {Marrone}, D.~P. and {McDonald}, M. and {McMahon}, J.~J. and {Meyer}, S.~S. and {Mocanu}, L. and {Mohr}, J.~J. and {Murray}, S.~S. and {Padin}, S. and {Pryke}, C. and {Reichardt}, C.~L. and {Rest}, A. and {Ruel}, J. and {Ruhl}, J.~E. and {Saliwanchik}, B.~R. and {Saro}, A. and {Sayre}, J.~T. and {Schaffer}, K.~K. and {Schrabback}, T. and {Shirokoff}, E. and {Song}, J. and {Spieler}, H.~G. and {Stanford}, S.~A. and {Staniszewski}, Z. and {Stark}, A.~A. and {Story}, K.~T. and {Stubbs}, C.~W. and {Vanderlinde}, K. and {Vieira}, J.~D. and {Vikhlinin}, A. and {Williamson}, R. and {Zahn}, O. and {Zenteno}, A.},
        title = "{Galaxy Clusters Discovered via the Sunyaev-Zel'dovich Effect in the 2500-Square-Degree SPT-SZ Survey}",
      journal = {\apjs},
         year = 2015,
        month = feb,
       volume = {216},
       number = {2},
          eid = {27},
        pages = {27},
          doi = {10.1088/0067-0049/216/2/27},
archivePrefix = {arXiv},
       eprint = {1409.0850},
 primaryClass = {astro-ph.CO},
       adsurl = {https://ui.adsabs.harvard.edu/abs/2015ApJS..216...27B}
}

@ARTICLE{Bleem:2020:ECS,
       author = {{Bleem}, L.~E. and {Bocquet}, S. and {Stalder}, B. and {Gladders}, M.~D. and {Ade}, P.~A.~R. and {Allen}, S.~W. and {Anderson}, A.~J. and {Annis}, J. and {Ashby}, M.~L.~N. and {Austermann}, J.~E. and {Avila}, S. and {Avva}, J.~S. and {Bayliss}, M. and {Beall}, J.~A. and {Bechtol}, K. and {Bender}, A.~N. and {Benson}, B.~A. and {Bertin}, E. and {Bianchini}, F. and {Blake}, C. and {Brodwin}, M. and {Brooks}, D. and {Buckley-Geer}, E. and {Burke}, D.~L. and {Carlstrom}, J.~E. and {Rosell}, A. Carnero and {Carrasco Kind}, M. and {Carretero}, J. and {Chang}, C.~L. and {Chiang}, H.~C. and {Citron}, R. and {Moran}, C. Corbett and {Costanzi}, M. and {Crawford}, T.~M. and {Crites}, A.~T. and {da Costa}, L.~N. and {de Haan}, T. and {De Vicente}, J. and {Desai}, S. and {Diehl}, H.~T. and {Dietrich}, J.~P. and {Dobbs}, M.~A. and {Eifler}, T.~F. and {Everett}, W. and {Flaugher}, B. and {Floyd}, B. and {Frieman}, J. and {Gallicchio}, J. and {Garc{\'\i}a-Bellido}, J. and {George}, E.~M. and {Gerdes}, D.~W. and {Gilbert}, A. and {Gruen}, D. and {Gruendl}, R.~A. and {Gschwend}, J. and {Gupta}, N. and {Gutierrez}, G. and {Halverson}, N.~W. and {Harrington}, N. and {Henning}, J.~W. and {Heymans}, C. and {Holder}, G.~P. and {Hollowood}, D.~L. and {Holzapfel}, W.~L. and {Honscheid}, K. and {Hrubes}, J.~D. and {Huang}, N. and {Hubmayr}, J. and {Irwin}, K.~D. and {James}, D.~J. and {Jeltema}, T. and {Joudaki}, S. and {Khullar}, G. and {Klein}, M. and {Knox}, L. and {Kuropatkin}, N. and {Lee}, A.~T. and {Li}, D. and {Lidman}, C. and {Lowitz}, A. and {MacCrann}, N. and {Mahler}, G. and {Maia}, M.~A.~G. and {Marshall}, J.~L. and {McDonald}, M. and {McMahon}, J.~J. and {Melchior}, P. and {Menanteau}, F. and {Meyer}, S.~S. and {Miquel}, R. and {Mocanu}, L.~M. and {Mohr}, J.~J. and {Montgomery}, J. and {Nadolski}, A. and {Natoli}, T. and {Nibarger}, J.~P. and {Noble}, G. and {Novosad}, V. and {Padin}, S. and {Palmese}, A. and {Parkinson}, D. and {Patil}, S. and {Paz-Chinch{\'o}n}, F. and {Plazas}, A.~A. and {Pryke}, C. and {Ramachandra}, N.~S. and {Reichardt}, C.~L. and {Remolina Gonz{\'a}lez}, J.~D. and {Romer}, A.~K. and {Roodman}, A. and {Ruhl}, J.~E. and {Rykoff}, E.~S. and {Saliwanchik}, B.~R. and {Sanchez}, E. and {Saro}, A. and {Sayre}, J.~T. and {Schaffer}, K.~K. and {Schrabback}, T. and {Serrano}, S. and {Sharon}, K. and {Sievers}, C. and {Smecher}, G. and {Smith}, M. and {Soares-Santos}, M. and {Stark}, A.~A. and {Story}, K.~T. and {Suchyta}, E. and {Tarle}, G. and {Tucker}, C. and {Vanderlinde}, K. and {Veach}, T. and {Vieira}, J.~D. and {Wang}, G. and {Weller}, J. and {Whitehorn}, N. and {Wu}, W.~L.~K. and {Yefremenko}, V. and {Zhang}, Y.},
        title = "{The SPTpol Extended Cluster Survey}",
      journal = {\apjs},
         year = 2020,
        month = mar,
       volume = {247},
       number = {1},
          eid = {25},
        pages = {25},
          doi = {10.3847/1538-4365/ab6993},
archivePrefix = {arXiv},
       eprint = {1910.04121},
 primaryClass = {astro-ph.CO},
       adsurl = {https://ui.adsabs.harvard.edu/abs/2020ApJS..247...25B}
}

@ARTICLE{Gschwend:2018:LineA,
       author = {{Gschwend}, J. and {Rossel}, A.~C. and {Ogando}, R.~L.~C. and {Neto}, A.~F. and {Maia}, M.~A.~G. and {da Costa}, L.~N. and {Lima}, M. and {Pellegrini}, P. and {Campisano}, R. and {Singulani}, C. and {Adean}, C. and {Benoist}, C. and {Aguena}, M. and {Carrasco Kind}, M. and {Davis}, T.~M. and {de Vicente}, J. and {Hartley}, W.~G. and {Hoyle}, B. and {Palmese}, A. and {Sadeh}, I. and {Abbott}, T.~M.~C. and {Abdalla}, F.~B. and {Allam}, S. and {Annis}, J. and {Asorey}, J. and {Brooks}, D. and {Calcino}, J. and {Carollo}, D. and {Castander}, F.~J. and {D'Andrea}, C.~B. and {Desai}, S. and {Evrard}, A.~E. and {Fosalba}, P. and {Frieman}, J. and {Garc{\'\i}a-Bellido}, J. and {Glazebrook}, K. and {Gerdes}, D.~W. and {Gruendl}, R.~A. and {Gutierrez}, G. and {Hinton}, S. and {Hollowood}, D.~L. and {Honscheid}, K. and {Hoormann}, J.~K. and {James}, D.~J. and {Kuehn}, K. and {Kuropatkin}, N. and {Lahav}, O. and {Lewis}, G. and {Lidman}, C. and {Lin}, H. and {Macaulay}, E. and {Marshall}, J. and {Melchior}, P. and {Miquel}, R. and {M{\"o}ller}, A. and {Plazas}, A.~A. and {Sanchez}, E. and {Santiago}, B. and {Scarpine}, V. and {Schindler}, R.~H. and {Sevilla-Noarbe}, I. and {Smith}, M. and {Sobreira}, F. and {Sommer}, N.~E. and {Suchyta}, E. and {Swanson}, M.~E.~C. and {Tarle}, G. and {Tucker}, B.~E. and {Tucker}, D.~L. and {Uddin}, S. and {Walker}, A.~R.},
        title = "{DES science portal: Computing photometric redshifts}",
      journal = {Astronomy and Computing},
         year = 2018,
        month = oct,
       volume = {25},
        pages = {58-80},
          doi = {10.1016/j.ascom.2018.08.008},
archivePrefix = {arXiv},
       eprint = {1708.05643},
 primaryClass = {astro-ph.GA},
       adsurl = {https://ui.adsabs.harvard.edu/abs/2018A&C....25...58G}
}

@ARTICLE{Rykoff:2012:OpticalRichness,
       author = {{Rykoff}, E.~S. and {Koester}, B.~P. and {Rozo}, E. and
                 {Annis}, J. and {Evrard}, A.~E. and {Hansen}, S.~M. and
                 {Hao}, J. and {Johnston}, D.~E. and {McKay}, T.~A. and
                 {Wechsler}, R.~H.},
        title = "{Robust Optical Richness Estimation with Reduced Scatter}",
      journal = {\apj},
         year = 2012,
        month = feb,
       volume = {746},
       number = {2},
          eid = {178},
        pages = {178},
          doi = {10.1088/0004-637X/746/2/178},
archivePrefix = {arXiv},
       eprint = {1104.2089},
 primaryClass = {astro-ph.CO},
       adsurl = {https://ui.adsabs.harvard.edu/abs/2012ApJ...746..178R}
}

@ARTICLE{Rykoff:2014:AlgorithmCatalog,
       author = {{Rykoff}, E.~S. and {Rozo}, E. and {Busha}, M.~T. and
                 {Cunha}, C.~E. and {Finoguenov}, A. and {Evrard}, A. and
                 {Hao}, J. and {Koester}, B.~P. and {Leauthaud}, A. and
                 {Nord}, B. and {Pierre}, M. and {Reddick}, R. and
                 {Sadibekova}, T. and {Sheldon}, E.~S. and {Wechsler}, R.~H.},
        title = "{redMaPPer. I. Algorithm and SDSS DR8 Catalog}",
      journal = {\apj},
         year = 2014,
        month = apr,
       volume = {785},
       number = {2},
          eid = {104},
        pages = {104},
          doi = {10.1088/0004-637X/785/2/104},
archivePrefix = {arXiv},
       eprint = {1303.3562},
 primaryClass = {astro-ph.CO},
       adsurl = {https://ui.adsabs.harvard.edu/abs/2014ApJ...785..104R}
}

@ARTICLE{Bulbul:2024:eROSITA,
       author = {{Bulbul}, E. and {Liu}, A. and {Kluge}, M. and {Zhang}, X. and {Sanders}, J.~S. and {Bahar}, Y.~E. and {Ghirardini}, V. and {Artis}, E. and {Seppi}, R. and {Garrel}, C. and {Ramos-Ceja}, M.~E. and {Comparat}, J. and {Balzer}, F. and {B{\"o}ckmann}, K. and {Br{\"u}ggen}, M. and {Clerc}, N. and {Dennerl}, K. and {Dolag}, K. and {Freyberg}, M. and {Grandis}, S. and {Gruen}, D. and {Kleinebreil}, F. and {Krippendorf}, S. and {Lamer}, G. and {Merloni}, A. and {Migkas}, K. and {Nandra}, K. and {Pacaud}, F. and {Predehl}, P. and {Reiprich}, T.~H. and {Schrabback}, T. and {Veronica}, A. and {Weller}, J. and {Zelmer}, S.},
        title = "{The SRG/eROSITA All-Sky Survey. The first catalog of galaxy clusters and groups in the Western Galactic Hemisphere}",
      journal = {\aap},
         year = 2024,
        month = may,
       volume = {685},
          eid = {A106},
        pages = {A106},
          doi = {10.1051/0004-6361/202348264},
archivePrefix = {arXiv},
       eprint = {2402.08452},
 primaryClass = {astro-ph.CO},
       adsurl = {https://ui.adsabs.harvard.edu/abs/2024A&A...685A.106B}
}

@ARTICLE{Schlegel:1998:Dust,
       author = {{Schlegel}, David J. and {Finkbeiner}, Douglas P. and {Davis}, Marc},
        title = "{Maps of Dust Infrared Emission for Use in Estimation of Reddening and Cosmic Microwave Background Radiation Foregrounds}",
      journal = {\apj},
         year = 1998,
        month = jun,
       volume = {500},
       number = {2},
        pages = {525-553},
          doi = {10.1086/305772},
archivePrefix = {arXiv},
       eprint = {astro-ph/9710327},
 primaryClass = {astro-ph},
       adsurl = {https://ui.adsabs.harvard.edu/abs/1998ApJ...500..525S}
}

@ARTICLE{DESDR2,
       author = {{Abbott}, T.~M.~C. and {Adam{\'o}w}, M. and {Aguena}, M. and {Allam}, S. and {Amon}, A. and {Annis}, J. and {Avila}, S. and {Bacon}, D. and {Banerji}, M. and {Bechtol}, K. and {Becker}, M.~R. and {Bernstein}, G.~M. and {Bertin}, E. and {Bhargava}, S. and {Bridle}, S.~L. and {Brooks}, D. and {Burke}, D.~L. and {Carnero Rosell}, A. and {Carrasco Kind}, M. and {Carretero}, J. and {Castander}, F.~J. and {Cawthon}, R. and {Chang}, C. and {Choi}, A. and {Conselice}, C. and {Costanzi}, M. and {Crocce}, M. and {da Costa}, L.~N. and {Davis}, T.~M. and {De Vicente}, J. and {DeRose}, J. and {Desai}, S. and {Diehl}, H.~T. and {Dietrich}, J.~P. and {Drlica-Wagner}, A. and {Eckert}, K. and {Elvin-Poole}, J. and {Everett}, S. and {Evrard}, A.~E. and {Ferrero}, I. and {Fert{\'e}}, A. and {Flaugher}, B. and {Fosalba}, P. and {Friedel}, D. and {Frieman}, J. and {Garc{\'\i}a-Bellido}, J. and {Gaztanaga}, E. and {Gelman}, L. and {Gerdes}, D.~W. and {Giannantonio}, T. and {Gill}, M.~S.~S. and {Gruen}, D. and {Gruendl}, R.~A. and {Gschwend}, J. and {Gutierrez}, G. and {Hartley}, W.~G. and {Hinton}, S.~R. and {Hollowood}, D.~L. and {Honscheid}, K. and {Huterer}, D. and {James}, D.~J. and {Jeltema}, T. and {Johnson}, M.~D. and {Kent}, S. and {Kron}, R. and {Kuehn}, K. and {Kuropatkin}, N. and {Lahav}, O. and {Li}, T.~S. and {Lidman}, C. and {Lin}, H. and {MacCrann}, N. and {Maia}, M.~A.~G. and {Manning}, T.~A. and {Maloney}, J.~D. and {March}, M. and {Marshall}, J.~L. and {Martini}, P. and {Melchior}, P. and {Menanteau}, F. and {Miquel}, R. and {Morgan}, R. and {Myles}, J. and {Neilsen}, E. and {Ogando}, R.~L.~C. and {Palmese}, A. and {Paz-Chinch{\'o}n}, F. and {Petravick}, D. and {Pieres}, A. and {Plazas}, A.~A. and {Pond}, C. and {Rodriguez-Monroy}, M. and {Romer}, A.~K. and {Roodman}, A. and {Rykoff}, E.~S. and {Sako}, M. and {Sanchez}, E. and {Santiago}, B. and {Scarpine}, V. and {Serrano}, S. and {Sevilla-Noarbe}, I. and {Smith}, J. Allyn and {Smith}, M. and {Soares-Santos}, M. and {Suchyta}, E. and {Swanson}, M.~E.~C. and {Tarle}, G. and {Thomas}, D. and {To}, C. and {Tremblay}, P.~E. and {Troxel}, M.~A. and {Tucker}, D.~L. and {Turner}, D.~J. and {Varga}, T.~N. and {Walker}, A.~R. and {Wechsler}, R.~H. and {Weller}, J. and {Wester}, W. and {Wilkinson}, R.~D. and {Yanny}, B. and {Zhang}, Y. and {Nikutta}, R. and {Fitzpatrick}, M. and {Jacques}, A. and {Scott}, A. and {Olsen}, K. and {Huang}, L. and {Herrera}, D. and {Juneau}, S. and {Nidever}, D. and {Weaver}, B.~A. and {Adean}, C. and {Correia}, V. and {de Freitas}, M. and {Freitas}, F.~N. and {Singulani}, C. and {Vila-Verde}, G. and {Linea Science Server}},
        title = "{The Dark Energy Survey Data Release 2}",
      journal = {\apjs},
         year = 2021,
        month = aug,
       volume = {255},
       number = {2},
          eid = {20},
        pages = {20},
          doi = {10.3847/1538-4365/ac00b3},
archivePrefix = {arXiv},
       eprint = {2101.05765},
 primaryClass = {astro-ph.IM},
       adsurl = {https://ui.adsabs.harvard.edu/abs/2021ApJS..255...20A}
}

@ARTICLE{Rozo:2016:RedmagicSV,
       author = {{Rozo}, E. and {Rykoff}, E.~S. and {Abate}, A. and {Bonnett}, C. and {Crocce}, M. and {Davis}, C. and {Hoyle}, B. and {Leistedt}, B. and {Peiris}, H.~V. and {Wechsler}, R.~H. and {Abbott}, T. and {Abdalla}, F.~B. and {Banerji}, M. and {Bauer}, A.~H. and {Benoit-L{\'e}vy}, A. and {Bernstein}, G.~M. and {Bertin}, E. and {Brooks}, D. and {Buckley-Geer}, E. and {Burke}, D.~L. and {Capozzi}, D. and {Rosell}, A. Carnero and {Carollo}, D. and {Kind}, M. Carrasco and {Carretero}, J. and {Castander}, F.~J. and {Childress}, M.~J. and {Cunha}, C.~E. and {D'Andrea}, C.~B. and {Davis}, T. and {DePoy}, D.~L. and {Desai}, S. and {Diehl}, H.~T. and {Dietrich}, J.~P. and {Doel}, P. and {Eifler}, T.~F. and {Evrard}, A.~E. and {Neto}, A. Fausti and {Flaugher}, B. and {Fosalba}, P. and {Frieman}, J. and {Gaztanaga}, E. and {Gerdes}, D.~W. and {Glazebrook}, K. and {Gruen}, D. and {Gruendl}, R.~A. and {Honscheid}, K. and {James}, D.~J. and {Jarvis}, M. and {Kim}, A.~G. and {Kuehn}, K. and {Kuropatkin}, N. and {Lahav}, O. and {Lidman}, C. and {Lima}, M. and {Maia}, M.~A.~G. and {March}, M. and {Martini}, P. and {Melchior}, P. and {Miller}, C.~J. and {Miquel}, R. and {Mohr}, J.~J. and {Nichol}, R.~C. and {Nord}, B. and {O'Neill}, C.~R. and {Ogando}, R. and {Plazas}, A.~A. and {Romer}, A.~K. and {Roodman}, A. and {Sako}, M. and {Sanchez}, E. and {Santiago}, B. and {Schubnell}, M. and {Sevilla-Noarbe}, I. and {Smith}, R.~C. and {Soares-Santos}, M. and {Sobreira}, F. and {Suchyta}, E. and {Swanson}, M.~E.~C. and {Thaler}, J. and {Thomas}, D. and {Uddin}, S. and {Vikram}, V. and {Walker}, A.~R. and {Wester}, W. and {Zhang}, Y. and {da Costa}, L.~N.},
        title = "{redMaGiC: selecting luminous red galaxies from the DES Science Verification data}",
      journal = {\mnras},
         year = 2016,
        month = sep,
       volume = {461},
       number = {2},
        pages = {1431-1450},
          doi = {10.1093/mnras/stw1281},
archivePrefix = {arXiv},
       eprint = {1507.05460},
 primaryClass = {astro-ph.IM},
       adsurl = {https://ui.adsabs.harvard.edu/abs/2016MNRAS.461.1431R}
}

@ARTICLE{Rykoff:2016:DESCatalog,
       author = {{Rykoff}, E.~S. and {Rozo}, E. and {Hollowood}, D. and
                 {Bermeo-Hernandez}, A. and {Jeltema}, T. and {Mayers}, J. and
                 {Romer}, A.~K. and {Rooney}, P. and {Saro}, A. and
                 {Vergara Cervantes}, C. and {Wechsler}, R.~H. and
                 {Wilcox}, H. and {Abbott}, T.~M.~C. and {Abdalla}, F.~B. and
                 {Allam}, S. and {Annis}, J. and others},
        title = "{The RedMaPPer Galaxy Cluster Catalog From DES Science Verification Data}",
      journal = {\apjs},
         year = 2016,
        month = may,
       volume = {224},
       number = {1},
          eid = {1},
        pages = {1},
          doi = {10.3847/0067-0049/224/1/1},
archivePrefix = {arXiv},
       eprint = {1601.00621},
 primaryClass = {astro-ph.CO},
       adsurl = {https://ui.adsabs.harvard.edu/abs/2016ApJS..224....1R}
}

@ARTICLE{Vanderlinde_2010,
   title={GALAXY CLUSTERS SELECTED WITH THE SUNYAEV-ZEL’DOVICH EFFECT FROM 2008 SOUTH POLE TELESCOPE OBSERVATIONS},
   volume={722},
   ISSN={1538-4357},
   url={http://dx.doi.org/10.1088/0004-637X/722/2/1180},
   DOI={10.1088/0004-637x/722/2/1180},
   number={2},
   journal={The Astrophysical Journal},
   publisher={American Astronomical Society},
   author={Vanderlinde, K. and Crawford, T. M. and de Haan, T. and Dudley, J. P. and Shaw, L. and Ade, P. A. R. and Aird, K. A. and Benson, B. A. and Bleem, L. E. and Brodwin, M. and Carlstrom, J. E. and Chang, C. L. and Crites, A. T. and Desai, S. and Dobbs, M. A. and Foley, R. J. and George, E. M. and Gladders, M. D. and Hall, N. R. and Halverson, N. W. and High, F. W. and Holder, G. P. and Holzapfel, W. L. and Hrubes, J. D. and Joy, M. and Keisler, R. and Knox, L. and Lee, A. T. and Leitch, E. M. and Loehr, A. and Lueker, M. and Marrone, D. P. and McMahon, J. J. and Mehl, J. and Meyer, S. S. and Mohr, J. J. and Montroy, T. E. and Ngeow, C.-C. and Padin, S. and Plagge, T. and Pryke, C. and Reichardt, C. L. and Rest, A. and Ruel, J. and Ruhl, J. E. and Schaffer, K. K. and Shirokoff, E. and Song, J. and Spieler, H. G. and Stalder, B. and Staniszewski, Z. and Stark, A. A. and Stubbs, C. W. and van Engelen, A. and Vieira, J. D. and Williamson, R. and Yang, Y. and Zahn, O. and Zenteno, A.},
   year={2010},
   month=Sept, pages={1180–1196} }

@ARTICLE{Jimeno:2018:redMaPPerGas,
       author = {{Jimeno}, Pablo and {Diego}, Jose M. and {Broadhurst}, Tom and
                 {De Martino}, Ivan and {Lazkoz}, Ruth},
        title = "{Planck/SDSS cluster mass and gas scaling relations for a volume-complete redMaPPer sample}",
      journal = {\mnras},
         year = 2018,
        month = jul,
       volume = {478},
       number = {1},
        pages = {638-650},
          doi = {10.1093/mnras/sty987},
archivePrefix = {arXiv},
       eprint = {1706.00395},
 primaryClass = {astro-ph.CO},
}

@ARTICLE{Popesso:2026:GasFraction,
       author = {{Popesso}, P. and {Biviano}, A. and {Marini}, I. and {Dolag}, K. and
                 {Vladutescu-Zopp}, S. and {Csizi}, B. and {Biffi}, V. and
                 {Lamer}, G. and {Robotham}, A. and {Bravo}, M. and
                 {Lovisari}, L. and {Ettori}, S. and {Angelinelli}, M. and
                 {Driver}, S. and {Toptun}, V. and {Dev}, A. and {Mazengo}, D. and
                 {Merloni}, A. and {Comparat}, J. and {Ponti}, G. and
                 {Mroczkowski}, T. and {Bulbul}, E. and {Grandis}, S. and {Bahar}, E.},
        title = "{The hot gas mass fraction in halos: From Milky Way-like groups to massive clusters}",
      journal = {\aap},
         year = 2026,
        month = mar,
       volume = {707},
          eid = {A362},
        pages = {A362},
          doi = {10.1051/0004-6361/202453256},
archivePrefix = {arXiv},
       eprint = {2411.16555},
 primaryClass = {astro-ph.GA},
}

@ARTICLE{Zenteno:2025:Derositas,
       author = {{Zenteno}, A. and {Kluge}, M. and {Kharkrang}, R. and {Hernandez-Lang}, D. and {Damke}, G. and {Saro}, A. and {Monteiro-Oliveira}, R. and {Carrasco}, E.~R. and {Salvato}, M. and {Comparat}, J. and {Fabricius}, M. and {Snigula}, J. and {Arevalo}, P. and {Cuevas}, H. and {Nilo Castellon}, J.~L. and {Ramirez}, A. and {V{\'e}liz Astudillo}, S. and {Landriau}, M. and {Myers}, A.~D. and {Schlafly}, E. and {Valdes}, F. and {Weaver}, B.~A. and {Mohr}, J.~J. and {Grandis}, S. and {Klein}, M. and {Liu}, A. and {Bulbul}, E. and {Zhang}, X. and {Sanders}, J.~S. and {Bahar}, Y.~E. and {Ghirardini}, V. and {Ramos}, M.~E. and {Balzer}, F.},
        title = "{The dynamical state of eROSITA clusters and its impact on the brightest cluster galaxy luminosity}",
      journal = {\aap},
         year = 2025,
        month = jun,
       volume = {698},
          eid = {A171},
        pages = {A171},
          doi = {10.1051/0004-6361/202452440},
archivePrefix = {arXiv},
       eprint = {2503.21066},
 primaryClass = {astro-ph.CO},
       adsurl = {https://ui.adsabs.harvard.edu/abs/2025A&A...698A.171Z}
}

@ARTICLE{Wu:2019:ClusterWLCov,
       author = {{Wu}, Hao-Yi and {Weinberg}, David H. and {Salcedo}, Andr{\'e}s N. and {Wibking}, Benjamin D. and {Zu}, Ying},
        title = "{Covariance matrices for galaxy cluster weak lensing: from virial regime to uncorrelated large-scale structure}",
      journal = {\mnras},
         year = 2019,
        month = dec,
       volume = {490},
       number = {2},
        pages = {2606-2626},
          doi = {10.1093/mnras/stz2617},
archivePrefix = {arXiv},
       eprint = {1907.06611},
 primaryClass = {astro-ph.CO},
       adsurl = {https://ui.adsabs.harvard.edu/abs/2019MNRAS.490.2606W}
}

@BOOK{Fisher:1925,
       author = {{Fisher}, Ronald A.},
        title = "{Statistical Methods for Research Workers}",
         year = 1925,
    publisher = {Oliver and Boyd},
      address = {Edinburgh},
       adsurl = {https://ui.adsabs.harvard.edu/abs/1925sma..book.....F}
}

@article{mclennan1925origin,
  author  = {McLennan, John C. and Shrum, Gordon M.},
  title   = {On the Origin of the Auroral Green Line in the Spectrum of the Night Sky},
  journal = {Proceedings of the Royal Society of London. Series A, Containing Papers of a Mathematical and Physical Character},
  volume  = {108},
  number  = {748},
  pages   = {501--512},
  year    = {1925},
  publisher={The Royal Society}
}

@article{Messier1781,
  author     = {Messier, Charles},
  title      = {{Catalogue des N{\'e}buleuses et des Amas d'{\'E}toiles}},
  journal    = {Connoissance des Temps, ou Connoissance des mouvemens c{\'e}lestes,
                pour l'ann{\'e}e bissextile 1784},
  year       = {1781},
  pages      = {227--267},
  note       = {Paris: Imprimerie Royale},
  adsbibcode = {1781cote.rept..227M},
  url        = {https://ui.adsabs.harvard.edu/abs/1781cote.rept..227M/abstract}
}

%%%%%%%%%%%%%%%%%%%%%%%%%%%%%%%%%%%%%%%%%%%%%%%%%%

%%%%%%%%%%%%%%%%% APPENDICES %%%%%%%%%%%%%%%%%%%%%

\appendix 

\section{Cluster Bayesian population model}\label{appx:tSZModel}

In Section \ref{sec:Validation:Scaling:SZ}, we perform a joint analysis of the richness and tSZ observables through a Bayesian population model. In this Appendix, we detail this model. We first write down the generating function of our cluster observables, starting from halo mass, $M$, and redshift, $z$. Our mass definition follows $\Mtwohc$ as noted above. The mean mass--observable relations are given by,
\begin{equation}\label{eq:lambdatrue_relation}
    \langle \ln \lambdatrue | M, z\rangle = \ln\bigg[A_\lambda \bigg(\frac{M}{3 \times 10^{14} M_\odot h^{-1}}\bigg)^{B_\lambda} \bigg(\frac{1 + z}{1 + 0.6}\bigg)^{C_\lambda}\bigg]
\end{equation}
\begin{equation}\label{eq:zeta_relation}
    \langle \ln \zeta | M, z\rangle = \ln\bigg[\gamma A_{\rm SZ} \bigg(\frac{M}{3 \times 10^{14}M_\odot h^{-1}}\bigg)^{B_{\rm SZ}} \bigg(\frac{E(z)}{E(0.6)}\bigg)^{C_{\rm SZ}}\bigg]
\end{equation}
\noindent
where $\lambdatrue$ and $\zeta$ are the true richness and SZ significance, respectively, of the cluster. The scaling  of $A_{\rm SZ}$ by $\gamma$ in Equation \eqref{eq:zeta_relation} allows for any (relative) spatial variation in noise levels of a given survey. We use two sub-regions for SPT (the SPT-SZ and SPT-ECS regions), and three sub-regions for ACT; see Appendix \ref{appx:SZdepth} for the definition of the latter regions. The intrinsic scatter and correlation between the true properties are captured by the covariance matrix, $\Sigma_{\rm multi}$, of the log-observables, 
\begin{equation}\label{eqn:cov_mat_relation}
    \begin{bmatrix}
    \sigma^2_{\ln \zeta} & \rho \sigma_{\ln\zeta}\sigma_{\ln \lambdatrue} \\
    \rho \sigma_{\ln\zeta}\sigma_{\ln \lambdatrue} & \sigma^2_{\ln\lambdatrue}
    \end{bmatrix}
\end{equation}
defined by the intrinsic scatter, $\sigma_{\ln \lambdatrue}$ and $\sigma_{\ln \zeta}$, and a correlation coefficient, $\rho$. For simplicity, we will refer to the two scatter parameters as $\sigma_\lambda$ and $\sigma_{\rm SZ}$. The likelihood of the two intrinsic observables is given by,
\vspace{10pt}
\begin{equation}
P\left(
    \begin{bmatrix}
    \ln \zeta \\
    \ln \lambdatrue
    \end{bmatrix}
\,\middle|\, M,z,\vec{p} \right)
=
\mathcal{N}\left(
    \begin{bmatrix}
    \langle \ln \zeta \rangle \\
    \langle \ln \lambdatrue \rangle
    \end{bmatrix},
\Sigma_{\rm multi}\right).
\end{equation}
where the vector $\vec{p}$ contains the scaling relation parameters in Equation \eqref{eq:lambdatrue_relation} and Equation \eqref{eq:zeta_relation}, as well as the covariance matrix parameters above. It also includes the cosmological parameters, which are all fixed to the values in \citet{Planck:2020:Cosmo} in our work.
The intrinsic cluster properties can be converted into the survey observables as,
\begin{equation}
    P(\ln\lambda | \ln\lambdatrue, M) = \mathcal{N} \left( \ln \lambdatrue , e^{-\langle \ln \lambdatrue | M\rangle} \right)
\end{equation}
\begin{equation}
    P(\xi | \zeta) = \mathcal{N} \left( \sqrt{\zeta^2 + 3}, 1 \right).
\end{equation}
For richness, the observed $\lambda$ is related to the intrinsic quantity, $\lambdatrue$, through a simple Gaussian distribution in log-space, with a variance given by the Poisson noise \citep{Costanzi:2021:DESxSPT}. For the tSZ, the observed detection significance, $\xi$, is biased high relative to the true significance, $\zeta$ \citep{Vanderlinde_2010}, as the observable $\xi$ is obtained by maximizing the detection statistic over two sky coordinates and a filtering scale, $\theta_c$. The maximization step increases the inferred SNR relative to the truth.

\begin{widetext}
Our likelihood is constructed using these two observables, $\xi$ and $\lambda$. As mentioned before, we use a Bayesian population likelihood. We start by writing down the differential number density of halos,
\begin{equation}
\frac{{\rm d}^4 n}{{\rm d}\lambda \, {\rm d} \xi \, {\rm d} z \, {\rm d} M} = \iint {\rm d} \lambdatrue \, {\rm d} \zeta \frac{{\rm d}n}{{\rm d} M}\, P \left( \ln\lambda \middle | \ln \lambdatrue \right) P \left( \xi \middle | \zeta \right) 
P\!\left(
\begin{bmatrix}
\ln \zeta \\
\ln \lambdatrue
\end{bmatrix}
\,\middle|\, M,z,\vec{p}
\right)
\end{equation}
\noindent
The abundance is marginalized over the intermediate latent variables we introduce in our model, so that it only depends on the observables, redshift, and mass. Here, ${\rm d}n/{\rm d}M$ is the halo mass function (HMF), which is evaluated using the model of \citet{Tinker:2008:HMF} at the fiducial cosmology.

Following \citet{bocquet2024sptclustersdeshst}, we assume the cluster population is governed by Poisson realizations. Our observable space is $\mathcal{O} =\left( \lambda,  \xi, z \right)$, and if we consider infinitesimally small bins, we obtain our final likelihood,
\begin{equation}\label{eqn:scaling:PoissonPopulation}
    \ln\mathcal{L}(\vec{p}) = \sum_i^{N_{\rm cl}} \ln\int_{M_{\text{min}}}^{M_{\text{max}}} {\rm d}M \left.\frac{{\rm d}^4 n(\vec{p})}{\rm d\lambda \, \rm d\xi \, {\rm d}z \, {\rm d}M}\right\vert_{\mathcal{O}_i} - \sum_j \Omega_j \int_{M_{\text{min}}}^{M_{\text{max}}} {\rm d}M \int_{z_{\text{min}}}^{z_{\rm max}} {\rm d}z \frac{{\rm d} V}{ {\rm d}z \,{\rm d}\Omega} T(z)\int_{\lambda_{\rm min}}^\infty {\rm d}\lambda \, \int_{\xi_{\rm min}}^\infty {\rm d}\xi  \, \frac{{\rm d}^4 n(\vec{p})}{{\rm d}\lambda \, {\rm d}\xi \, {\rm d}z \, {\rm d}M}
\end{equation}
\noindent
where $N_{\rm cl}$ is the number of clusters, $\Omega_j$ is the area of the $j^{\rm th}$ sub-region of the survey, $\frac{{\rm d}V}{ {\rm d}z \,{\rm d}\Omega}$ is the differential volume element as a function of redshift and solid angle (evaluated at the fiducial cosmology), $T(z)$ is the fractional change in effective area as a function of redshift (Appendix \ref{appx:SurveyArea}). The survey selection function is captured in the second term in Equation \eqref{eqn:scaling:PoissonPopulation}, through the limits on the integrals over $z, \lambda, \xi$. For DES and DECADE we set the minimum optical richness to $\lambda > 20$. For ACT, the minimum detection significance is set to $\xi > 4$, while for SPT we use $\xi>4.5$ for SPT-SZ and $\xi>5$ for SPT-ECS, following \citet{bocquet2024sptclustersdeshst}.
\end{widetext}

The likelihood in Equation \ref{eqn:scaling:PoissonPopulation} can be extended to multiple datasets, assuming they are distinct non-overlapping patches on the sky and therefore uncorrelated, with the following expression
\begin{equation}
    \ln \mathcal{L}_{\rm tot}(\vec{p}) = \sum_{i}^{N_{\rm survey}} \ln \mathcal{L}_{i}(\vec{p}).
\end{equation}
We sample the likelihood using Hamiltonian Monte-Carlo (HMC) sampling as implemented in \textsc{PyMC} \citep{PyMC}. The HMC is an attractive approach as we sample over the $\Mtwohc$, $\lambdatrue$, and $\zeta$ of individual objects in the joint dataset.\footnote{Equation \eqref{eqn:scaling:PoissonPopulation} is the marginalized form, which integrates over variables that cannot be directly observed, while our \textsc{PyMC} implementation uses an equivalent latent-variable formulation that samples these variables instead. The latter approach has the added advantage of providing posteriors on the latent variables, such as halo mass.} Thus, the number of sampling variables is roughly three times the number of objects in the analysis, and is $\mathcal{O}(10^3)$. Standard Monte-Carlo methods are inefficient at handling such a high-dimensional sampling problem, and so, we rely on the efficient HMC implementations in \textsc{PyMC} for this work.

\section{Impact of survey inhomogeneity}\label{appx:Inhomog}

As detailed extensively in \citet{paper1} and \citet{paper3}, the \decade data are more spatially inhomogeneous than DES; the survey depth, image quality and other survey properties (SP) vary significantly across the sky due to the survey being an amalgamation of many public datasets (rather than a singular, dedicated observing program). See Figure 1 in \citet{paper3} for an example of this inhomogeneity relative to that of DES. One may be concerned that the properties/calibrations of our cluster catalogs will incur some spatial variations as a result. \redmapper is particularly well-suited to tackle such effects as its cluster finding algorithm explicitly accounts for a variety of spatially varying observational conditions or SPs; particularly, the magnitude limit/depth of the sky (see Section \ref{sec:ClusterFinding:Redmapper} or \citealt{Rykoff:2016:redmapper}). We now empirically show that the redshift and mass calibration do not depend on these quantities.

Throughout, we use the following maps: airmass, seeing (or PSF), exposure time, sky brightness, magnitude limit, stellar density as determined by \textit{Gaia}, and Galactic extinction from \citet{Schlegel:1998:Dust}. The former five maps are available for all four bands, $griz$, and we use the $i$-band maps for simplicity. We have confirmed that our results are consistent if we use the maps from the other bands.

\subsection{Redshift and richness estimates}\label{appx:Inhomog:redshift}

\begin{figure*}
    \centering
    \includegraphics[width=2\columnwidth]{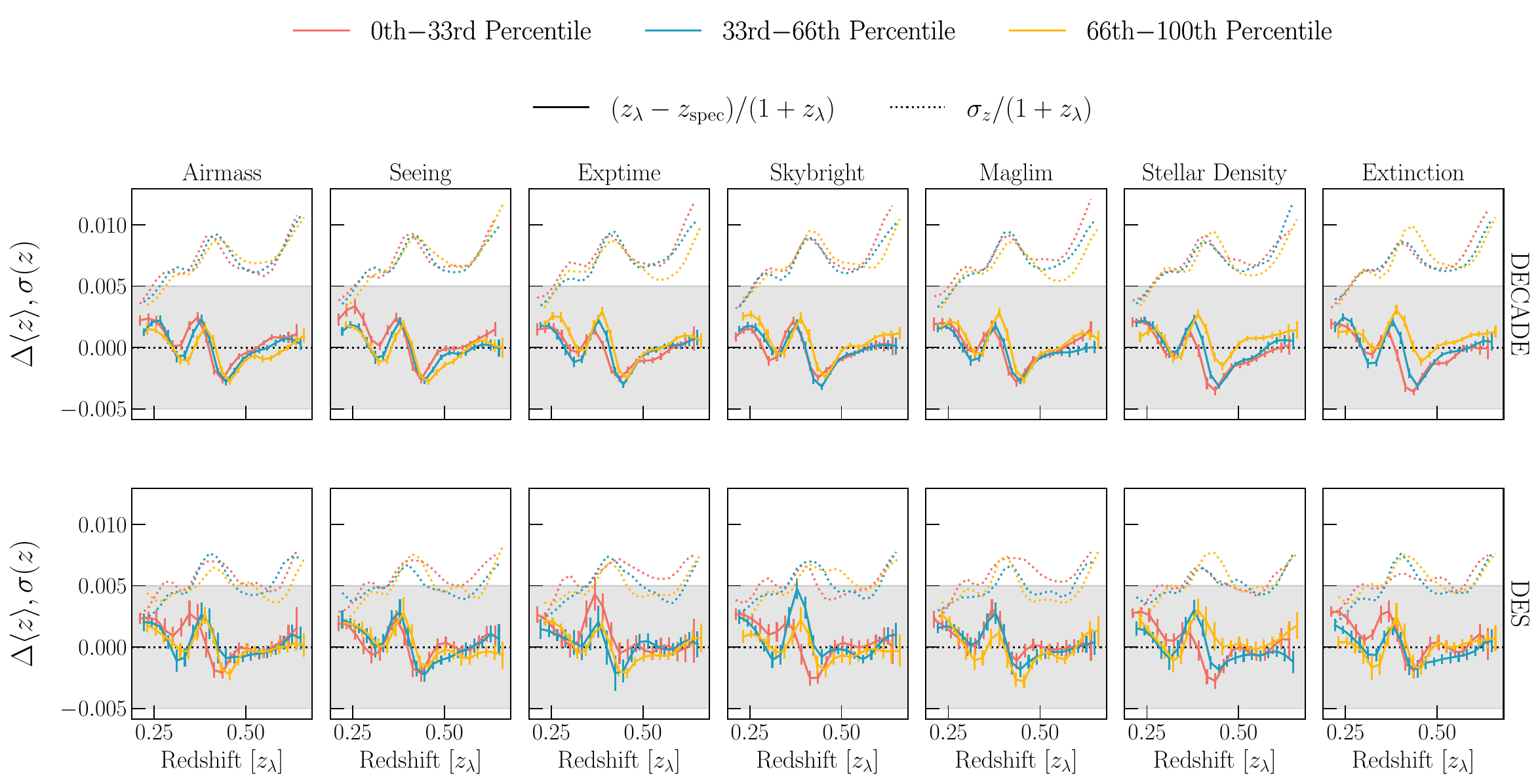}
    \caption{The inferred bias and scatter in the photometric redshifts from \redmapper (relative to matched spectroscopic estimates) when splitting the sample into three quantiles based on a survey observing property. Some quantiles show shifts in the redshift bias estimates, but these shifts are within uncertainties or within $\Delta = 0.001$ of each other. The scatter in the relation is similarly stable across the different quantiles. In summary, the \redmapper redshift estimates do not exhibit significant correlations with the SPs.}
    \label{fig:redshift_inhomog}
\end{figure*}

\begin{figure*}
    \centering
    \includegraphics[width=2\columnwidth]{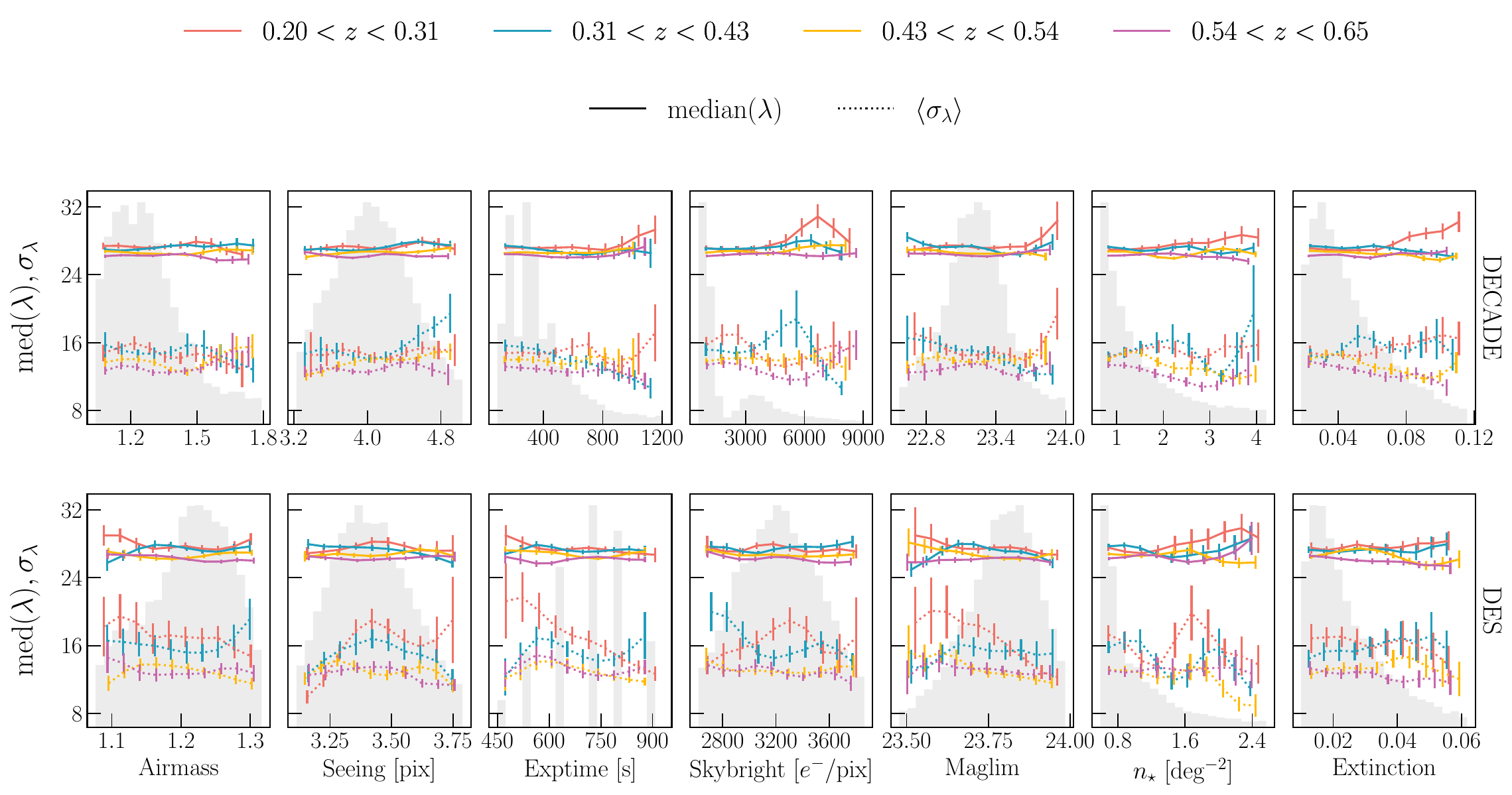}
    \caption{The median richness of the catalog, as a function of survey properties, in four redshift bins. We show the distribution of SPs in the background. The variations in the median richness are negligible, especially when compared to the richness errors reported by \redmapper. The largest deviations are $\Delta \lambda \approx \pm 5$, which is around one-third of the uncertainties quoted on $\lambda$. We conclude that the final richness estimates from \redmapper adequately remove correlations with SPs.}
    \label{fig:richness_inhomog}
\end{figure*}

We first check the sensitivity of our redshift estimates to inhomogeneity in the SP maps and present the results in Figure \ref{fig:redshift_inhomog}. We generate variants of the redshift accuracy panel from Figure \ref{fig:Catalog}, but after splitting the catalog into subsamples based on a given SP map. We perform the split in terciles, selecting pixels in a HEALPix map of $\texttt{NSIDE = 4096}$ based on the observing condition value in the pixel. We find no significant change in the redshift performance in any of the cases, for both DES and DECADE. In some cases, there are shifts of at most $\Delta \langle z \rangle \approx 0.002$, which are subdominant to the uncertainties. In most cases, the shifts are much smaller. The measured scatter between the \redmapper and spectroscopic redshift estimates is also stable across the quantiles. The same is observed when using the reported \redmapper uncertainties, but we omit it from Figure \ref{fig:redshift_inhomog} for brevity. In summary, the accuracy and precision of our redshift estimates are consistent across sky regions with significantly different SPs.

Next, we check the correlation of cluster richness with SPs. We bin clusters into 10 bins based on the SP value at their location, and compute the median richness as a function of the SP. We do this for four redshift bins. Figure \ref{fig:richness_inhomog} presents the trends in the median richness and the median of the uncertainties reported by \redmapper. The distributions of SP values are shown as the gray histograms in the background. In general, there is close to no trend between richness and SPs. We find changes of $\Delta \lambda = \pm 5$ in the tails of the SP distribution, whereas most of the sample exhibits shifts closer to $\Delta \lambda =  \pm 1$. The uncertainties in cluster richness are somewhat correlated with a few properties (particularly seeing and depth-related SPs such as exposure time and magnitude limit) but the trend is only weakly detected, is not present at all redshifts, and is well below the average richness scatter. In conclusion, we find the richness estimates have weak-to-no trends with the SP maps. We also note that the distribution of SP values, shown via the gray histograms in the panels, are qualitatively different between DECADE (top) and DES (bottom) due to the previously discussed differences in how the surveys are constructed (Section \ref{sec:data:photometric}).

\subsection{Weak lensing profiles}\label{appx:Inhomog:DeltaSigma}

In Figure \ref{fig:DeltaSigma} we presented our measurements for $\Delta \Sigma$, the lensing profiles around clusters in a given richness and redshift bin. This measurement can also incur systematic biases due to inhomogeneous SPs. The impact of lensing correlations was extensively checked in \citet{paper1} and \citet{paper3}, and these works found no trends between the lensing signal and SPs. However, the cluster lensing measurement probes somewhat different scales and can have different sensitivities. Thus, we redo this check in the specific analysis configuration used in this work.

For each bin in richness and redshift, we split the sample into halves based on some SP (a high and low variant), and measure the average lensing profile for each half. The statistical consistency between the two is reported in Table \ref{tab:DsigmaSys} as a set of $p$-values. We report results for three key SP maps --- magnitude limit, seeing, and Galactic extinction --- and note the others show similar results (but are omitted for the sake of brevity). All comparisons result in $p \gtrsim 0.01$, indicating statistical consistency. Thus, the lensing-based mass calibration of clusters can be performed across the whole survey at once and does not need to account for area-dependent deviations in the SP values.

\begin{table*}
    \centering
    \begin{tabular}{r@{\;}c@{\;}c@{\;}c@{\;}c|cccc|cccc|cccc}
    \hline
    \multirow{2}{*}{Survey property} & &  \multirow{2}{*}{Survey} & & \multirow{2}{*}{Obs.} & \multicolumn{4}{c}{$0.2 < z < 0.35$} & \multicolumn{4}{c}{$0.35 < z < 0.5$} & \multicolumn{4}{c}{$0.5 < z < 0.65$}\\
    & & & & & Bin 1 & Bin 2 & Bin 3 & Bin 4 & Bin 1 & Bin 2 & Bin 3 & Bin 4 & Bin 1 & Bin 2 & Bin 3 & Bin 4\\
    \hline
    \multirow{4}{*}{ Magnitude Limit } & \rdelim\{{4}{*} & \multirow{2}{*}{ DECADE } & \rdelim\{{2}{*} & $\lambda$ & 0.064 & 0.065 & 0.026 & 0.011 & 0.35 & 0.65 & 0.95 & 0.18 & 0.79 & 0.42 & 0.40 & 0.049 \\
     & & & & $\lambda_{\rm cyl}$ & 0.053 & 0.99 & 0.21 & 0.39 & 0.47 & 0.85 & 0.98 & 0.98 & 0.76 & 0.21 & 0.45 & 0.38 \\[5pt]
     &  & \multirow{2}{*}{ DES } & \rdelim\{{2}{*} & $\lambda$ & 0.79 & 0.47 & 0.19 & 0.55 & 0.87 & 0.093 & 0.62 & 0.78 & 0.35 & 0.57 & 0.57 & 0.055 \\
     & & & & $\lambda_{\rm cyl}$ & 0.50 & 0.62 & 0.090 & 0.072 & 0.67 & 0.36 & 0.38 & 0.28 & 0.072 & 0.54 & 0.10 & 0.35 \\[10pt]

    \multirow{4}{*}{ Seeing } & \rdelim\{{4}{*} & \multirow{2}{*}{ DECADE } & \rdelim\{{2}{*} & $\lambda$ & 0.68 & 0.95 & 0.86 & 0.35 & 0.065 & 0.48 & 0.77 & 0.23 & 0.17 & 0.073 & 0.46 & 0.98 \\
     & & & & $\lambda_{\rm cyl}$ & 0.78 & 0.92 & 0.81 & 0.32 & 0.010 & 0.32 & 0.42 & 0.45 & 0.048 & 0.70 & 0.14 & 0.82 \\[5pt]
     &  & \multirow{2}{*}{ DES } & \rdelim\{{2}{*} & $\lambda$ & 0.054 & 0.71 & 0.0053 & 0.86 & 0.23 & 0.19 & 0.41 & 0.68 & 0.17 & 0.74 & 0.086 & 0.35 \\
     & & & & $\lambda_{\rm cyl}$ & 0.030 & 0.59 & 0.40 & 0.41 & 0.72 & 0.84 & 0.43 & 0.14 & 0.17 & 0.94 & 0.062 & 0.012 \\[10pt]

    \multirow{4}{*}{ Galactic Extinction } & \rdelim\{{4}{*} & \multirow{2}{*}{ DECADE } & \rdelim\{{2}{*} & $\lambda$ & 0.39 & 0.21 & 0.95 & 0.38 & 0.78 & 0.22 & 0.055 & 0.14 & 0.27 & 0.82 & 0.24 & 0.64 \\
     & & & & $\lambda_{\rm cyl}$ & 0.13 & 0.52 & 0.54 & 0.032 & 0.25 & 0.19 & 0.29 & 0.036 & 0.73 & 0.71 & 0.074 & 0.093 \\[5pt]
     &  & \multirow{2}{*}{ DES } & \rdelim\{{2}{*} & $\lambda$ & 0.76 & 0.51 & 0.26 & 0.19 & 0.76 & 0.33 & 0.77 & 0.25 & 0.16 & 0.17 & 0.23 & 0.012 \\
     & & & & $\lambda_{\rm cyl}$ & 0.62 & 0.81 & 0.11 & 0.20 & 0.86 & 0.37 & 0.69 & 0.10 & 0.77 & 0.79 & 0.42 & 0.014\\
    \hline
    \end{tabular}
    \caption{The $p$-values indicating the consistency of $\Delta \Sigma$ profiles between two subsamples, per redshift and richness bin. The subsamples are defined as regions of the sky that have high/low values for a given SP. The uncertainties for each subsample are obtained from jackknife subsampling and are added in quadrature to obtain the uncertainties used in the final $p$-value estimate. There is good agreement in all cases, with $p$-values consistent to better than $3\sigma$, \ie $p > 0.0015$. We only show three observing conditions for brevity, and have found similar results for sky brightness, airmass, and stellar density. Note that this consistency persists even when binning our clusters by $\lambdacyl$ instead of $\lambda$.}
    \label{tab:DsigmaSys}
\end{table*}

\section{Comparisons to other richness estimators}\label{appx:RichnessCompare}

\subsection{Cylindrical richness}\label{appx:RichnessCompare:Cylinder}

\begin{figure}
    \centering
    \includegraphics[width=\columnwidth]{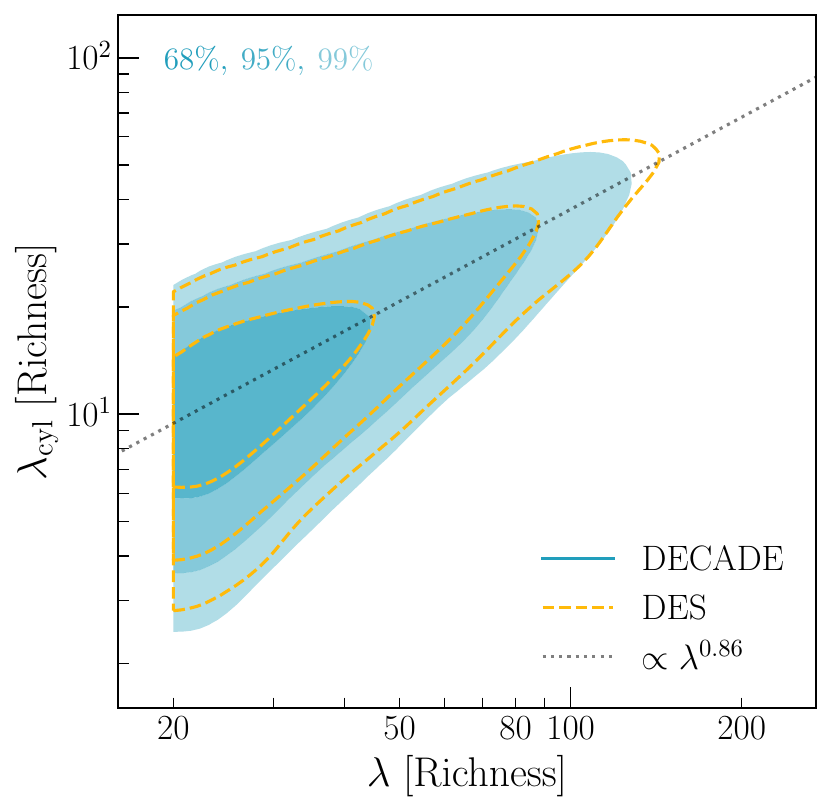}
    \caption{Similar to Figure \ref{fig:Xrayscaling} but for the $\lambdacyl$-$\lambda$ relation for DES and DECADE. The dotted line shows the slope of the best-fit linear model to the log-log relation.}
    \label{fig:CylindValid}
\end{figure}

We have utilized two different optical observables for our catalogs: the fiducial $\lambda$ (richness) defined in \redmapper, as well as the cylindrical richness, $\lambdacyl$, estimated using an LRG sample (see Section \ref{sec:ClusterFinding:cylinder}). In Figure \ref{fig:CylindValid}, we show the $\lambda$--$\lambdacyl$ scaling relation, as well as the level contours of the 2D distribution spanned by the two properties. We do so for both DES and DECADE, though the two are essentially identical in their behavior. We fit a line to the log-log relation and find $\lambdacyl \propto \lambda^{0.86}$, which is fairly close to the one-to-one relation. In particular, $\lambdacyl$ has lower values at all $\lambda$ in part because it only uses galaxies with $L > 0.5 \Lstar$, while \redmapper uses all galaxies down to $L > 0.2 \Lstar$.

\subsection{IDL \redmapper richness}\label{appx:RichnessCompare:IDL}

As noted in Section \ref{sec:MassCalibration:Scaling}, the IDL-based implementation of \redmapper has some differences from the Python-based version. As a result, the richness estimators are not identical. Therefore, the mass--richness relations from earlier works in the literature need not agree with the ones extracted in this work. To facilitate comparisons between these results, we build an approximate relation connecting the richness estimates from the two versions. We do so by abundance matching the clusters in our Python-based DES Y6 catalog to those of the IDL-based DES Y3 catalog\footnote{\url{https://des.ncsa.illinois.edu/releases/y3a2/Y3key-cluster}} from \citet{DES:2025:6x2pt_Y6}.

The abundance matching is performed in bins of redshift, spanning $0.1 < z < 0.8$ and of width $\Delta z = 0.1$. Within each redshift bin, we abundance match the clusters from the two catalogs to obtain a catalog with both IDL-based and Python-based richness estimates. We then fit a simple relation to these quantities, and find,
\begin{equation}\label{eqn:IDLvsPython}
   \frac{\lambda_{\rm Y6}}{\lambda_{\rm Y3}} = 0.84 + 0.09 \ln\bigg(\frac{\lambda_{\rm Y3}}{40}\bigg) + 0.17 (z - 0.4).
\end{equation}
where the fit is performed using the Y6 redshift estimate. The Python-based estimate is 16\% lower, on average, compared to the IDL-based one, and this ratio is fairly richness-independent, since the slope is around $0.09$. The abundance matching also traces any other differences between DES Y3 and DES Y6, including image data quality and modifications to the image processing pipelines, if they impact the richness estimates. Accounting for such changes is a benefit as it makes our comparisons in Section \ref{sec:MassCalibration:Scaling} more accurate.

Equation \eqref{eqn:IDLvsPython} is used to approximately convert the mass--richness relations derived from DES Y3 data into ones that reflect the updated richness estimator used in this work. Through this, we can easily compare our results to those from the literature. Equation \eqref{eqn:IDLvsPython} should not be considered a precise calibration of the differences between the estimators (such a calibration requires running the two estimators on the same truth catalog), and is instead an approximate, empirical calibration to facilitate general comparisons to existing results.

\section{Consistency with DES Year 3 data}\label{appx:Y3Compare}

\begin{figure}
    \centering
    \includegraphics[width=\columnwidth]{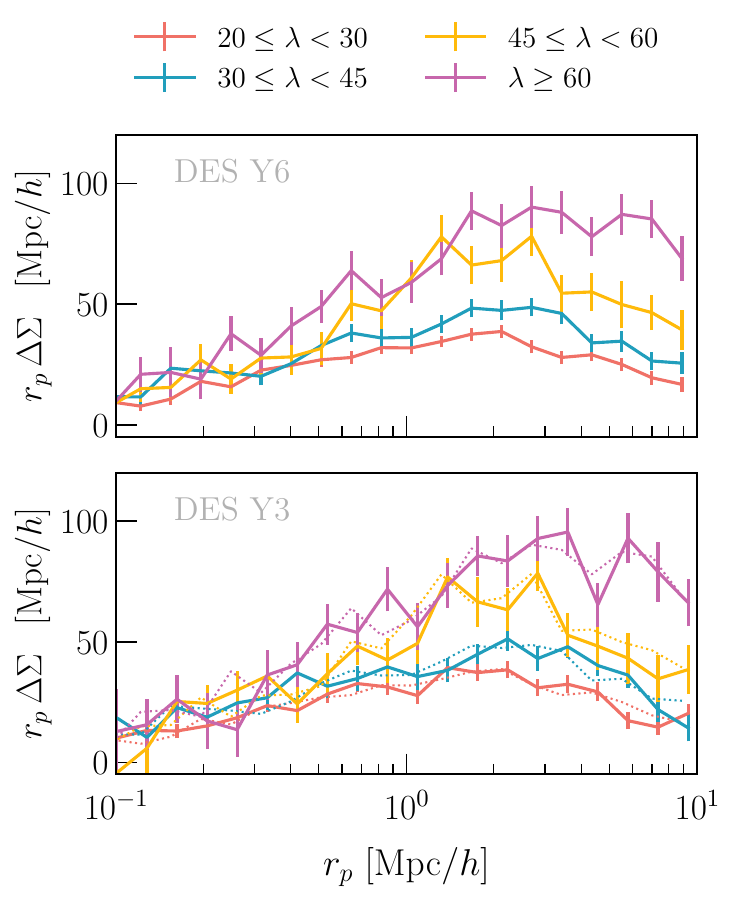}
    \caption{The lensing profiles of the clusters identified in DES Y6 data (top), compared to those found in DES Y3 (bottom), for the $0.2 < z < 0.35$ bin. In both cases, lensing is measured using the DES Y3 source-galaxy catalog. The richness bins for the DES Y3 sample are set by abundance matching to the bins defined for the DES Y6 sample. The bottom panel shows the Y6 profiles (from the top panel) as dotted lines to enable easier comparisons. The two sets of profiles are statistically consistent for all richness bins ($p > 0.1$).}
    \label{fig:Y3vsY6}
\end{figure}

In our main analysis above, we perform a number of validations of our \redmapper catalog by comparing the DES subset to the DECADE subset, and by cross-matching each to cluster measurements from the mm-wave and X-ray data. It is also worthwhile to compare our DES Y6-based catalog to that published from the DES Y3 data \citep{DES:2025:6x2pt_Y6}. However, as previously mentioned, the Y3 catalog used an earlier IDL-based version of \redmapper that did not have the same implementation choices as the latest Python-based version used here, so we do not see precise agreement in the richness estimates of the clusters from DES Y3 and DES Y6. A more practical comparison then is to check that the lensing profiles, which serve as a proxy for halo mass, are consistent for the two catalogs. This is precisely what we do here. Figure \ref{fig:Y3vsY6} shows the profiles for the first redshift bin, estimated using the same pipeline as Section \ref{sec:MassCalibration:WL}. In the range $0.1 \mpc/h < r_p < 10 \mpc/h$, the two sets of profiles are statistically consistent with $p > 0.1$ for all redshift and richness bins; we do not show the other redshift bins for brevity. We conclude that the DES \redmapper catalog produced here is consistent with that of DES Y3 \citep{DES:2025:6x2pt_Y6}.

\section{Effective area}\label{appx:SurveyArea}

As previously discussed, the quality of imaging data in a survey is not constant and has variations from image to image. These often emerge as spatial variations in the survey data quality that can imprint onto the objects detected/characterized in these images. We have discussed some of the potential effects of such variations on cluster finding and calibration, and validated above that these effects are minimal (see Appendix \ref{appx:Inhomog}). One other effect from such spatial variations is the effective survey area. For a given redshift, not all regions of a survey have the requisite data quality to detect clusters at high confidence. As a result, the effective area of the survey will have a redshift dependence. In an ideal survey, this curve is a Heaviside function whose value takes the area of the survey at low-enough redshifts and drops to zero at high redshifts (in practice, the transition would be smoothed over due to the effect of noise). In an inhomogeneous survey, the transition between these two regimes becomes wider.

\begin{figure}
    \centering
    \includegraphics[width=\columnwidth]{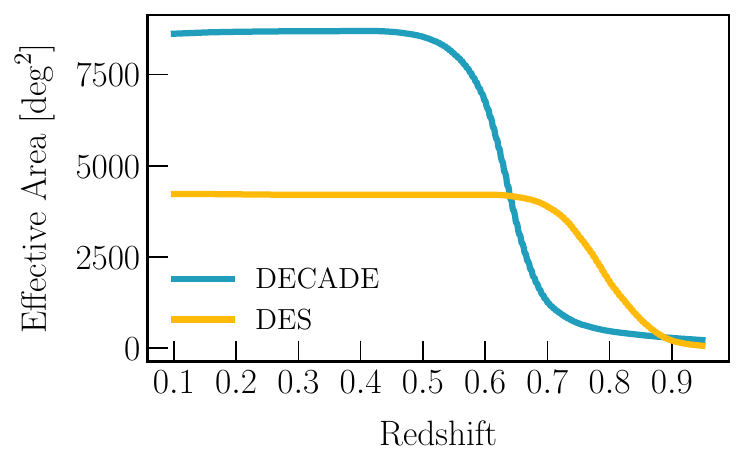}
    \caption{The effective area as a function of redshift, as estimated by \redmapper. The area goes down at higher redshifts as only subsets of the survey have enough depth to observe higher-redshift objects. The effective area in DECADE drops at lower redshifts than DES as the DECADE data are shallower.}
    \label{fig:EffArea}
\end{figure}

Figure \ref{fig:EffArea} shows the estimate for DECADE and DES, as computed within \redmapper using the same procedure employed for making random catalogs (Section \ref{sec:ClusterFinding:Redmapper}). All regions of DES Y6 detect clusters up to $z \approx 0.7$, whereas only a fraction of the DECADE footprint can detect clusters at that redshift. In our analysis of the cluster population (Section \ref{sec:Validation:Scaling:SZ} and Appendix \ref{appx:tSZModel}), we follow \citet{Costanzi:2021:DESxSPT}, who use cluster detections up to $z = 0.65$ but incorporate the effective area estimates into their modeling of the redshift evolution of the sample. The term, $T(z)$, in Equation \eqref{eqn:scaling:PoissonPopulation} accounts for this effect.

\section{Depth variations in mm-wave data}\label{appx:SZdepth}

As discussed in Section \ref{sec:data:multiwavelength} and later in Section \ref{sec:Validation:Scaling:SZ}, the mm-wave datasets used in this work do not have perfectly uniform noise properties across the sky. This is particularly important as those datasets use the signal-to-noise, $\xi$, as their observable. So, a spatially varying noise translates to a spatially varying $\xi-\Mtwohc$ relation. \citet{Bleem:2020:ECS} discuss in detail the noise properties of the different SPT subfields, including between the SPT-SZ and SPT-ECS surveys that we employ in this work. Table 1 in \citet{Bleem:2020:ECS} reports the relative noise levels across these various subfields, which is then utilized in models of the cluster scaling relations \citep{bocquet2024sptclustersdeshst}. In our work, we follow this approach but at a coarser-grained level. We split the data into two subsets --- the SPT-SZ and SPT-ECS regions --- and assume that the noise level within each survey is sufficiently uniform for the purposes of this analysis. Thus, we use a single $\gamma$ factor (see Equation \ref{eq:zeta_relation}) for the full SPT-ECS survey, and its value captures its noise level relative to that of SPT-SZ.

\begin{figure}
    \centering
    \includegraphics[width=\linewidth]{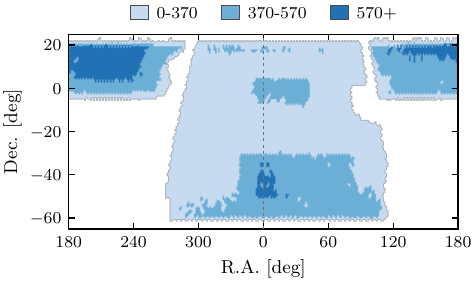}
    \caption{The ACT DR6 random catalog binned into three distinct regions after applying a HEALPix pixelization at $\texttt{NSIDE = 32}$. Regions are defined by the local density of random-catalog objects per pixel. See Figure \ref{fig:ACT_mocks_hist}.}
    \label{fig:ACT_mocks_map}
\end{figure}

\begin{figure}
    \centering
    \includegraphics[width=\linewidth]{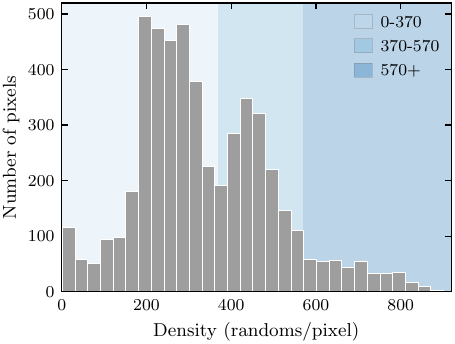}
    \caption{Density distribution of ACT random-catalog objects per
    HEALPix pixel, defined on a map of $\texttt{NSIDE = 32}$. Region boundaries are set at densities of
    $[0, 370)$, $[370, 570)$, and $[570, \infty)$, with amplitude rescaling factors $\gamma_0$, $\gamma_1$, and $\gamma_2$ assigned to each region, respectively. We fix $\gamma_1 = 1$ as our reference and rescale the remaining regions relative to it. Note that the shape of this distribution depends on the pixelization resolution. While we define regions by identifying subcomponents in the multimodal distributions, this choice can be varied. In the limit that there is no actual variation in the noise levels between these regions, the free parameters $\gamma_0$ and $\gamma_2$ will recover values consistent with $\gamma_i = 1$, at the cost of slightly weaker constraints on the scaling relation parameters.}
    \label{fig:ACT_mocks_hist}
\end{figure}

The ACT DR6 data has significantly more inhomogeneous noise properties as it reflects the variety of observing strategies within ACT over the past decade. So, we must marginalize over these relative noise factors of the survey. We take an approximate for in our implementation, where we first define a few distinct sub-regions of the ACT data. We then utilize the randoms catalog provided in the ACT DR6 public release to perform these field definitions, as the randoms trace the overall completeness of the survey. We generate a HEALPix count map of the random catalog, and use the counts per pixel to assign the pixels into a given sub-region. Figure \ref{fig:ACT_mocks_map} shows the spatial distribution of the different regions. Some structure can be clearly identified: the BOSS North and Deep 56 regions of ACT DR4 have lower noise relative to other parts of the footprint \citep[][see their Figure 3]{Madhavacheril:2020:ComptonY} as they have been observed for longer than the other parts of the survey. In general, the noise variations have a more complicated geometry \citep[][see their Figure 9]{Hilton:2026:ACTDR6}. We use the one-point distribution of the random-catalog object counts to determine the regions to split the data into. Figure \ref{fig:ACT_mocks_hist} shows this distribution, marking the three bins we use. The edges of the three regions are roughly the local-minima points demarcating adjacent components of a multimodal distribution.

\section{Posterior joint distributions}\label{appx:FullTriangle}

For completeness, we show the full posteriors for all parameters, and for all four combinations listed in Table \ref{tab:posterior_summary}. The different transparencies show the $1\sigma$ and $2\sigma$ regions. See Equations \eqref{eq:zeta_relation}, \eqref{eq:lambdatrue_relation}, and \eqref{eqn:cov_mat_relation} in Appendix \ref{appx:tSZModel} for more a detailed description of these parameters.

\begin{figure*}
    \centering
    \includegraphics[width=2\columnwidth]{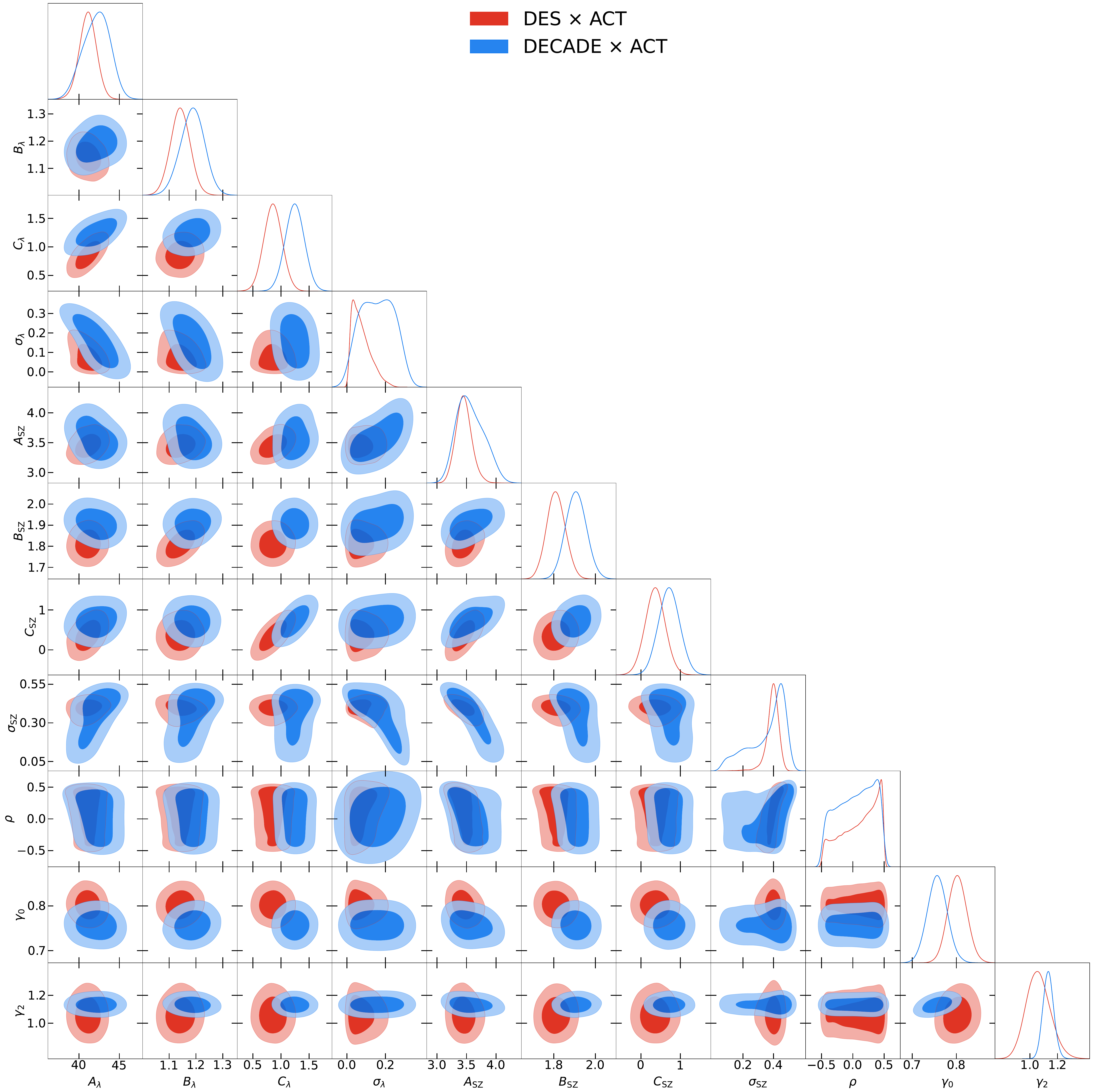}
    \caption{Posterior constraints on the joint scaling-relation parameters
    for the DES$\times$ACT (red) and DECADE$\times$ACT (blue) fiducial samples,
    $z \in [0.20, 0.65]$ and $\xi > 4$. Shaded contours show the 68\% and
    95\% credible regions. The constraints on $\gamma_0$ and $\gamma_2$ are expected to differ between the DES-based and DECADE-based analysis as they correspond to noise levels in different patches of the sky.}
    \label{fig:ACT_posteriors}
\end{figure*}

\begin{figure*}
    \centering
    \includegraphics[width=2\columnwidth]{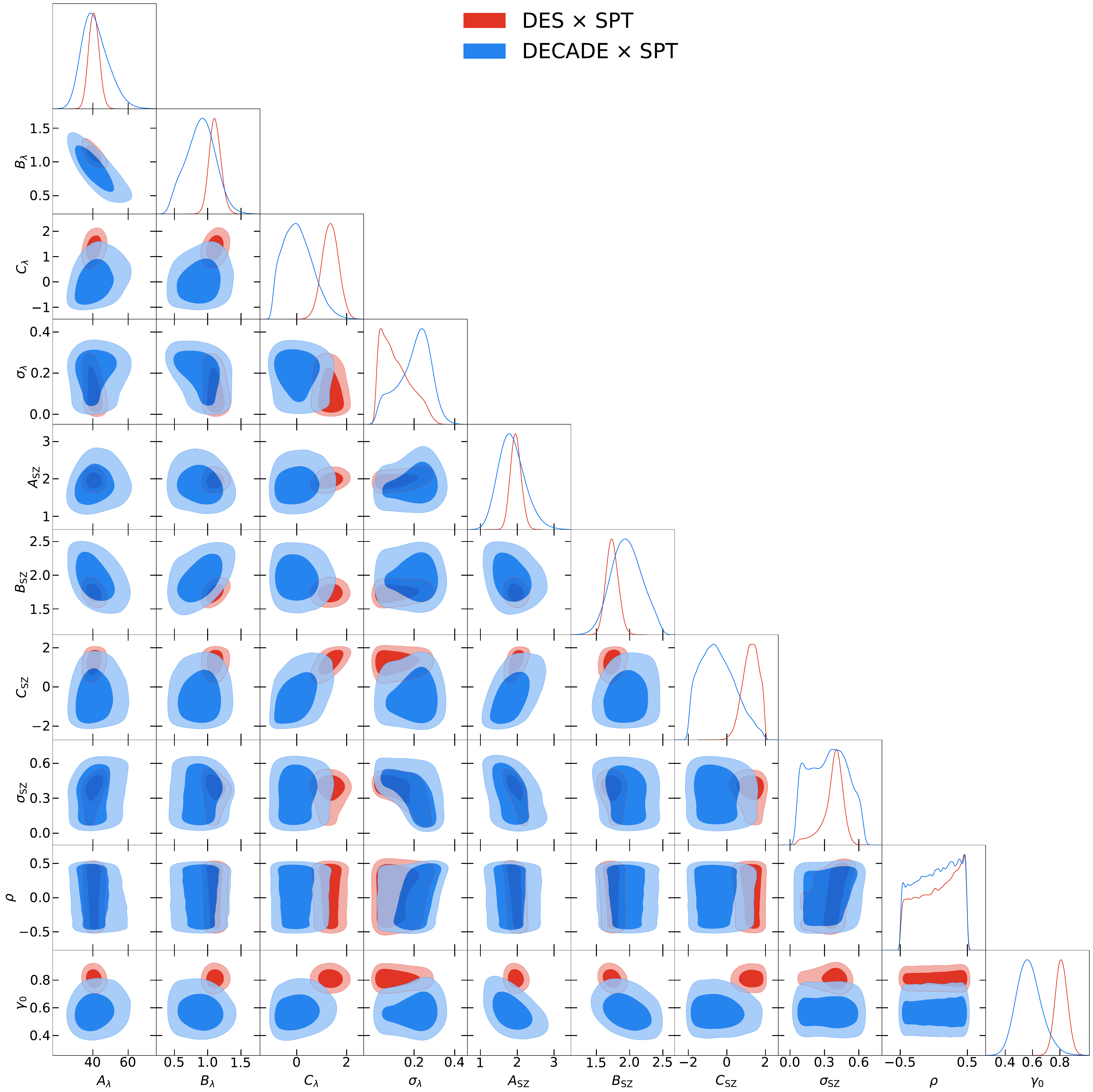}
    \caption{Posterior constraints on the joint scaling-relation parameters for the
    DES$\times$SPT (red) and DECADE$\times$SPT (blue) fiducial samples,
    $z \in [0.20, 0.65]$, $\xi > 5.0$ (ECS) and $\xi > 4.5$
    (SPT-SZ). Shaded contours show the 68\% and 95\% credible
    regions. The DECADE $\times$ SPT dataset has relatively few clusters, and the posteriors are less well constrained. The constraints on $\gamma_0$ are expected to differ between the DES-based and DECADE-based analysis as they correspond to noise levels in different patches of the sky.}
    \label{fig:SPT_posteriors}
\end{figure*}

% Don't change these lines
% \bsp	% typesetting comment
\label{lastpage}

\end{document}